\documentclass[12pt]{article}
\usepackage{amsmath}
\usepackage{graphicx,psfrag,epsf}
\usepackage{enumerate}
\usepackage{natbib}
\usepackage{url} 

\usepackage{amsthm}
\usepackage{amsmath}
\usepackage{amssymb}
\usepackage{comment}
\usepackage{multirow}
\usepackage{multicol}
\usepackage{dcolumn}
\usepackage[table]{xcolor}
\usepackage{booktabs}
\usepackage{rotating}
\usepackage{bbm}
\usepackage{natbib}
\usepackage[colorlinks,citecolor=blue,urlcolor=blue,filecolor=blue,backref=page]{hyperref}
\usepackage{graphicx}
\usepackage{xr}
\usepackage{float}
\usepackage{makecell}
\usepackage{caption}
\usepackage{subcaption}
\usepackage{setspace}

\setcitestyle{round}

\numberwithin{equation}{section}
\theoremstyle{plain}
\newtheorem{thm}{Theorem}[section]

\newtheorem{definition}{Definition}[section]

\DeclareMathOperator{\diag}{diag}

\newcommand\dd{\mathop{}\!\mathrm{d}}

\newcommand{\normal}{\mathsf{N}}
\newcommand{\DP}{\mathsf{DP}}

\newcommand{\bet}{\mathsf{beta}}

\newcommand{\E}{\mathsf{E}}
\renewcommand{\Pr}{\mathsf{Pr}}

\newcommand{\bfbeta}{\boldsymbol{\beta}}
\newcommand{\bfgamma}{\boldsymbol{\gamma}}

\newcommand{\bfepsilon}{\boldsymbol{\epsilon}}
\newcommand{\bfxi}{\boldsymbol{\xi}}

\newcommand{\bfSigma}{\boldsymbol{\Sigma}}
\newcommand{\bfGamma}{\boldsymbol{\Gamma}}

\newcommand{\bfOmega}{\boldsymbol{\Omega}}
\newcommand{\bfLambda}{\boldsymbol{\Lambda}}
\newcommand{\bfzero}{\boldsymbol{0}}
\newcommand{\bfones}{\boldsymbol{1}}

\newcommand{\bfy}{\boldsymbol{y}}
\newcommand{\bfx}{\boldsymbol{x}}
\newcommand{\bfg}{\boldsymbol{g}}
\newcommand{\bfX}{\boldsymbol{X}}
\newcommand{\bfG}{\boldsymbol{G}}
\newcommand{\bfI}{\boldsymbol{I}}
\newcommand{\bm}{ {\boldsymbol m} }
\newcommand{\bV}{ {\boldsymbol V} }

\newcommand{\reals}{\mathbb{R}}
\newcommand{\naturals}{\mathbb{N}}
\newcommand{\indic}{\mathbb{I}}

\newcommand{\bgam}{\bfgamma}
\newcommand{\betag}{\bfbeta_{\bfgamma}}

\newcommand{\blind}{1}

\begin{document}

\def\spacingset#1{\renewcommand{\baselinestretch}%
{#1}\small\normalsize} \spacingset{1}


\if1\blind
{
  \title{\bf Dirichlet process mixtures of block $g$ priors for model selection and prediction in linear models}
  \author{Anupreet Porwal\\
    Google Inc.\\
    and \\
    Abel Rodriguez\hspace{.2cm} \\
    Department of Statistics, University of Washington}
  \maketitle
} \fi

\if0\blind
{
  \bigskip
  \bigskip
  \bigskip
  \begin{center}
    {\LARGE\bf Dirichlet process mixtures of block $g$ priors for model selection and prediction in linear models}
\end{center}
  \medskip
} \fi

\begin{abstract}
This paper introduces Dirichlet process mixtures of block $g$ priors for model selection and prediction in linear models. These priors are extensions of traditional mixtures of $g$ priors that allow for differential shrinkage for various (data-selected) blocks of parameters while fully accounting for the predictors' correlation structure, providing a bridge between the literatures on model selection and continuous shrinkage priors. We show that Dirichlet process mixtures of block $g$ priors are consistent in various senses and, in particular, that they avoid the conditional Lindley ``paradox'' highlighted by \cite{som2016conditional}. 
Further, we develop a Markov chain Monte Carlo algorithm for posterior inference that requires only minimal \textit{ad-hoc} tuning.  Finally, we investigate the empirical performance of the prior in various real and simulated datasets.  In the presence of a small number of very large effects, Dirichlet process mixtures of block $g$ priors lead to higher power for detecting smaller but significant effects without only a minimal increase in the number of false discoveries. 
\end{abstract}

\noindent%
\vfill

\newpage
\spacingset{1.9} 
\addtolength{\textheight}{.5in}%

\section{Introduction}


Model selection and model averaging are foundational tasks in statistics and machine learning.  Associated Bayesian procedures typically rely on the computation of Bayes factors and posterior model probabilities, whose properties are heavily dependent of the choice of priors associated with the parameters of each model.  This feature makes model selection and model averaging tasks difficult in situations where standard ``objective'' or ``default'' priors (such as the reference prior, \citealp{berger2009formal}) are improper.  This is true even in well-studied settings, such as Gaussian linear models.

The literature on noninformative priors and default Bayes factors for model selection for (generalized) linear models is extensive.  Examples include $g$-priors \citep{Zellner1986}, mixtures of $g$-priors \citep{zellner1980posterior,liang2008mixtures}, unit information priors \citep{kass1995reference}, intrinsic Bayes factors \citep{berger1996linear}, non-local priors \citep{johnson2010use,johnson2012bayesian} and power-expected-posterior priors \citep{fouskakis2015power,porwal2023laplace}, among other approaches.  See \cite{Forte2018} and \cite{consonni2018prior} for recent reviews.

\cite{bayarri2012criteria} describes a series of desiderata for default priors used for model selection and model averaging, with a particular focus on problems involving multiple linear regression.  These include various forms of consistency and invariance, as well as predictive matching.  More recently,  \cite{som2014paradoxes} and \cite{som2016conditional} suggested  additional criteria related to the behavior of the Bayes factor as a subset of the significant coefficients grow to infinity.  This setting is important because it serves as a proxy for situations in which effects sizes vary dramatically across covariates.  Such situations arise often in practice, and they are arguably the kind of problem in which well-designed statistical methods can make a real difference. 
\cite{som2014paradoxes} and \cite{som2016conditional} show that mixtures of $g$ priors \citep{liang2008mixtures} fail to satisfy their new criteria (a behavior they call \textit{the conditional Lindley paradox}), and introduce mixtures of block $g$ priors, which address the issue by  assigning different shrinkage parameters to preselected groups of coefficients.

In the absence of prior information to drive the \textit{a priori} selection of the blocks, the methodology of \cite{som2014paradoxes} is difficult to implement in practice. Furthermore, the prior introduced in \cite{som2014paradoxes} assumes that the blocks of coefficients are independent a priori.  When there is strong colinearity between covariates associated with ``large'' and ``small'' coefficients, the independence assumption can lead to loss of efficiency.  Our first contribution in this paper is to develop Dirichlet process (DP) mixtures of block $g$ priors that allow for differential shrinkage across coefficients while fully accounting for the observed correlations among predictors and treating the blocks of covariates as an unknown parameter that must be inferred from the data.  Similar approaches have been suggested in the literature at least as early as in \cite{liang2008mixtures} but, to the best of our knowledge, they have not been pursued before, perhaps because of perceived computational challenges.

Because of our focus on differential shrinkage, the literature on continuous shrinkage priors is also relevant to our discussion. 
Examples of continuous shrinkage priors include the Student $t$-prior \citep{tipping2001sparse}, the Bayesian Lasso \citep{park2008bayesian, hans2009bayesian}, the Horseshoe prior \citep{carvalho2010horseshoe}, the Normal-Gamma prior \citep{griffin2005alternative,brown2010inference}, { semiparametric multiple-shrinkage priors \citep{maclehose2010bayesian}}, the Bayesian adaptive Lasso \citep{leng2014bayesian}, the Dirichlet-Laplace prior \citep{bhattacharya2015dirichlet}, global-local shrinkage priors \citep{polson2012local}, the Beta-prime prior \citep{bai2018beta}, the Horseshoe-pit prior \citep{denti2021horseshoe}, the group Inverse-Gamma Gamma prior of \cite{boss2023group}, and global-local-tail priors \citep{lee2020tail} among others.  Continuous shrinkage priors tend to have computational advantages, can be connected to penalized likelihood methods, and are very effective in predictive settings.  However, because they place  probability zero on any one value of the parameter space, variable selection can be performed only by either looking at the coverage of posterior credible intervals or by thresholding the posterior distributions of the coefficients (e.g., see \citealp{li2017variable}). Both of these procedures tend to work best in settings where enough prior information is available to establish practical significance.  For this reason, the literature on continuous shrinkage priors is often considered as distinct from that on priors for model selection.  A second contribution of this paper is to show that DP mixtures of $g$ priors provide a unifying framework for these two strands of the literature, with canonical methods in each of the two corresponding to special cases of ours.

The remainder of the paper is organized as follows.  Section \ref{se:motivation} introduces our notation and reviews the conditional Lindley paradox. Section \ref{sec:DPblock} introduces our proposed methodology and reviews its connections with the broader literature.  In Section \ref{sec:properties}, we investigate the properties of the prior and the associated Bayes factors, with a particular emphasis on the criteria introduced in \cite{bayarri2012criteria} and \cite{som2016conditional}.  Section \ref{sec:computation} discusses the computational implementation of our model.
Section \ref{sec:simulations} and Section \ref{sec:realdata} illustrate the performance of our methodology in both simulated and real datasets.  Finally, Section \ref{sec:discussion} discusses future directions for research.

\section{Motivation:  Bayesian variable selection and mixtures of $g$-priors}\label{se:motivation}

Consider a collection of linear models for the observed response vector $\bfy = (y_1, \ldots, y_n)^T$ based on the $n \times p$ (centered) design matrix $\bfX$. 
The collection of models is indexed by the binary vector $\bfgamma = (\gamma_1, \ldots, \gamma_p)$, $\gamma_j \in \{ 0,1 \}$, so that 
$$
\mathcal{M}_{\bfgamma} : \bfy = \bfones_n \beta_0 + \bfX_{\bfgamma} \bfbeta_{\bfgamma} + \bfepsilon ,
$$
where $\bfepsilon \sim \normal_n(\bfzero, \sigma^2\bfI_n)$, the $n$-th variate normal distribuion with mean $\bfzero$ and covariance matrix proportional to the $n \times n$ identity matrix $\bfI_n$, $
\bfones_n$ is the $n$-dimensional vector of ones, $\beta_0$ is an unknown intercept, $\bfX_{\bfgamma}$ denotes the submatrix of $\bfX$ consisting on the columns for which $\gamma_j = 1$, $\bfbeta$ is the vector of unknown regression coefficients, and $\bfbeta_{\bfgamma}$ is the subvector of $\bfbeta = (\beta_1, \ldots, \beta_p)^T$ corresponding to the entries for which $\gamma_j =1$.

We are interested in model comparison problems among these $2^p$ models, as well as estimation and prediction under model uncertainty.  The classical Bayesian solution to these problems involves the computation of Bayes factors of the form 
$$
BF_{\bfgamma, \bfgamma'} (\bfy)= \frac{\int f(\bfy \mid \beta_0, \bfbeta_{\bfgamma}, \sigma^2, \bfgamma) f(\beta_0, \bfbeta_{\bfgamma}, \sigma^2 \mid \bfgamma) \dd \beta_0 \dd \bfbeta_{\bfgamma} \dd \sigma^2}{\int f(\bfy \mid \beta_0, \bfbeta_{\bfgamma'}, \sigma^2, \bfgamma') f(\beta_0, \bfbeta_{\bfgamma'}, \sigma^2 \mid \bfgamma') \dd \beta_0 \dd \bfbeta_{\bfgamma'} \dd \sigma^2}
$$
for an appropriate model-specific prior $f(\beta_0, \bfbeta_{\bfgamma}, \sigma^2 \mid \bfgamma)$, which 
is often factorized as 
$f(\beta_0, \bfbeta_{\bfgamma}, \sigma^2 \mid \bfgamma) = f(\beta_0, \sigma^2) f(\bfbeta_{\bfgamma} \mid \sigma^2, \bfgamma)$.
The parameters $(\beta_0, \sigma^2)$ are usually assigned the reference prior $f(\beta_0, \sigma^2) \propto \frac{1}{\sigma^2}$ (e.g., see \citealp{berger1998bayes}).  A common choice for $f(\bfbeta_{\bfgamma} \mid \sigma^2, \bfgamma)$ is the so-called mixture of $g$-priors \citep{liang2008mixtures}
$$
f(\bfbeta_{\bfgamma} \mid \sigma^2, \bfgamma) = \int
\phi \left( \bfbeta_{\bfgamma} \mid \bfzero, g\sigma^2 \left\{ \bfX^T_{\bfgamma}\bfX_{\bfgamma} \right\}^{-1} \right)
f(g \mid \bfgamma)\dd g ,
$$
where $\phi$ denotes the density of the multivariate normal distribution and $f(g \mid \bfgamma)$ is a suitable hyperprior. 
When $f(g \mid \bfgamma)$ is chosen carefully (e.g., an appropriately scaled member of the Compound Confluent Hypergeometric distribution introduced in   \citealp{gordy1998generalization}), and under mild regularity conditions, 
Bayes factors based on mixtures of $g$ priors have various appealing theoretical properties like model selection consistency and information consistency \citep{liang2008mixtures,bayarri2012criteria}.

\begin{figure}[hbt!]
    \centering
    \includegraphics[width = 0.75\linewidth]{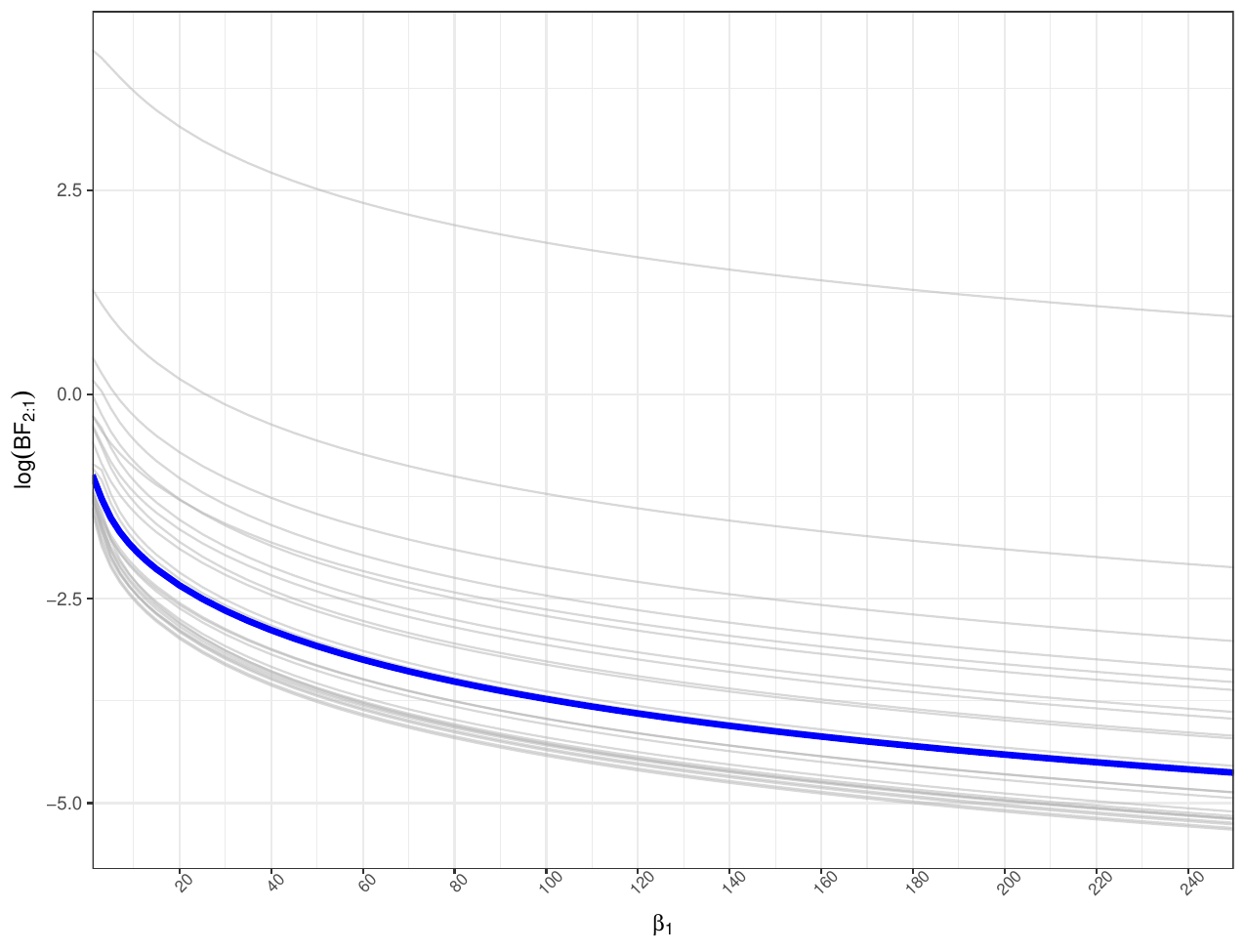}
    \caption{Empirical illustration of the conditional Lindley paradox.  Thin grey lines correspond to 100 simulated datasets, while the thick blue line corresponds to the average.}
\label{fig:clphg}
\end{figure}

In spite of these strong theoretical guarantees, procedures based on mixtures of $g$ priors do suffer from some undesirable properties.  For example,  \cite{som2014paradoxes} and \cite{som2016conditional} showed that Bayes factor based on mixtures of $g$ priors suffer from the \textit{conditional Lindley paradox}.  Roughly speaking, this ``paradox'' states that, when comparing nested models, if at least one of the regression coefficients common to both models is large relative to other coefficients present only in the bigger model, the Bayes factor will place too much weight on the smaller model irrespective of the data generating model. 
Consider the two models,
\begin{align}\label{eq:clpmod}
    \mathcal{M}_{0}: & \bfy = \boldsymbol{1}_n\beta_0+\bfX_{1}\bfbeta_1+\bfepsilon, \quad   \mathcal{M}_{a}: \bfy = \boldsymbol{1}_n\beta_0 + \bfX_{1}\bfbeta_1+\bfX_{2}\bfbeta_2+\bfepsilon,
\end{align}
%
where $\bfX_1$ and $\bfX_2$ are $n\times p_1$ and $n\times p_2$ dimensional matrices such that $\bfX_1^T\bfX_2 = \bfzero$, $\bfbeta_1$ and $\bfbeta_2$ are $p_1$ and $p - p_1$ dimensional vectors, and $\bfepsilon$ is the observational noise.  Further, for fixed $n$, $p_1$, $p$, $\bfX_1$, $\bfX_2$, $\beta_0$, $\bfbeta_2 \ne \bfzero$ and $\bfepsilon$, consider a sequence of vectors $\{ \bfbeta_1(N) : N \in \naturals\}$ and the associated sequence $\{ \bfy(N) : N \in \naturals\}$ such that $\bfy(N) = \bfones_n\beta_0 + \bfX_{1}\bfbeta_1(N)+\bfX_{2}\bfbeta_2+\bfepsilon$.  
\cite{som2016conditional} showed that, if $\left\| \bfbeta_1 (N) \right\| \rightarrow \infty$ as $N\rightarrow \infty$, then, for the Bayes Factor {for comparing $\mathcal{M}_a$ and $\mathcal{M}_0$,} $BF_{a,0}(\bfy)$, based on the hyper-$g/n$ distribution $f(g) = (1 + g/n)^{-a/2}$ \citep{liang2008mixtures}, we have $BF_{a,0} \left(\bfy(N)\right)\rightarrow 0$, irrespective of $\bfX_1$, $\bfX_2$, $\beta_0$, $\bfbeta_2$ and $\epsilon$.  To illustrate the paradox, we present in Figure \ref{fig:clphg} the behavior of $\log BF_{a,0}\left(\bfy(N)\right)$ for 100 randomly constructed triads $(\bfX_1, \bfX_2, \bfepsilon)$ where $n=100$, $p=2$, $p_1=1$, $\beta_0=0.5, \beta_2=1$.  We see that, in every case, $\log BF_{a,0}\left(\bfy(N)\right)$ seems to decrease towards $-\infty$.    

The conditional Lindley paradox is a consequence of the use of a common shrinkage factor $g$ and cannot be solved through alternative choices {of the prior on a single shrinkage coefficient $g$}.  This is because, as some coefficients grow, the estimate of the common $g$ also must grow.  The result is that small but non-zero coefficient end up being shrunk towards zero.  
To remedy this, \cite{som2014paradoxes} propose priors that allow for different shrinkage coefficients for various blocks of parameters. In the case of $\mathcal{M}_0$ and $\mathcal{M}_a$ above, these are
\begin{align}\label{eq:sompriormodel0}
    \bfbeta_1 \mid g, \sigma^2 , \mathcal{M}_0 \sim \normal \left( \bfzero , \sigma^2 g \left\{\bfX_1^T\bfX_1\right\}^{-1}  \right)
\end{align}
and
\begin{align}\label{eq:sompriormodel1}
\begin{pmatrix}
    \bfbeta_1 \\
    \bfbeta_2    
\end{pmatrix} \mid g_1, g_2, \sigma^2 , \mathcal{M}_a \sim \normal \left( \begin{pmatrix}
    \bfzero \\
    \bfzero
\end{pmatrix}, \sigma^2
\begin{pmatrix}
    g_1 \left\{ \bfX_1^T\bfX_1 \right\}^{-1} & \bfzero \\
    \bfzero & g_2 \left\{ \bfX_2^T\bfX_2 \right\}^{-1}
\end{pmatrix}
\right) ,
\end{align}
where $g$, $g_1$ and $g_2$ are independent and identically distributed (i.i.d.), e.g., from a hyper-$g/n$ distribution.  \cite{som2016conditional} showed that, under this block $g$-prior, the limit of $BF_{1,0}\left(\bfy(N)\right)$ as $N$ grows has a strictly positive lower bound, therefore avoiding the conditional Lindley paradox.  Note, however, that implementing this strategy requires that we specify up front which groups of parameters will be assigned a common shrinkage parameter.  This is problematic because the structure of the blocks can have a very big impact on the performance of the methods.  Indeed, if  at least one of the components of $\bfbeta_2$  goes to infinite as well, so that $0 < \lim_{N \to \infty} \frac{ \| \bfbeta_1\|^2}{ \| \bfbeta_2\|^2} = d < \infty$
, then $\lim_{N \to \infty} BF_{a,0}(\bfy(N)) = 0$.


\section{A new class of priors:  Dirichlet process mixtures of block $g$ priors}\label{sec:DPblock}

In this paper we consider mixtures of priors of the form
\begin{align}\label{eq:manygprior}
\bfbeta_{\bfgamma} \mid g_1, \ldots, g_{p_{\bfgamma}}, \sigma^2, \bfgamma \sim \normal \left(\bfzero, \sigma^2 \bfG_{\bfgamma}^{1/2} \bfSigma_{\bfgamma} \bfG_{\bfgamma}^{1/2} \right) ,
\end{align}
where $p_{\bfgamma} = \sum_{j=1}^{p} \gamma_j$, $\bfSigma_{\bfgamma}$ is a known covariance matrix that might depend on model $\bfgamma$, $\bfG^{1/2}_{\bfgamma} = \diag\{ g_1^{1/2}, \ldots, g_{p_{\bfgamma}}^{1/2} \}$, and $g_1, \ldots, g_{p_{\bfgamma}}$ are identically distributed. 
The associated marginal likelihood conditional on $\bfgamma$ and $g_1, \ldots, g_{p_{\bfgamma}}$ is given by:
\begin{align}\label{eq:marlikcond}
    f(\bfy \mid \bfgamma,  g_1, \ldots, g_{p_{\bfgamma}}) = 
    \frac{\Gamma\left(\frac{n-1}{2} \right)}{ \pi^{\frac{n-1}{2}} \sqrt{n}} \left|  \bfOmega_{\bfgamma} \right|^{-1/2}
    \left[ \bfy^T \bfOmega_{\bfgamma}^{-1} \bfy - n \bar{\bfy}^2 \right]^{-\frac{n-1}{2}}.
\end{align}
where $\bfOmega_{\bfgamma} = \bfI_n + \bfX_{\bfgamma}\bfG_{\bfgamma}^{1/2}  \bfSigma_{\bfgamma} \bfG_{\bfgamma}^{1/2} \bfX_{\bfgamma}^T$ (see Section \ref{ap:eq:marlikcond} of the supplementary materials).
The differential-shrinkage $g$ prior can be obtained by setting $\bfSigma_{\bfgamma} = \left\{ \bfX_{\bfgamma}^T\bfX_{\bfgamma}\right\}^{-1}$. Indeed, note that the standard $g$ prior is then obtained by further setting $g_1 = g_2 = \cdots = g_{p_{\bfgamma}} = g$.

A natural approach to modeling the $g_j$s is to assign them a parametric family that is flexible enough to encompass various tail behaviors.  One example is 
\begin{align}\label{eq:betaprime}
f(g \mid \tau^2, a, b) &= \frac{\Gamma(a + b + 2)}{\tau^2 \Gamma(a+1) \Gamma(b+1)} g^{b} \left( 1 + \frac{g}{\tau^2} \right)^{-a-b-2} ,       & g &> 0,
\end{align}
where $\Gamma(z) = \int_0^{\infty} t^{z-1} \exp\left\{ -t \right\} \dd t$ denotes the standard Gamma function.  Note that this family is defined for $a,b > -1$ and $\tau^2 > 0$ and corresponds to a Beta prior on $g/(\tau^2 + g)$.  It includes, for example,  the hyper-$g/n$ prior (which corresponds to $-1/2 \le a \le 0$, $b=0$ and $\tau^2=n$), as well as the half Cauchy distribution that underlies the Horseshoe prior \citep{carvalho2010horseshoe} (which corresponds to $a=b=-1/2$ and assigning $\tau^2$ a half-Cauchy distribution).  Hence, borrowing from the literature on continuous shrinkage priors, we call priors of this type ``global-local'' $g$ priors, where $g_1, \ldots, g_{p_{\gamma}}$ are ``local'' shrinkage parameter and $\tau^2$ is a ``global'' shrinkage parameter (which can be either known or unknown).

One potential challenge of the approach just outlined is the need to estimate what is, potentially, a very large number of different shrinkage factors.  This is not only computationally costly, but the data is likely to have limited information about each of them, especially in high-correlation settings.  Another challenge is that the performance of the procedure can be affected by the choice of the parametric family used to model the $g_j$s.  {In particular, recent work by  \cite{piironen2017sparsity} and \cite{lee2020tail} indicates that, in the context of continuous shrinkage priors, the optimal tail behavior of $p(g_j \mid \tau^2)$ might depend on the level of sparsity among coefficients.}

Both of these challenges can be addressed through the use of a nonparametric specification for the distribution of the $g_j$s based on the Dirichlet process \citep{ferguson1973bayesian}. A random distribution $H$ is said to follow a Dirichlet process prior with centering measure $H_0$ and concentration parameter $\alpha$, denoted $H \mid H_0,  \alpha \sim \DP(\alpha, H_0)$, if it admits a representation of the form
\begin{align}\label{eq:DPconstructive}
H(\cdot) = \sum_{k=1}^{\infty} w_k \delta_{g^{*}_k} (\cdot),    
\end{align}
where $\delta_{a}$ denotes a point mass at $a$, $g^{*}_1, g^{*}_2, \ldots$ is an i.i.d.\ sequence with $g^{*}_k \sim H_0$, and $w_k=v_k \prod_{l<k} (1-v_l)$, with $v_1, v_2, \ldots$ another i.i.d.\ sequence with $v_k \sim \bet(1, \alpha)$ \citep{sethuraman1994constructive}.  Because the samples from a Dirichlet process are almost surely discrete distributions, if $g_1, \ldots, g_c$ is an i.i.d.\ sample from a random $H \mid H_0, \alpha \sim \DP(\alpha, H_0)$, there is a positive probability of ties among the $g_i$s.  In fact, their joint distribution can be described through two sequences, $\tilde{g}_1, \tilde{g}_2, \ldots$ i.i.d.\ such that $\tilde{g}_k \sim H_0$, and $\xi_1, \ldots, \xi_c$ such that $\xi_1 = 1$ and 
\begin{align*}
    \xi_j \mid 
    \xi_{j-1}, \ldots, \xi_1, \alpha &\sim \sum_{k=1}^{K^{j-1}} \frac{m_k^{j}}{j + \alpha - 1} \delta_{k}+ \frac{\alpha}{j + \alpha - 1} \delta_{K^{j-1} + 1} ,  & j \ge 2 ,
\end{align*}
where $K^{j-1} = \max_{j' < j}\{ \xi_{j'}\}$ 
and $m_k^{j} = \sum_{j' < j} \indic (\xi_{j'} = k )$ \citep{blackwell1973ferguson}.  The value of $g_j$ can then be recovered from those of $\tilde{g}_1, \tilde{g}_2, \ldots$ and $\xi_1, \ldots, \xi_c$ through the relationship $g_j = \tilde{g}_{\xi_j}$.  The vector $\bfxi$ defines a partition $\rho = \{S_1, \ldots, S_{K}\}$ of the set $\mathcal{I} = \{1, \ldots, c\}$ such that $\cup_{k=1}^{K} S_k = \mathcal{I}$, $S_k \cap S_{k'} = \emptyset$ for $k \ne k'$, and $|S_k| = m_k$ is the number of elements in $S_k$, so that $i \in S_k$ if and only if $\xi_i = k$ and
$
f(\rho \mid \alpha) = \frac{\Gamma(\alpha)}{\Gamma(\alpha + c)} \alpha^{K} \prod_{k=1}^{K} \Gamma(m_k)
$.

In our setting, we let $g_j \mid H$ i.i.d. for $j=1,\ldots, p_{\bfgamma}$ and $H \mid \alpha \sim \DP(\alpha, H_0)$, 
where $H_0$ is the distribution associated with \eqref{eq:betaprime}, and $\alpha$ is assigned the {parameterization-invariant} prior introduced in \cite{rodriguez2013jeffreys}, which has density $f(\alpha \mid \bfgamma) = \sqrt{\frac{1}{\alpha}\sum_{j=1}^{p_{\bfgamma}-1} \frac{j}{(\alpha + j)^2} }$.  {The use of this prior circumvents the need to elicit prior information about the number of groups into which the coefficients are expected to be clustered.} The resulting prior,
\begin{multline}\label{eq:dpblockg}
    p(\bfbeta_{\bfgamma} \mid a, b, \tau^2, \bfgamma) = \int \phi\left(\bfbeta_{\bfgamma} \mid \bfzero, \sigma^2 \bfG_{\bfgamma}^{1/2} 
    \left\{ \bfX_{\bfgamma}^T \bfX_{\bfgamma}\right\}^{-1} \bfG_{\bfgamma}^{1/2} \right) \\
    f\left(\tilde{\bfg} \mid \rho, \gamma, \tau^2=n, a, b \right)
    %
    f(\rho \mid \gamma, \alpha) f(\alpha \mid \bfgamma) \dd \tilde{\bfg} 
    \dd \rho \dd \alpha ,
    %
\end{multline}
is a \textit{Dirichlet process mixture of block $g$ priors}.

Because there might be ties among the $g_j$s, the model implicitly defines a partition of the coefficients in which those assigned to the same group share a common shrinkage factor.  
The concentration parameter $\alpha$ controls the prior distribution on the partitions, with $\alpha \to 0$ leading to the standard (mixture of) $g$ priors, and $\alpha \to \infty$ leading back to  global-local $g$ priors defined above where each coefficient is assigned its own shrinkage factor. Since the model treats both $\alpha$ and $\rho$ as unknown, the model is able to learn an appropriate partition of the coefficients as it performs model selection.  Furthermore, the use of a nonparametric prior for $H$ implies that the model is potentially capable of learning from the data the shape of the distribution of the shrinkage factors,
which can alleviate concerns about the specific choice of the hyperparameters $a$, $b$ and $\tau^2$.

\subsection{Unifying continuous shrinkage and variable selection priors}\label{sec:litrelationship}

Traditionally, frameworks based on continuous shrinkage priors have recognized that differential shrinkage might be needed to attain optimal performance, but until recently (e.g., \citealp{boss2023group}) they have tended to downplay the need to account for co-linearity among covariates.  On the other hand, the  literature on priors for model selection has, from the very beginning, acknowledged the need to account for colinearity, but has been slower to recognize the need for differential shrinkage, perhaps because of the computational challenges involved. Our approach provide a unifying framework for thinking about these two strands of the literature.  To see this, consider a slightly less general spike-and-slab version of our prior where $\bfbeta \mid \sigma^2, \tilde{g}_1, \ldots, \tilde{g}_p, \bfgamma \sim \normal\left( \bfzero, \sigma^2  \tilde{\bfG}^{1/2} \left\{ \bfGamma \bfSigma^{-1} \bfGamma
\right\}^{-} \tilde{\bfG}^{1/2}  \right)$, 
$\bfGamma = \diag\{ \gamma_1, \ldots, \gamma_p \}$, $\tilde{\bfG} = \diag\{ \tilde{g}_{\xi_1}, \ldots, \tilde{g}_{\xi_p}  \}$, and $A^{-}$ represents the Moore-Penrose inverse of $A$.  Different choices of $\bfSigma$ and of priors on $\bfgamma$ and $\rho$ lead to various well-known procedures.  For example, as we noted before, the DP mixture of block $g$ priors includes the standard $g$ prior and the ``global-local'' block $g$ prior as special (limit) cases.  Furthermore, when $\bfX$ is orthogonal and the grouping variable $\rho$ is treated as known, it also includes the block $g$ prior of \cite{som2014paradoxes} as a special case.

On the other hand, if we fix the model to $\bfgamma = (1, 1, \ldots, 1)$, either $\bfX$ is orthonormal or we take $\bfSigma = \bfI$, 
and $\tau^2$ is given a hyperprior, the DP mixture of block $g$ priors corresponds to the Horseshoe Pit mixture prior of \cite{denti2021horseshoe}.  {As a consequence, it also includes a number of traditional continuous shrinkage priors such as those in \cite{park2008bayesian}, \cite{brown2010inference}, \cite{carvalho2010horseshoe}, \cite{leng2014bayesian},  \cite{bhattacharya2015dirichlet} and \cite{bai2018beta}. 
Also, under the full model and block orthogonality and a known group structure, 
our framework also includes 
the Group Inverse Gamma Gamma shrinkage prior of 
\cite{boss2023group}.}

\section{Properties of DP mixtures of block $g$ prior}\label{sec:properties}

\subsection{Tail behavior}\label{sec:tail}

Since, marginally, samples from a distribution generated by a Dirichlet process follow the baseline measure (e.g., see \citealp{blackwell1973ferguson} or \citealp{antoniak1974mixtures}), the marginal distribution for the $l$-th entry of $\bfbeta_{\bfgamma}$ under a DP mixture of block $g$ priors is given by
$f(\bfbeta_{\bfgamma,l} \mid \tau^2, a, b, \gamma) = \int \normal(\bfbeta_{\bfgamma,l} \mid 0, g  \kappa_{\gamma,l,l} \sigma^2 ) f(g \mid \tau^2, a, b) \dd g$,
where $\kappa_{\gamma,l,l}$ is the $l$-th diagonal entry of $\left\{  \bfX_{\bfgamma}^T \bfX_{\bfgamma} \right\}^{-1}$. 
The tail behavior of this type of marginal distributions was studied in \cite{boss2023group} (see their Theorem 2.1).  In particular, the index of regular variation of the marginal prior is $\omega = -2b-3$, i.e., 
$$
\lim_{\bfbeta_{\bfgamma,l} \to \infty} \frac{f(t \bfbeta_{\bfgamma,l} \mid \tau^2, a, b, \gamma)}{f(\bfbeta_{\bfgamma,l} \mid \tau^2, a, b, \gamma)} = t^{-2b-3}.
$$
This implies that the our prior has heavy (polynomial) tails {and point estimators derived from our procedures is robust,} in the sense of having bounded influence in the case of likelihood-prior conflict.  However, note that the contour plots associated with the DP mixture of block $g$ priors are not elliptical, unlike those of the standard $g$ prior (please see Section \ref{ap:scatterplotsbivariagte} of the supplementary materials).  {These non-elliptical contours enable non-uniform shrinkage by the DP block $g$ priors, which in turn is what enables them to skirt the conditional Lindley paradox}.  

\subsection{Information consistency of Bayes factors}\label{sec:infconsistency}

For fixed $n$, $p$ and $\bfX$ that is full rank, consider a sequence of observations $\bfy(1), \bfy(2), \ldots$ such that $\| \hat{\bfbeta}_{\bfgamma}(N) \| \to \infty$ and $N \to \infty$, where $\hat{\bfbeta}_{\bfgamma}(N) = \left[\bfX_{\bfgamma}^{T}\bfX_{\bfgamma}\right]^{-1} \bfX_{\bfgamma}^T \bfy(N)$ is the maximum likelihood estimator or $\bfbeta_{\bfgamma}$ based on $\bfy(N)$.  The Bayes factor $BF_{\bfgamma, \bfzero}$ is information consistent if $BF_{\bfgamma, \bfzero} \left(\bfy(N)\right) \to \infty$ as $N \to \infty$.  

Bayes factors under standard mixtures of $g$ priors are known to be information consistent under appropriate conditions on the prior on $g$.  The following theorem, which is analogous to Theorem 2 in \cite{liang2008mixtures}, establishes general conditions on the joint prior on $g_1, \ldots, g_{p_{\bfgamma}}$ that ensure information consistency for general mixtures of block $g$ priors.

\begin{thm}\label{th:infoconsistency1}
%
    %
    %
    %
Let $\nu_{+}$ be the largest eigenvalue of $\bfX_{\bfgamma}^T\bfX_{\bfgamma}$ and $\lambda_{-}(\bfG_{\bfgamma})$ be the smallest eigenvalue of $\bfX_{\bfgamma}^T \bfX_{\bfgamma}  - \left[  \{ \bfX_{\bfgamma}^T \bfX_{\bfgamma} \}^{-1} + \bfG_{\bfgamma}^{1/2}\{ \bfX_{\bfgamma}^T \bfX_{\bfgamma} \}^{-1}\bfG_{\bfgamma}^{1/2}\right]^{-1}$.  The mixing prior $f(g_1 , \ldots , g_{p_{\bfgamma}})$ leads to Bayes factors that are information consistent if 
\begin{multline}\label{eq:infconsiscond}
   \int \left| \bfX_{\bfgamma}^T \bfX_{\bfgamma} + \bfG_{\bfgamma}^{1/2} \bfX_{\bfgamma}^T \bfX_{\bfgamma} \bfG_{\bfgamma}^{1/2} \right|^{-1/2}  \\
   \left[ 1 - \frac{\lambda_{-}(\bfG_{\bfgamma})}{\nu_{+}}  \right]^{-\frac{n-1}{2}} f(g_1 , \ldots , g_{p_{\bfgamma}}) \dd g_1 \ldots \dd g_{p_{\bfgamma}} = \infty .
\end{multline} 
for all $p_{\bfgamma} \le p$.
\end{thm}
The proof is included in Section \ref{ap:informationconsistency} of the supplementary materials. A slightly simpler condition that applies to  DP mixtures of block $g$ priors is the following:

\begin{thm}\label{th:infoconsistency2}
  A sufficient condition for the DP mixture of block $g$ priors to lead to Bayes factors that are information consistent is for the density of the centering measure, $f(g \mid \tau^2, a, b)$, to satisfy
  $
  \int (1 + g_j)^{(n-1-p_{\bfgamma})/2} f(g_j \mid \tau^2, a, b) \dd g_j = \infty
  $
   for all $p_{\bfgamma} \le p$.
\end{thm}
See Section \ref{ap:informationconsistency} of the supplementary materials.  This is the same condition in Theorem 2 of \cite{liang2008mixtures}.  Hence, this result just indicates that any mixing distribution for $g$ that leads to information-consistent Bayes factors under a standard mixture of $g$ priors also leads to information consistent Bayes factors under the DP mixture of block $g$ priors.


\subsection{Information consistency of block structures}\label{sec:blockconsistency}

One of the key motivations to consider DP mixtures of block $g$ priors is the desire to avoid having to decide a priori on an appropriate partition for the covariates. In this Section, we show that, when the design matrix $\bfX$ is orthogonal and the true coefficients have very different sizes, our prior assigns coefficients of different sizes to separate clusters with high probability.  Before proceeding with our main result, we need to introduce the concept of refinement of a partition (sometimes called a fragmentation, e.g., see \citealp{bertoin2006random}).

\begin{definition}[Refinement of a partition] 
Let $\rho=\{S_1,\dots,S_K\}$ and $\rho'=\{S_1',\dots,S_{K'}^{'}\}$ denote two  partition of a set $\mathcal{I} = \{1,\ldots, c\}$ with $K$ and $K'$ unique blocks respectively, such that $1\leq K' \leq K \leq c$. Then, $\rho$ is said to be a refinement of $\rho'$, denoted by $\rho\prec \rho'$, if and only if for every $S_k \in \rho$ there exist a $S_{j}^{'}\in \rho'$ such that $S_k\subseteq S_{j}^{'}$.  
\end{definition}

\begin{thm}\label{th:partition}
Let $\bfX$ be a full rank, centered, orthogonal design matrix of size $n \times p$, and $\bfX_1$ and $\bfX_2$ be two non-overlapping submatrices of sizes $n\times p_1$ and $n\times p_2$ with $p_1 > 0$, $p_2>0$ and $p_1 + p_2 \le p$. Denote by $\mathcal{I}_1 = \{ j^{(1)}_1, \ldots, j^{(1)}_{p_1} \}$ the set of indexes associated with the columns of $\bfX$ included in $\bfX_1$ and $\mathcal{I}_2 = \{ j^{(2)}_1, \ldots, j^{(2)}_{p_2} \}$ the columns associated with $\bfX_2$, so that $\mathcal{I} = \mathcal{I}_1 \cup \mathcal{I}_2$ and $\mathcal{I}_1 \cap \mathcal{I}_2 = \emptyset$.
Consider now an asymptotic regime where, for fixed $n$, $p_1$, $p_2$, $\bfX_1$, $\bfX_2$, $\beta_0$, $\bfbeta_2$ and $\bfepsilon$, a sequence of observations $\{ \bfy(N) : N \in \naturals\}$ is generated as $\bfy(N) = \bfones_n\beta_0 + \bfX_{1}\bfbeta_1(N)+\bfX_{2}\bfbeta_2+\bfepsilon$, where $\{ \bfbeta_1(N) : N \in \naturals\}$ is a sequence such that, $\beta_{j}^{2}(N) \sim \mathcal{O}(N)$ for all $j \in \mathcal{I}_1$.  If  $\rho_0 = \{ \mathcal{I}_1, \mathcal{I}_2 \}$, then 
    \begin{enumerate}
    \item For $\rho \nprec \rho_{0}$, $\lim_{N \to \infty} \frac{f\left(\bfy(N) \mid \rho\right)}{f\left(\bfy(N) \mid \rho_{0}\right)} = 0$, and
    \item For $\rho \prec \rho_{0}$, $\lim_{N \to \infty}\frac{f\left(\bfy(N) \mid \rho \right)}{f\left(\bfy(N) \mid \rho_{0} \right)} = c_{\rho}$, with $0 < c_{\rho} < \infty$.
\end{enumerate}
\end{thm}

The proof can be seen in Section \ref{ap:cluseringprop} of the supplementary materials.

\subsection{Conditional Lindley paradox}\label{sec:CLP}

In addition to being important on its own right, the previous result allows us to show that, when the design matrix $\bfX$ is orthogonal, the Bayes factors based on  DP mixtures of block $g$ priors avoid the conditional Lindley paradox.

\begin{thm}\label{th:conLindleyparadox}
Let $\bfX$ be a full rank, centered, orthogonal design matrix of size $n \times p$, and $\bfX_1$ and $\bfX_2$ be two non-overlapping submatrices of sizes $n\times p_1$ and $n\times p_2$ with $p_1 > 0$, $p_2>0$ and $p_1 + p_2 = p$, and consider the pair of models $\mathcal{M}_{\bfgamma_0}: \bfy = \boldsymbol{1}_n\beta_0+\bfX_{1}\bfbeta_1+\bfepsilon$ and $\mathcal{M}_{\bfgamma_{a}}: \bfy = \boldsymbol{1}_n\beta_0 + \bfX_{1}\bfbeta_1+\bfX_{2}\bfbeta_2+\bfepsilon$.  Denote by $\mathcal{I}_1 = \{ j^{(1)}_1, \ldots, j^{(1)}_{p_1} \}$ the set of indexes associated with the columns of $\bfX$ included in $\bfX_1$ and $\mathcal{I}_2 = \{ j^{(2)}_1, \ldots, j^{(2)}_{p_2} \}$ the columns associated with $\bfX_2$, so that $\mathcal{I} = \mathcal{I}_1 \cup \mathcal{I}_2$ and $\mathcal{I}_1 \cap \mathcal{I}_2 = \emptyset$.
%

For fixed $n$, $p_1$, $p_2$, $\bfX_1$, $\bfX_2$, $\beta_0$, $\bfbeta_2$ and $\bfepsilon$, let $\{ \bfbeta_1(N) : N \in \naturals\}$ be a sequence such that $\beta_{j}^{2}(N) \sim \mathcal{O}(N)$ as $N \to \infty$ for all $j \in \mathcal{I}_1$, 
and $\{ \bfy(N) : N \in \naturals\}$ be the associated sequence generated by setting $\bfy(N) = \bfones_n\beta_0 + \bfX_{1}\bfbeta_1(N)+\bfX_{2}\bfbeta_2+\bfepsilon$.  Then, for the Bayes factor based on a DP mixture of block $g$ priors under the hyper-$g/n$ mixture distribution we have
%
$$
\lim_{N \to \infty} BF_{\bfgamma_a,\bfgamma_0} \left( \bfy(N) \right) > 0
$$
for any any $\{ \bfy(N) : N \in \naturals\}$ and any pair of models $\bfgamma_0$ and $\bfgamma_a$.  
\end{thm}

{The proof, which can be found in Section \ref{ap:condLindleyparadox} of the supplementary materials, only requires that the prior on the partition $\rho$ puts positive probability on at least one refinement of the true partition $\rho_0 = \{ \mathcal{I}_1, \mathcal{I}_2 \}$.  This is trivially true for the Dirichlet process, whether its concentration parameter is fixed or treated as unknown and given a prior distribution.  In fact, this is a very mild requirement that is also satisfied by the Bayes factor constructed under the global-local $g$ prior, 
and one that does not require that the partition model be able to identify the true partition $\rho_0$.}  On the other hand, orthogonality plays a key role in the proof of this theorem, just like block orthogonality is key to similar proofs in \cite{som2014paradoxes}.  Simulation studies suggest that the result hold in the non-orthogonal case.

\subsection{Model selection consistency}\label{sec:modelselcons}

Model selection consistency refers to the ability of the procedure to choose the correct model as the sample size grows.  DP mixtures of block $g$ priors are model selection consistent in the fixed $p$ regime (see Section \ref{ap:modelselconsistency} of the supplementary materials).

\begin{thm}\label{th:asympconsistency}
Assume that a sequence of observations $y_1, y_2, \ldots$ is generated from some model $\bfgamma_T \in \{ 0,1 \}^p$ (i.e., one of the models considered by our procedure), and that $p$ is fixed.  Also, assume the following regularity conditions:
\begin{itemize}
        \item[(i)] the column space $\mathcal{C}(\bfX)$ does not contain $\bfones_n$.
        \item[(ii)] The sequence of covariate vectors $\bfx_{1}, \bfx_{2}, \ldots$ are such that $\| \bfx_i \|^2$ is bounded by a constant for all $i=1,2,\ldots$.
        \item[(iii)] The smallest eigenvalue of $\bfX^T \bfX/n$ is lower bounded by a positive constant for all $n$.
    \end{itemize}
Then, under the DP mixture block $g$ priors with $\tau^2 = n$, $\lim_{n \to \infty}
\Pr(\bfgamma = \bfgamma_T \mid \bfy) = 1
$, as long as the prior on models satisfies $f(\bfgamma_T) > 0$.
\end{thm}

\subsection{Intrinsic consistency}\label{sec:intrinsicconsistency}

Under slightly more stringent regularity conditions than those required for model selection consistency,  DP mixtures of block $g$ priors are also intrinsically consistent.

\begin{thm}
    Assume that, as $n$ grows, the columns   $\bfx_1, \bfx_2, \ldots$ of the design matrix satisfy either of the following two conditions for a finite, positive definite matrix $\bfLambda$:
    \begin{itemize}
        \item[(i)] If $\bfx_1, \bfx_2, \ldots$ forms a deterministic sequence, then $\frac{1}{n} \bfX^T \bfX \underset{n \to \infty}{\longrightarrow} \bfLambda$.
        \item[(ii)] If $\bfx_1, \bfx_2, \ldots$ are random, then they are independent and identically distributed from a distribution with mean $\bfzero$ and covariance $\bfLambda$.
    \end{itemize}

    Then, the DP mixture of block $g$ priors with $\tau^2 = n$ converges to a proper, non-degenerate intrinsic prior of the form
\begin{multline*}
    f(\bfbeta_{\bfgamma} \mid a, b, \bfgamma) = \int \phi\left(\bfbeta_{\bfgamma} \mid \bfzero, \sigma^2 \bfG_{\bfgamma}^{1/2} 
    \bfLambda^{-1} \bfG_{\bfgamma}^{1/2} \right) \\
    f\left( \tilde{\bfg} \mid \rho, \gamma, \tau^2=n, a, b \right)
    %
    f(\rho \mid \bfgamma, \alpha) f(\alpha \mid \bfgamma) \dd \tilde{g}_1 \ldots \dd \tilde{g}_{p_{\bfgamma}} 
    \dd \rho \dd \alpha .
\end{multline*}\end{thm}

\section{Computation}\label{sec:computation}

It is possible to construct MCMC algorithms for model selection under the DP mixtures of block $g$ prior that require very minimal tuning. To do so, we take advantage of the conditional conjugacy of the priors and, when possible, we integrate out the intercept $\beta_0$, the vector of regression coefficients $\bfbeta_{\bfgamma}$ and/or the variance $\sigma^2$ when deriving conditional posteriors.  Additionally, we represent the shrinkage coefficients $g_1, \ldots, g_{p_{\bfgamma}}$ in terms of their unique values $\tilde{\bfg}_{\bfgamma} = (\tilde{g}_1, \ldots, \tilde{g}_{K_{\bfgamma}})$ and the group indicators $\bfxi_{\bfgamma} = (\xi_1, \ldots, \xi_{p_{\bfgamma}})$ (recall Section \ref{sec:DPblock}).  The resulting posterior takes the form
\begin{align*}
        f(\bfgamma, \tilde{\bfg}, \bfxi, \alpha \mid \bfy) \propto f(\bfy \mid \bfgamma,  \tilde{\bfg}, \bfxi)
     f \left(\tilde{\bfg} \mid \bfgamma \right) f(\bfxi \mid \bfgamma, \alpha) f(\alpha | \bfgamma) f(\bfgamma) ,
\end{align*}
where $ f(\bfy \mid \bfgamma,  \tilde{g}_1, \ldots, \tilde{g}_{K_{\bfgamma}}, \xi_1, \ldots, \xi_{p_{\bfgamma}})$ corresponds to \eqref{eq:marlikcond} with $\bfSigma_{\bfgamma} = \left\{ \bfX_{\bfgamma}^T\bfX_{\bfgamma} \right\}^{-1}$, and $p(\bfgamma)$ is an appropriate prior on the space of models, e.g., a Beta-Binomial prior
$f(\bfgamma) = \frac{\Gamma(c+d)}{\Gamma(c) \Gamma(d)} \frac{\Gamma(c + p_{\bfgamma}) \Gamma(d + p - p_{\bfgamma})}{\Gamma(c+d+p)}$.

Our MCMC algorithm then alternates sampling from the full conditionals $f(\bfgamma, \tilde{\bfg} , \bfxi  \mid \cdots)$, $f(\bfxi \mid \cdots)$, $f(\alpha \mid \cdots)$, $f(\tilde{\bfg} \mid \cdots)$, $f(\beta_0, \bfbeta_{\bfgamma} \mid \cdots)$ and $f(\sigma^2 \mid \cdots)$.  To sample from $f(\bfgamma, \tilde{\bfg}, \bfxi  \mid \cdots)$, we use a random walk Metropolis algorithm in which, at each iteration, we propose to either add one variable, remove one variable, or swap one variable currently in the model with one that is not.  If a variable is added to the model, the corresponding value of $\xi_i$ and, if necessary, a new value of $\tilde{g}_k$, are proposed from the prior distributions on these parameters.  This is technically a Reversible Jump MCMC step \citep{green1995reversible}, albeit a very simple one.  In spite of this simplicity, the algorithm seems to perform quite well. On the other hand, once we condition on the model, sampling from  $f(\bfxi \mid \cdots)$ can be accomplished using any of the collapsed samplers for non-conjugate Dirichlet process mixture models (e.g., see \citealp{neal2000markov}).  To sample $f(\tilde{\bfg} \mid \cdots)$ we resort to a slight variant of the slice sampler introduced in \cite{finegold2014robust} and  \cite{liu2012rejection}.  Finally, sampling from $f(\alpha \mid \cdots)$ is accomplished through the use of a random walk Metropolis-Hastings algorithm with log-Gaussian proposals.  This is the only step of the algorithm that requires tuning of hyperparameters. Details are provided in Section \ref{ap:MCMCM} of the supplementary materials.  
\if1\blind
{
An implementation of the code is available from 
\url{https://github.com/Anupreet-Porwal/DP-mix-block-g-prior}.
} \fi

\section{Simulation studies}\label{sec:simulations}

\subsection{Conditional Lindley paradox}\label{sec:sim1}

Our first simulation study replicates the setting used to construct Figure \ref{fig:clphg} and shows empirical evidence supporting the theoretical results discussed in Sections \ref{sec:blockconsistency} and \ref{sec:CLP}. We consider a total 150 simulations, each of which involves a sequence of datasets generated under model $\mathcal{M}_{a}: y_i = \beta_0 + \beta_1 x_{i,1} + \beta_2 x_{i,2} + \epsilon_i$ for $i=1,\ldots, 100$.  All elements of a given sequence of datasets share the same values of $\beta_0$, $\beta_1$, $\bfx_1 = (x_{1,1}, \ldots, x_{100,1})'$, $\bfx_2 = (x_{1,2}, \ldots, x_{100,2})'$ and $\bfepsilon = (\epsilon_1, \ldots, \epsilon_{100})'$.  In particular, we set $\beta_0=0.5$, $\beta_1=1$ and generate $\epsilon$ from a standard multivariate Gaussian distribution and each of the pairs $(x_{i,1}, x_{i,2})'$ from a zero-mean bivariate Gaussian distribution with unit marginal standard deviations and correlation $\eta$.  The  datasets within each sequence are then constructed by considering a grid of values for $\beta_2$ in the interval $[0,240]$.  We are interested in the behavior of $B_{a,0}$, the Bayes factor under the DP mixture of $g$ priors that uses the hyper-$g/n$ as the baseline measure comparing the true model $\mathcal{M}_a$ against the simpler model $\mathcal{M}_0 : y_i = \beta_0+\beta_1 x_{i,1} +\epsilon_i$. We also investigate the behavior of $\Pr(\xi_1 \ne \xi_2 \mid \bfy)$, the posterior probability that the model assigns different shrinkage parameters to each  variable in the model (recall Section \ref{sec:blockconsistency}).

\begin{figure}[hbt!]
     \centering
     \begin{subfigure}[b]{0.4\textwidth}
         \centering
         \includegraphics[width=\textwidth]{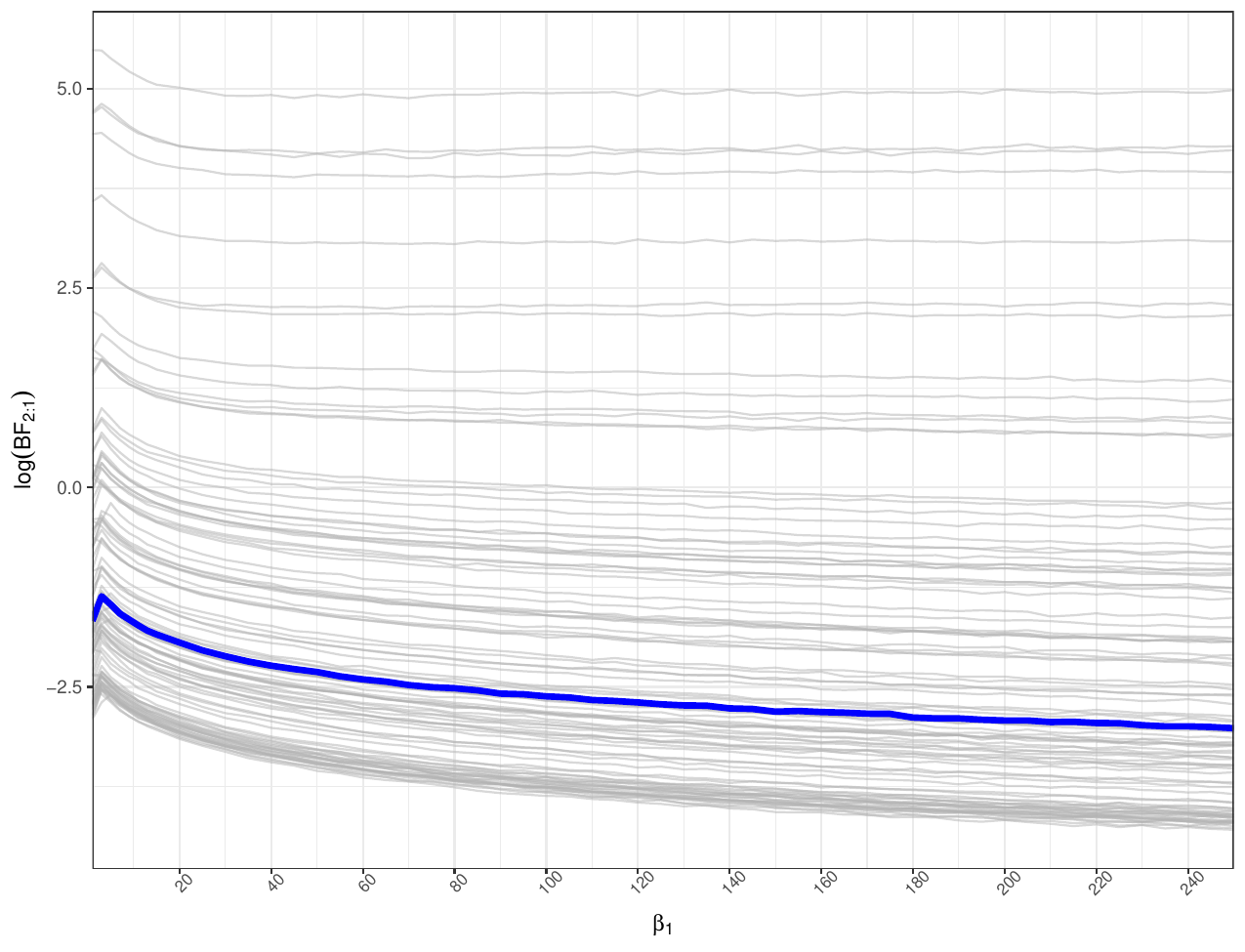}
         \caption{Log Bayes factor, $\eta = 0$}
     \end{subfigure}
     \hspace{3mm}
     \begin{subfigure}[b]{0.4\textwidth}
         \centering
         \includegraphics[width=\textwidth]{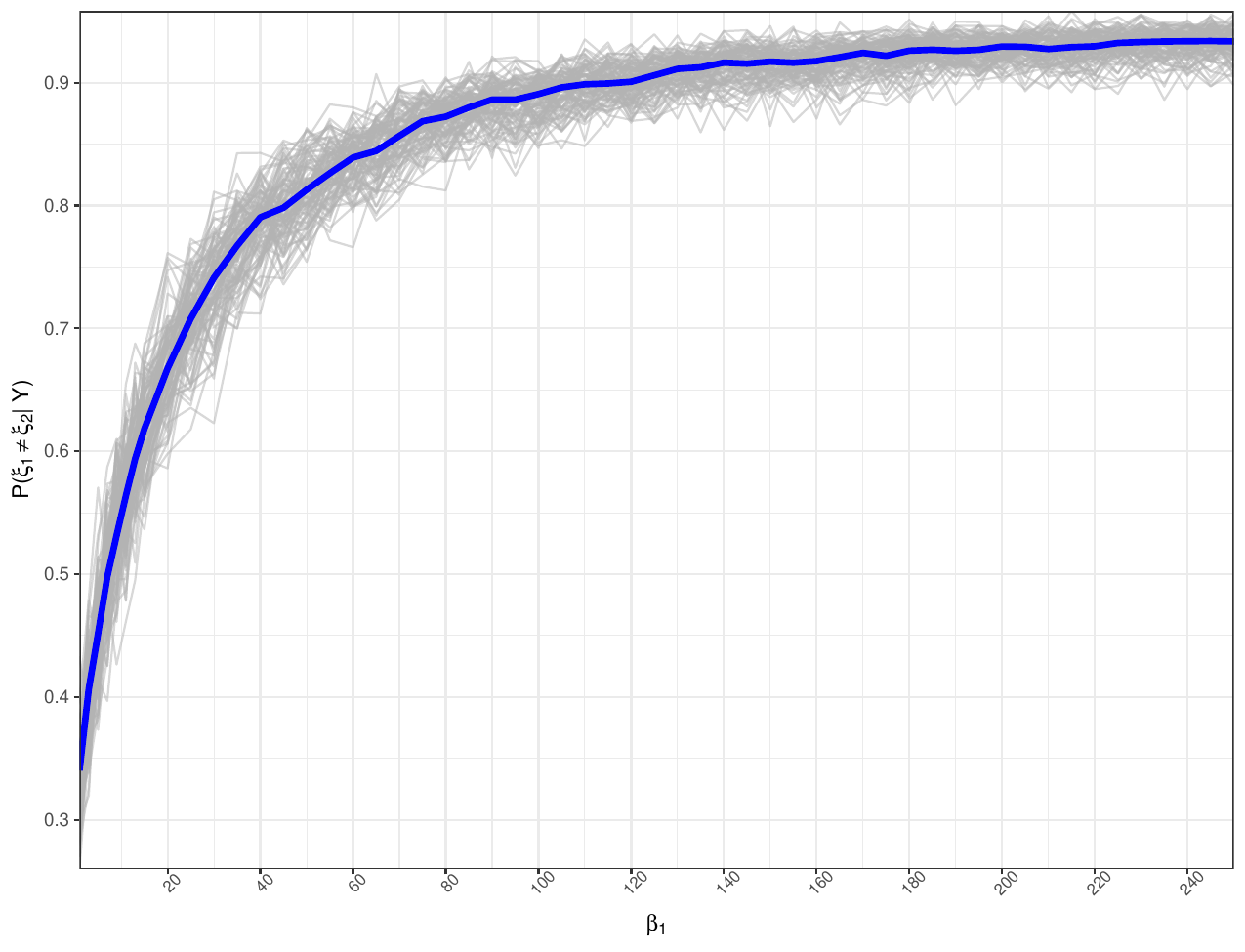}
         \caption{$\Pr(\xi_1 \ne \xi_2 \mid \bfy)$,,  $\eta = 0$}
     \end{subfigure} \\
     \begin{subfigure}[b]{0.4\textwidth}
         \centering
         \includegraphics[width=\textwidth]{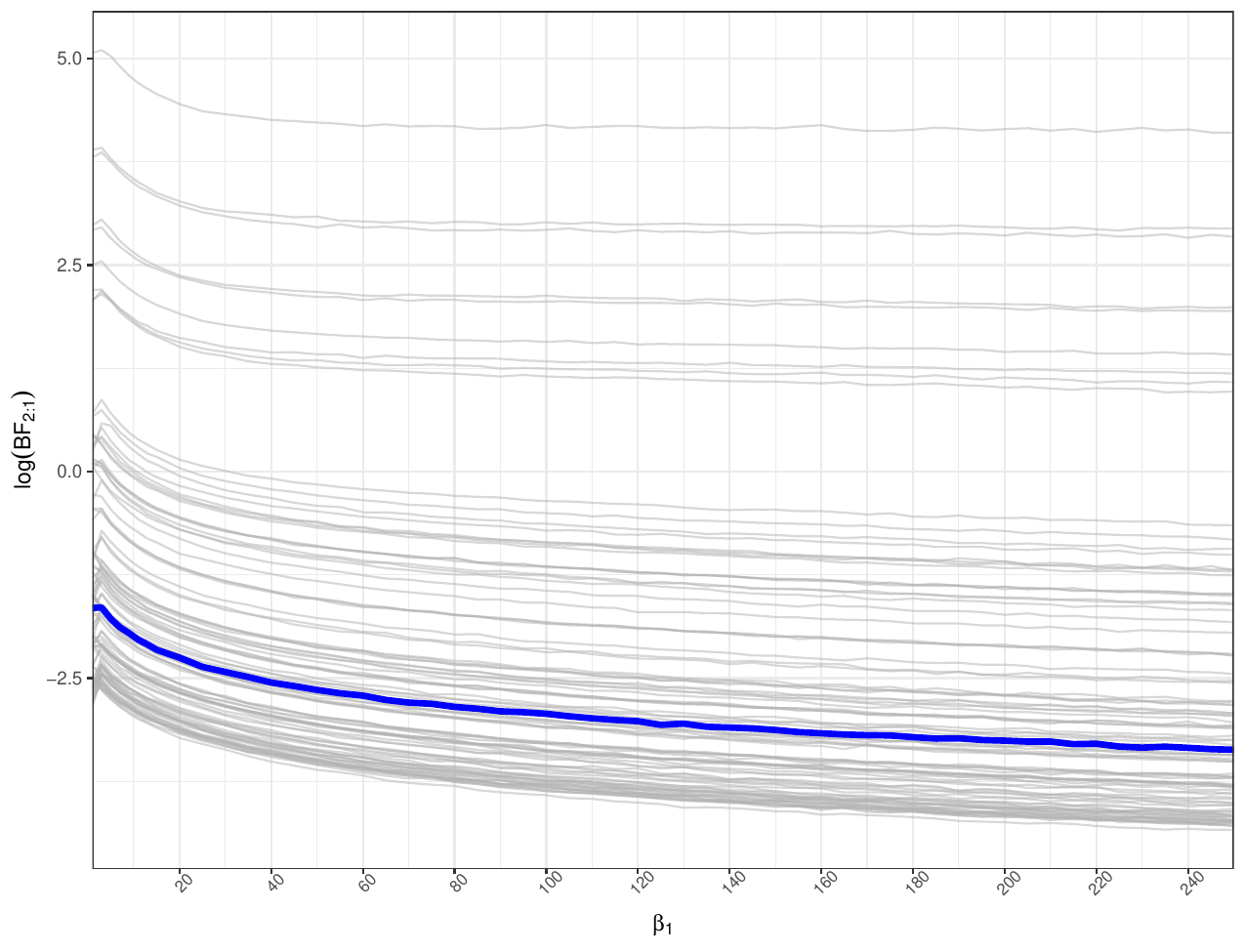}
         \caption{Log Bayes factor, $\eta = 0.5$}
     \end{subfigure}
     \hspace{3mm}
     \begin{subfigure}[b]{0.4\textwidth}
         \centering
         \includegraphics[width=\textwidth]{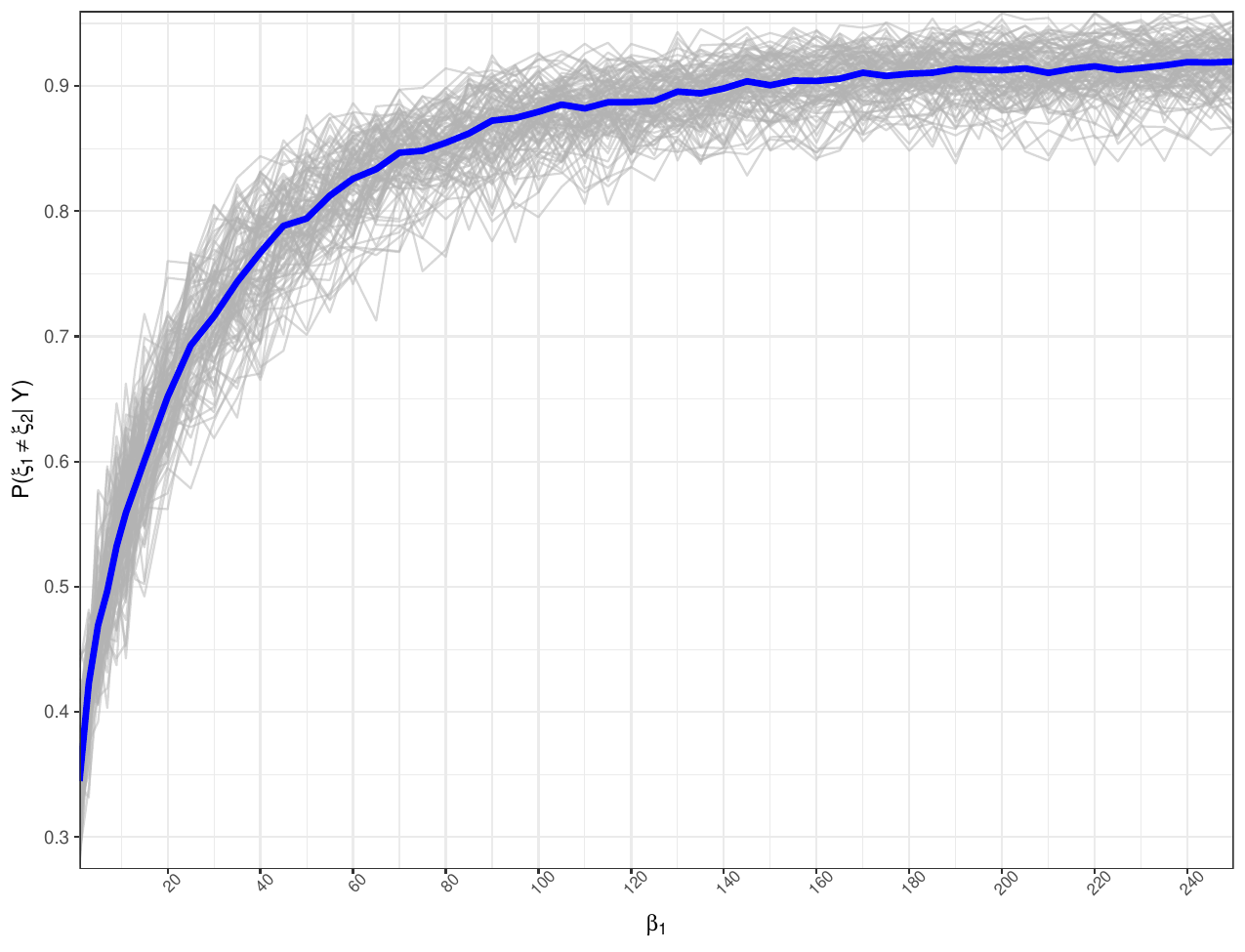}
          \caption{$\Pr(\xi_1 \ne \xi_2 \mid \bfy)$,  $\eta = 0.5$.}
     \end{subfigure}
    \caption{Behavior of $\log\left( B_{a,0}(\bfy)\right)$ (left column) and $\Pr(\xi_1 \ne \xi_2 \mid \bfy)$ (right column) under the DP mixture of block $g$ priors in our first simulation study.  Each thin grey line corresponds to one replicate of the simulation, while the thicker blue line corresponds to the mean curve.  Figures in the top row correspond to design matrices generated under $\eta=0$, while the bottom row corresponds to $\eta = 0.5$}\label{fig:clpDP}
\end{figure}

Figure \ref{fig:clpDP} shows the results of this simulation study for $\eta=0$ and $\eta=0.5$.  Compared to Figure \ref{fig:clphg}, the curves are somewhat noisy.  This is an artifact of the Monte Carlo noise introduced by our MCMC algorithm (the Bayes factor depicted in \ref{fig:clphg} is available in ``closed form'').  
With that caveat in mind, we note that the curves for $\log \left( B_{a,0}(\bfy) \right)$ decrease as $\beta_2$ increases but, unlike Figure \ref{fig:clphg}, both seem to stabilize towards an asymptote.  This agrees with the behavior predicted by Theorem \ref{th:conLindleyparadox}.  Similarly, and as predicted by Theorem \ref{th:partition}, we can see that $\Pr(\xi_1 \ne \xi_2 \mid \bfy)$ seems to converge to 1 as $\beta_2$ grows.

\subsection{Model selection, estimation and prediction performance}\label{sec:sim2}


We conducted a second simulation study to compare the model selection, estimation, and prediction performance of procedures based on DP mixtures of block $g$ priors with that of competing procedures.  {We considered three versions of DP mixtures of block $g$ priors:  one using a hyper-$g$ base measure (labeled ``DP block-$g$ ($\tau^2=1$)''), a ``unit information'' version using a hyper-$g/n$ base measure  (labeled ``DP block-$g$ ($\tau^2=n$)''), and one using a scaled hyper-$g$ base measure where the scale is in turn given a half-Cauchy hyperprior (labeled ``DP block-$g$ ($\tau^2 \sim \mathsf{HC}$) '').}  This second study, the setup of which is inspired by \cite{denti2021horseshoe}, assumes $n=500$ for all datasets and, as before, the vectors of covariates associated with the each observation are generated from a zero-mean multivariate normal distribution with unit marginal variances and correlation $\eta$ across all pairs of covariates.  We consider six scenarios that arise from combining three different values for the total number of covariates ($p=250$, $p=500$ and $p=750$) and two different levels of multicolinearity ($\eta=0$ and {$\eta=0.9$}).  
For all three values of $p$, $100$ of the coefficients are randomly sampled from a normal distribution with mean 0 and standard deviation 10 (we consider these ``large'' coefficients), $100$ are randomly sampled from a standard normal distribution (the ``small'' coefficients), and the remainder are set to 0 (the null coefficients). For each of the six scenarios, we generate 100 datasets.

In terms of competing approaches, we consider the following:  (a) a standard hyper-$g/n$ mixture of $g$ priors (which  we label ``g-prior'' in the sequel); (b) a version of the block $g$ prior of \cite{som2014paradoxes} with a hyper-$g/n$ hyperprior and known blocking structures where the covariates are allocated to $K=2$ groups:  one made of all the covariates associated with large coefficients plus half, randomly chosen variables associated with null coefficients, and another one made of the rest (labeled ``Som et al.\ ($\tau^2=n$, $K=2$)''); (c) a version of the block $g$ prior of \cite{som2014paradoxes} with $K=3$ fixed groups of covariates:  one made of all the covariates associated with large coefficients, one made of the covariates associated with small coefficients, and one made of the covariates associated with null coefficients (labeled ``Som et al.\ ($\tau^2=n$, $K=3$)''); {(d) three versions of ``global-local'' $g$-priors with distinct but identically distributed shrinkage parameters for each coefficient, one using a hyper-$g$ prior (labeled ``GL-$g$ ($\tau^2=1$)''), one using a hyper-$g/n$ prior (labeled ``GL-$g$ ($\tau^2=n$)''), and one using a scaled hyper-$g$ prior where the scale is given a half-Cauchy hyperprior (labeled ``GL-$g$ ($\tau^2 \sim \mathsf{HC}$) '');} (e) the adaptive Lasso (ALasso) of \citep{huang2008adaptive}; (f) the Horseshoe prior (\citealp{carvalho2010horseshoe}, labeled ``Horseshoe'' in the sequel) and (g) the Horseshoe-Pit prior (\citealp{denti2021horseshoe}, labeled ``HSM''). 
Computation under standard mixtures of $g$ priors relies on version 1.7.1 of the \texttt{R} package \texttt{BAS}, while computation under block $g$ priors with known blocking structures relies on a slight variation of our own code for the Diriclet mixtures of $g$ priors.  Computation for the adaptive Lasso relies on version 4.1.6 of the \texttt{R} package \texttt{glmnet}.  Computation under the Horseshoe prior relies on version 1.2 of the \texttt{R} package \texttt{bayesreg}, while computation under the Horseshoe-Pit prior relies on code from the author of that manuscript which, at the time of this writing, is available at \url{https://github.com/Fradenti/HorseshoeMix}.  {For all Bayesian procedures that require a prior on model space, we  assign $\bfgamma$ a Beta-Binomial prior with $c=d=1$ (e.g., see \citealp{scott2010bayes} and \citealp{porwal2022effect} and Section \ref{sec:computation} above).  
Furthermore, in order to avoid improper priors when $p \ge n$, the prior on models is constrained so that models for which $\sum_{k=1}^{p} \gamma_i > n-2$ receive zero probability.}

\begin{table}[hbt!]
    \centering
    \begin{subfloat}
    \centering 
    \tiny
    \scalebox{1}{
    \begin{tabular}{lccc|ccc} 
       &  \makecell{Power \\ (large coeffs)}  &
        \makecell{Power \\ (small coeffs)}  &
       \makecell{ Type I error \\(null coeffs)} &
              \makecell{Power \\ (large coeffs)} &
       \makecell{Power \\ (small coeffs)} &
       \makecell {Type I error \\ (null coeffs)}\\ \cline{1-7}
       & \multicolumn{6}{c}{$p = 250$} \\ \cline{1-7}
       & \multicolumn{3}{c|}{$\eta = 0$} & \multicolumn{3}{c}{$\eta = 0.9$} \\ \cline{1-7}

g-prior ($\tau^2=$n) & 0.986 & 0.856 & 0.005 & 0.949 & 0.506 & 0.004 \\ 
  Som et al. ($\tau^2=$n, K=2) & 0.986 & 0.913 & 0.037 & 0.962 & 0.850 & 0.219 \\ 
  Som et al. ($\tau^2=$n, K=3) & 0.992 & 1.000 & 1.000 & 0.980 & 1.000 & 1.000 \\ 
    \midrule
  GL-$g$ ($\tau^2=$1) & 1.000 & 1.000 & 1.000 & 1.000 & 1.000 & 1.000 \\ 
  GL-$g$ ($\tau^2=$n) & 0.990 & 0.908 & 0.054 & 0.969 & 0.694 & 0.046 \\ 
  GL-$g$ ($\tau^2\sim$HC) & 0.990 & 0.900 & 0.039 & 0.973 & 0.745 & 0.100 \\ 
  DP block-g ($\tau^2=$1) & 0.991 & 0.921 & 0.160 & 0.984 & 0.843 & 0.394 \\ 
  DP block-g ($\tau^2=$n) & 0.989 & 0.904 & 0.045 & 0.974 & 0.760 & 0.126 \\ 
  DP block-g ($\tau^2\sim$HC) & 0.990 & 0.906 & 0.049 & 0.977 & 0.788 & 0.211 \\ 
    \midrule
  ALasso & 0.956 & 0.569 & 0.011 & 0.873 & 0.352 & 0.157 \\ 
  Horseshoe & 0.989 & 0.889 & 0.027 & 0.965 & 0.656 & 0.022 \\ 
  HSM & 0.986 & 0.860 & 0.005 & 0.962 & 0.625 & 0.011 \\ 
\cline{1-7}
       & \multicolumn{6}{c}{$p = 500$} \\ \cline{1-7}
       & \multicolumn{3}{c|}{$\eta = 0$} &  \multicolumn{3}{c}{$\eta = 0.9$}  \\ \cline{1-7}
  g-prior ($\tau^2=$n) & 0.979 & 0.814 & 0.000 & 0.913 & 0.342 & 0.001 \\ 
  Som et al. ($\tau^2=$n, K=2) & 0.982 & 0.873 & 0.005 & 0.944 & 0.636 & 0.028 \\ 
  Som et al. ($\tau^2=$n, K=3) & 0.986 & 0.923 & 1.000 & 0.963 & 0.843 & 1.000 \\ 
      \midrule
  GL-$g$ ($\tau^2=$1) & 1.000 & 1.000 & 0.993 & 0.999 & 0.992 & 0.973 \\ 
  GL-$g$ ($\tau^2=$n) & 0.984 & 0.866 & 0.006 & 0.955 & 0.580 & 0.010 \\ 
  GL-$g$ ($\tau^2\sim$HC) & 0.984 & 0.865 & 0.005 & 0.957 & 0.597 & 0.014 \\ 
  DP block-g ($\tau^2=$1) & 0.985 & 0.873 & 0.014 & 0.961 & 0.636 & 0.040 \\ 
  DP block-g ($\tau^2=$n) & 0.984 & 0.867 & 0.006 & 0.957 & 0.612 & 0.016 \\ 
  DP block-g ($\tau^2\sim$HC) & 0.984 & 0.867 & 0.007 & 0.958 & 0.622 & 0.023 \\ 
    \midrule
  ALasso & 0.745 & 0.117 & 0.056 & 0.269 & 0.008 & 0.006 \\ 
  Horseshoe & 0.982 & 0.842 & 0.012 & 0.947 & 0.540 & 0.007 \\ 
  HSM & 0.981 & 0.831 & 0.001 & 0.943 & 0.506 & 0.001 \\ 

\cline{1-7}
       & \multicolumn{6}{c}{$p = 750$} \\ \cline{1-7}
       & \multicolumn{3}{c|}{$\eta = 0$}  & \multicolumn{3}{c}{$\eta = 0.9$}  \\ \cline{1-7}
  g-prior ($\tau^2=$n) & 0.959 & 0.607 & 0.000 & 0.887 & 0.212 & 0.000 \\ 
  Som et al. ($\tau^2=$n, K=2) & 0.982 & 0.863 & 0.003 & 0.941 & 0.576 & 0.015 \\ 
  Som et al. ($\tau^2=$n, K=3) & 0.985 & 0.884 & 0.488 & 0.957 & 0.703 & 0.563 \\ 
    \midrule
  GL-$g$ ($\tau^2=$1) & 0.992 & 0.922 & 0.468 & 0.976 & 0.792 & 0.464 \\ 
  GL-$g$ ($\tau^2=$n) & 0.985 & 0.849 & 0.003 & 0.946 & 0.506 & 0.005 \\ 
  GL-$g$ ($\tau^2\sim$HC) & 0.985 & 0.851 & 0.003 & 0.948 & 0.526 & 0.008 \\ 
  DP block-g ($\tau^2=$1) & 0.985 & 0.857 & 0.005 & 0.952 & 0.547 & 0.014 \\ 
  DP block-g ($\tau^2=$n) & 0.985 & 0.852 & 0.003 & 0.949 & 0.536 & 0.010 \\ 
  DP block-g ($\tau^2\sim$HC) & 0.985 & 0.853 & 0.004 & 0.951 & 0.548 & 0.013 \\ 
    \midrule
  ALasso & 0.722 & 0.106 & 0.054 & 0.242 & 0.007 & 0.004 \\ 
  Horseshoe & 0.980 & 0.800 & 0.005 & 0.940 & 0.456 & 0.004 \\ 
  HSM & 0.982 & 0.818 & 0.000 & 0.938 & 0.449 & 0.001 \\ 
 \cline{1-7}
    \end{tabular}}
    \end{subfloat}
    \vspace{2mm}
    \caption{Estimates of power for small (generated from a $\mathcal{N}(0,1)$ distribution) and large (generated from a $\mathcal{N}(0,10)$ distribution) coefficients, 
    and of type I error for null coefficients ($\beta=0$) in our second simulation study.  For the purpose of this table, coefficients are considered ``significant'' is their posterior inclusion probability is greater than 0.5.}\label{tab:errortypes09}
\end{table}

In the introduction, we motivated DP mixtures of block $g$ priors by arguing that it should lead to higher power for detecting smaller coefficients.  We also claimed that procedures that pre-select a blocking of the coefficients can be very sensitive to this choice. Evidence of these claims is presented in Table \ref{tab:errortypes09}, which shows the estimated power associated with identifying ``large'' (those generated from a zero-mean normal with standard deviation 10) and ``small'' (those generated from a standard normal distribution) coefficients, as well as the type I error (for the null coefficients) under the various procedures. For the purpose of this table, coefficients are considered ``significant'' if their posterior inclusion probability is greater than 0.5 (in the case of procedures based on variants of the $g$-prior), if they are included in the optimal model after applying generalized cross-validation to identify the optimal penalty parameter (for ALasso){, or if their 95\% posterior credible intervals do not cover zero (in the case of Horseshoe and HSM)}. First, we note that ALasso shows by far the lowest power to detect small coefficients. Even for large coefficients, the performance of ALasso degrades substantially in ``large $p$'' scenarios, {especially when variables are highly correlated}. The same is true of other penalized likelihood procedures we tried (results not shown). Next, we note that the standard $g$ prior tends to have lower power than the other Bayesian procedures we consider when it comes to detecting small coefficients. {Indeed, somewhat surprisingly, HSM and Horseshoe seem to have higher overall power than standard $g$-priors in this scenario with similar type-I errors, even though they were not originally designed for model selection. GL-$g$ procedures with $\tau^2=1$ also seem to perform poorly, yielding high power for detecting small signals, but also very high type I error rates.} Similarly to GL-$g$ with $\tau^2=1$, the procedure of \cite{som2014paradoxes} with $K=3$ tends to have the highest power, but it also comes with extremely high type I error rates. This is because, by parceling out the null coefficients into a separate cluster, the block $g$ prior ends up overfitting by learning a very small shrinkage factor for the coefficients in this block. The results for \cite{som2014paradoxes} with $K=2$ indicate that this issue can be addressed by having the null coefficients assigned to blocks that contain some significant coefficients, avoiding overfitting. This solution is, however, clearly impractical in real applications, as we do not know which coefficients are likely to be significant in the first place. { The best performing procedures correspond to GL-$g$ ($\tau^2=n$), GL-$g$ ($\tau^2\sim \mathsf{HC}$), and the three procedures based on Dirichlet process mixtures of block $g$-priors. These are all priors that can be understood as being either ``unit information'', or allowing for enough flexibility to learn the expected value of the $g_j$s from the data. For $\eta = 0$, all of these 5 procedures seem to yield very similar performance in terms of power and type I errors. However, for $\eta = 0.9$, procedures based on DP block-$g$ priors consistently show higher power for detecting small coefficients, at the price of a very small increase (in absolute terms) in the type I error.}
\begin{figure}[hbt!]
    \centering
    \includegraphics[width = 0.8\linewidth]{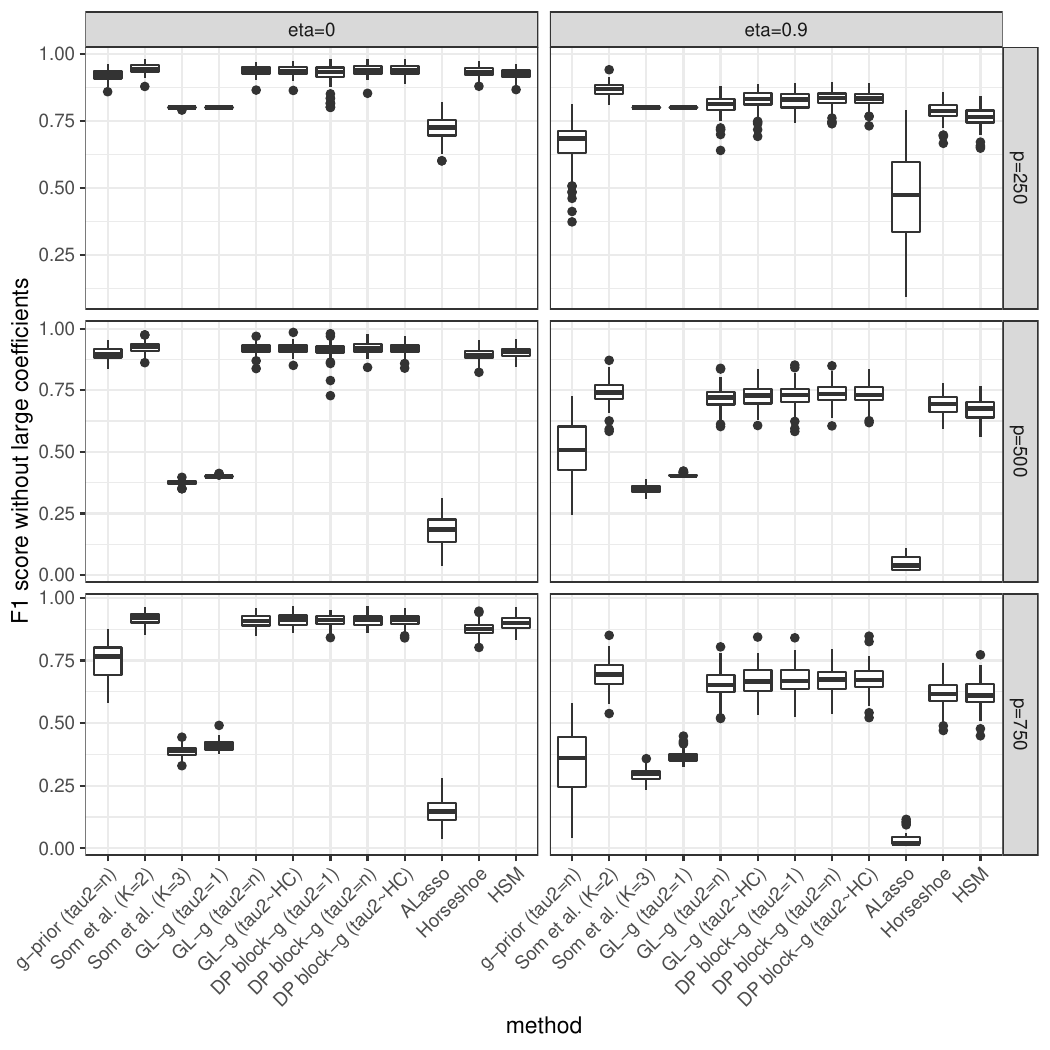}
    \caption{$F_1$ scores for model selection procedures for our second simulation study.  { These $F_1$ scores excludes the large coefficients in the calculation of precision and recall.}}
\label{fig:F1_sim}
\end{figure}

To investigate the false positive/false negative tradeoffs associated with the various methods, we present in Figure \ref{fig:F1_sim} boxplots across the various simulated datasets of $F_1$ for the various procedures. Recall that the $F_1$ score is defined as the harmonic mean of proportion of true positives among ``selected'' covariates (the precision) and the proportion of ``selected'' covariates among true positive covariates (the recall). The $F_1$ score ranges between 0 and 1, with a higher value indicating better model selection performance. { As would be expected from our discussion of power and false positive rates, ALasso is by far the worst performer, followed by Som et al.\ $(K=3)$ and procedures based on the GL-$g$ ($\tau^2=1$) prior. Procedures based on the standard $g$-prior perform somewhat better than these three, but they are still suboptimal, particularly when the covariates are highly correlated. The performance of the remaining procedures is very similar, but some subtle patterns are still visible. Som et al.\ $(K=2)$ seems to perform slightly better than the rest but, as discussed before, it relies on the unrealistic assumption that we are able to distinguish between large and small coefficients up front. Similarly, GL-$g$ ($\tau^2=n$), Horseshoe and HSM seem to slightly underperform, particularly when the correlation is high.}
%
%
\begin{figure}[hbt!]
    \centering
        \begin{subfigure}[b]{0.8\textwidth}
        \centering
        \includegraphics[width=\textwidth]{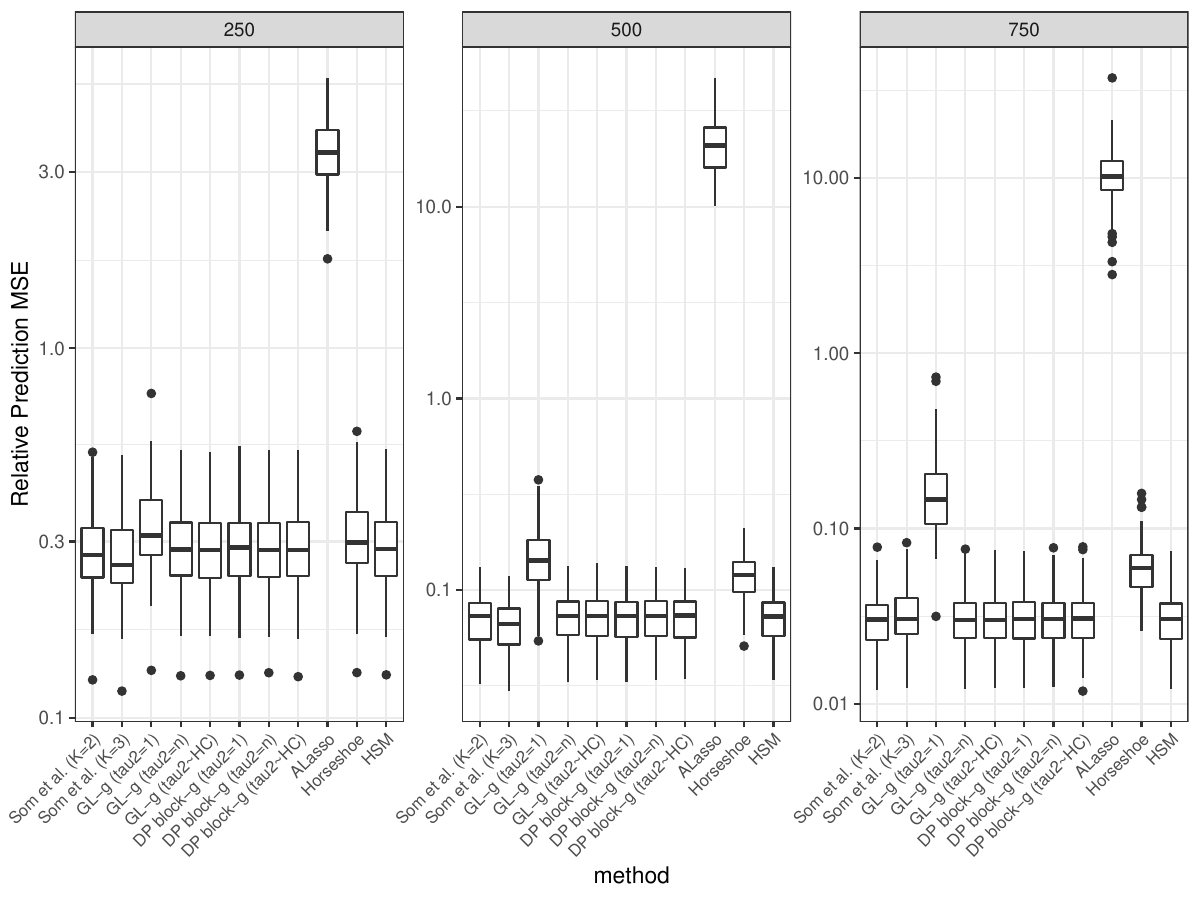}
        \caption{$\eta=0$}
        \label{fig:ozone_pip}
    \end{subfigure} \\
    \begin{subfigure}[b]{0.8\textwidth}
        \centering
        \includegraphics[width=\textwidth]{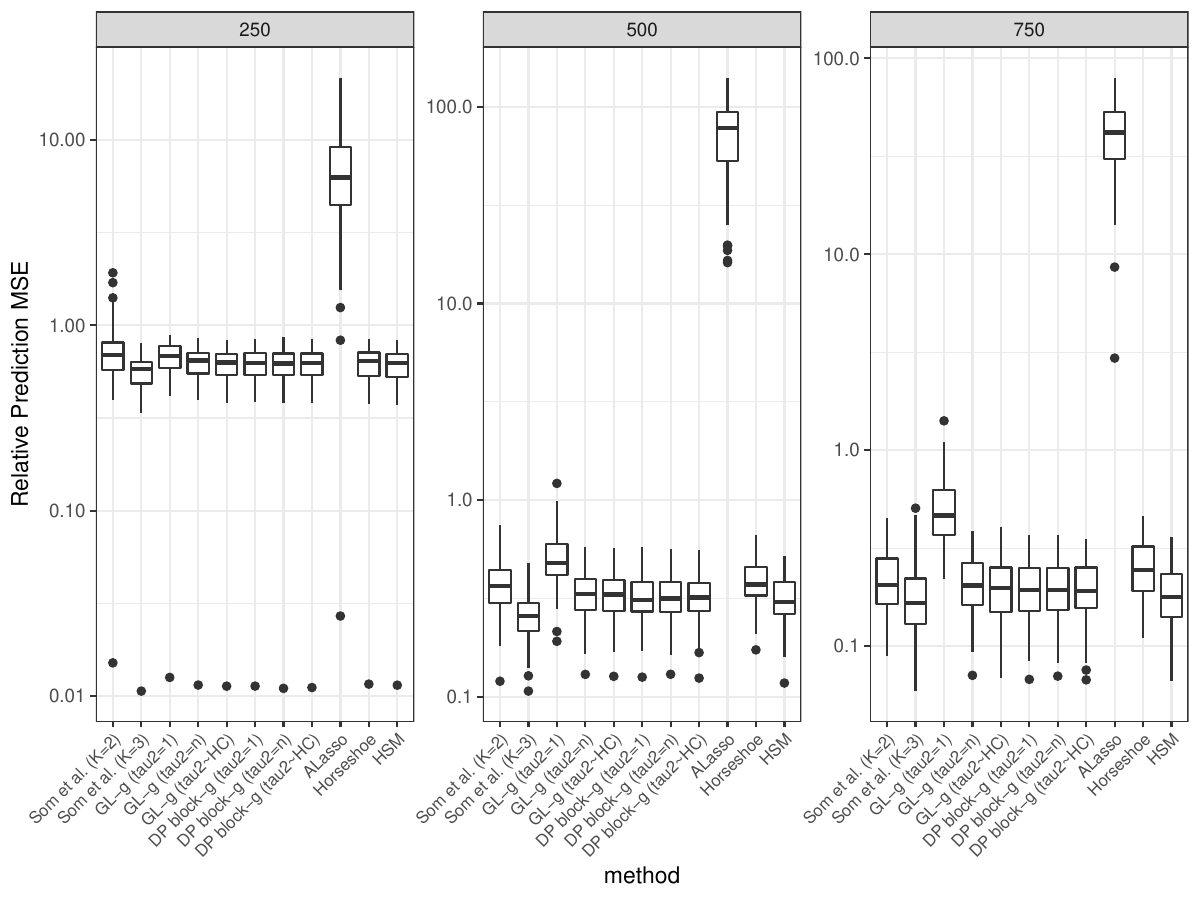}
        \caption{$\eta=0.9$}
        \label{fig:ozone_pip}
    \end{subfigure}
    \caption{Normalized prediction MSE for $\eta=0$ and $\eta = 0.9$.  Normalization is with respect to the prediction MSE under the standard $g$-prior.}
\label{fig:MSE_pred_rho0}
\end{figure}
Taken together, the results in Table \ref{tab:errortypes09} and Figure \ref{fig:F1_sim} highlight (a) the benefits of using differential shrinkage priors in the context of model selection, {(b) the need to either properly center the distribution of the shrinkage coefficients or, alternatively, allowing enough model flexibility to learn its center from its data,} and (c) the risks associated with the use of fixed rather than data-driven blocks in the development of model selection priors.

Finally, we present in Figure \ref{fig:MSE_pred_rho0} the prediction mean square error (MSE) for each of the procedures. 
To obtain these prediction MSEs, each dataset was augmented with a test set of 500 additional observations. Furthermore, in order to simply interpretation, we normalized all MSEs with respect to that under the $g$-prior for each dataset and graph the resulting ratios in a logarithmic scale. This means that values less than 1 correspond to methods with smaller (better) prediction MSE. Note that, with the exception of ALasso, all procedures consistently outperform $g$-priors in ``large $p$'' regimes, again illustrating the the advantages of differential shrinkage in Bayesian settings. Note that Horseshoe has worse performance than the other Bayesian procedures, particularly in ``large $p$'' scenarios. While perhaps surprising at first sight, this observation is consistent with the results of \cite{lee2020continuous}. Som et al.\ $(K=3)$ seems to perform slightly better than Som et al.\ $(K=2)$ in this evaluation, specially when $\eta = 0.9$, which is the opposite of what we observed when evaluating model selection performance. Similar to model selection performance, procedures based of the GL-$g$ ($\tau^2=1$) prior also have worse performance than the remaining GL-$g$ and DP block-$g$ procedures, all of which perform quite similarly in terms of predictions. { Additional results related to simulation study, including results 
for $\eta=0.5$, examples of posterior distributions over $(p_{\bfgamma}, K_{\bfgamma})$, and a sensitivity analysis 
for priors on the concentration parameter $\alpha$ and model space $\bfgamma$
can be seen in the supplementary materials.}

\section{The \texttt{ozone} dataset}\label{sec:realdata}

We further investigate the performance of DP mixtures of block $g$ prior using the \texttt{ozone} dataset introduced in \cite{breiman1985estimating} and later analyzed in \cite{casella2006objective} and \cite{liang2008mixtures}, among others. The dataset consists of daily measurements of the maximum ozone concentration near Los Angeles and eight meteorological variables. We consider regression models that might include all eight of these variables along with all possible interactions and squares, leading to up to 44 possible predictors. 
\begin{figure}[!h]
    \centering
    \begin{subfigure}[b]{0.49\textwidth}
        \centering
        \includegraphics[width=\textwidth]{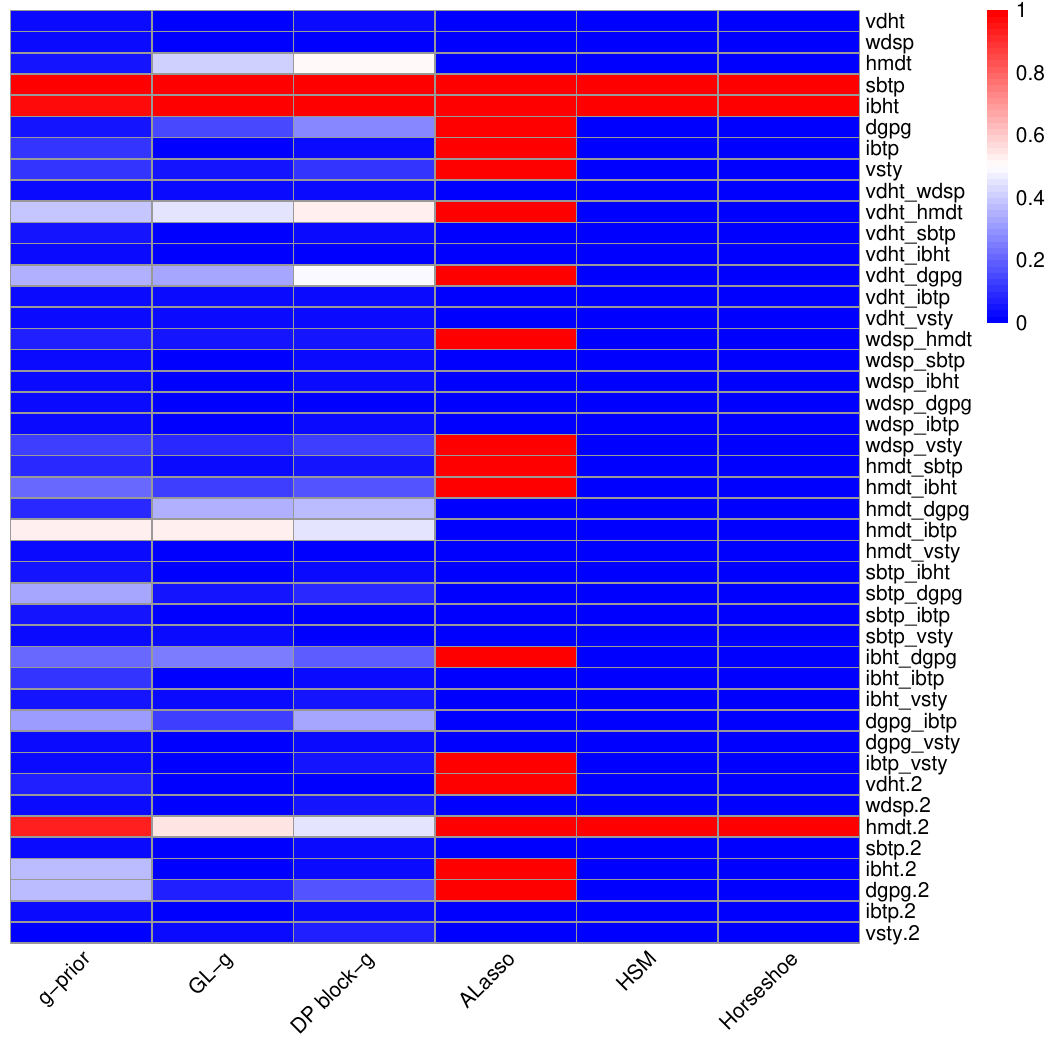}
        \caption{Posterior inclusion probabilities}
        \label{fig:ozone_pip}
    \end{subfigure}
    \hfill
    \begin{subfigure}[b]{0.49\textwidth}
        \centering
        \includegraphics[width=\textwidth]{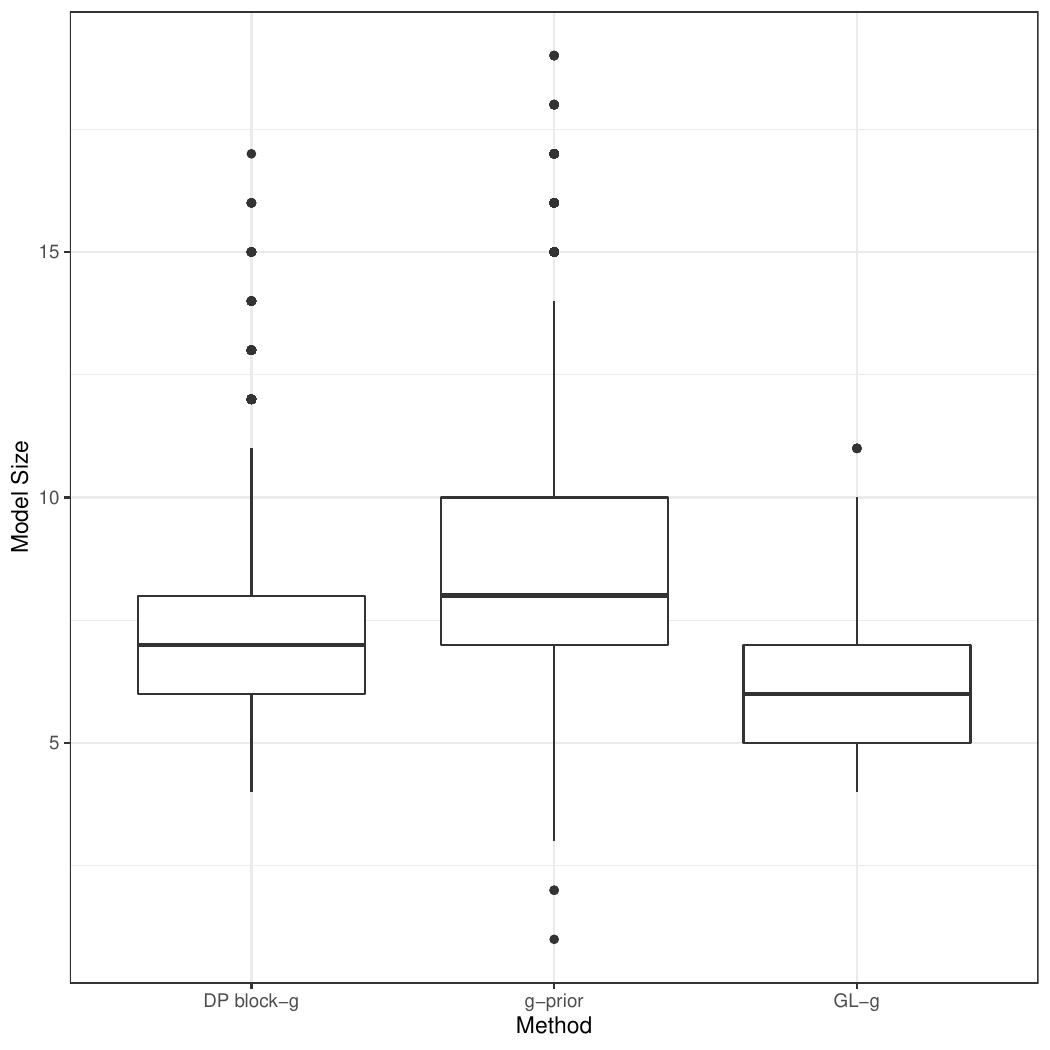}
        \caption{Model size}
        \label{fig:ozone_modsize}
    \end{subfigure}
     \caption{Posterior inclusion probabilities for individual variables and model sizes for various model selection procedures in the \texttt{ozone} dataset.}
    \label{fig:ozone_pip_plus_modsize}
\end{figure}

Figure \ref{fig:ozone_pip} shows the posterior inclusion probabilities (PIPs) for each of the predictors (i.e., $\Pr(\gamma_i = 1 \mid \bfy)$) under consideration for the various competing procedures described in Section \ref{sec:sim2}. In the case of ALasso, these are taken to be 1 if the variable is non-zero in the model fitted using optimal penalty parameter according to generalized crossvalidation. On the other hand, for Horseshoe and HSM, the PIPs are reported as 0 if the 95\% posterior credible interval for the variable includes 0, and as 1 otherwise. Note that ALasso is an outlier and tends to select a much larger number of variables (17) than any of the Bayesian procedures. On the other hand, there is fair bit of agreement in the PIPs among the various Bayesian procedures. For example, all of them agree in that \texttt{sbtp} (Sandburg Air Force Base temperature) and ibht (inversion base height at LAX) should be included in the model. There are, however, interesting differences as well. For example, HSM, Horseshoe and the standard $g$-prior all agree in including the square of \texttt{hmdt} (humidity) in the model, but not the main effect of \texttt{hmdt}. In contrast, GL-$g$ and DP block-$g$ assign moderate probabilities of inclusion to both the linear and quadratic terms associated with humidity instead. To complement these results, we show in Figure \ref{fig:ozone_pip} the posterior distribution of $p_{\bfgamma}$, the number of variables included in the model, for DP block-$g$, GL-$g$ and the standard $g$-prior. Interestingly, the standard $g$ prior tends to include the most variables (in some cases, as many as ALasso), while GL-$g$ tends to selects the most parsimonious models. As would be expected, DP block-$g$ is somewhere in between them.
\begin{figure}[!h]
    \centering
    \scalebox{0.8}{
    \includegraphics{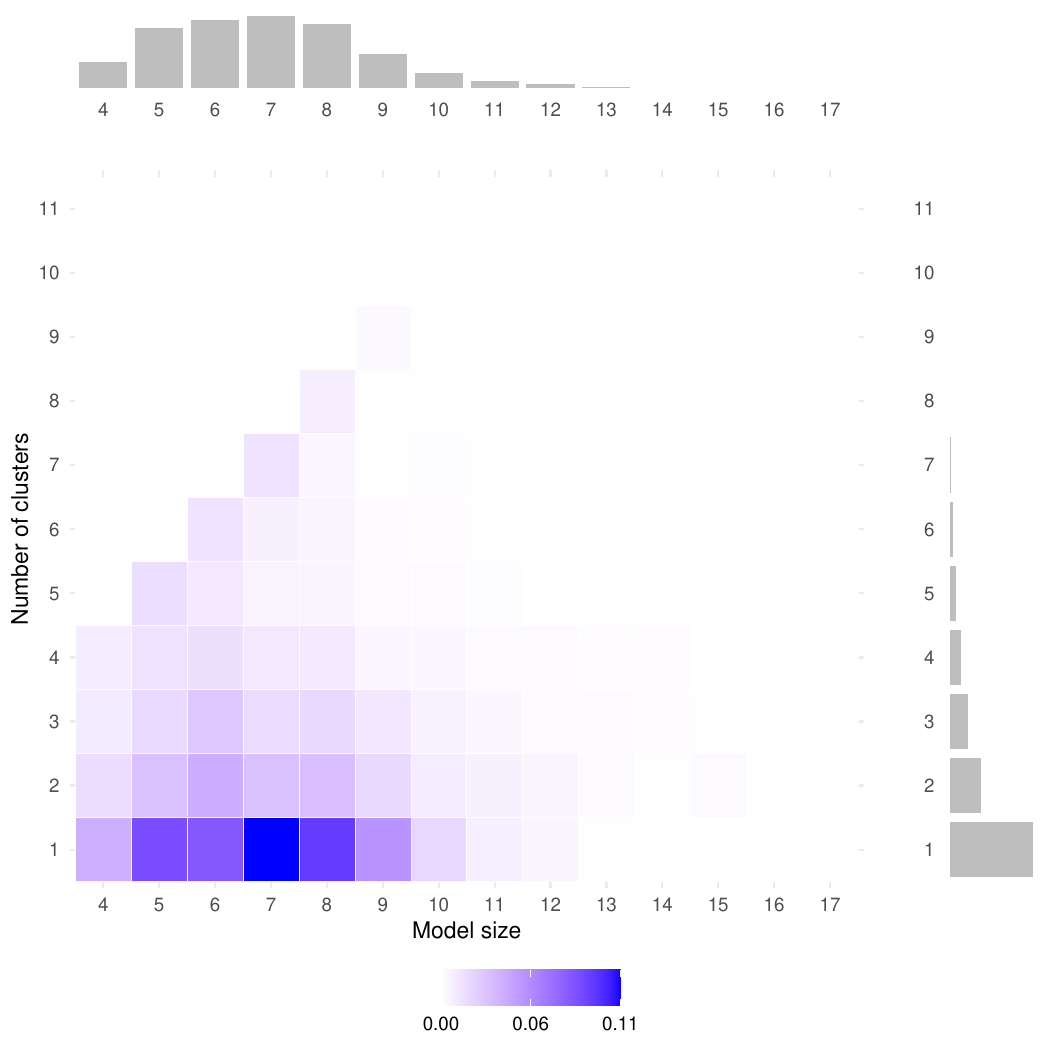}}
    \caption{Joint and marginal posterior distributions for $p_{\gamma}$ and $K_{\bfgamma}$ under the DP mixtures of block $g$ prior for the \texttt{ozone} dataset.}
    \label{fig:joint_p_K_ozone}
\end{figure}

To gain additional insight into the behavior of the various approaches, Figure \ref{fig:joint_p_K_ozone} shows the joint and marginal posterior distributions for $p_{\gamma}$ and $K_{\bfgamma}$ (the number of blocks in which the $p_{\gamma}$ included variables have been grouped) under the DP mixture of block $g$ priors. Note that the number of variables included by this procedure ranges between 4 and 17, with a clear mode at 7. The procedure also places moderate probability (around 0.49) on models that group these variable into more than one block of variables, but virtually no probability to any model with more than 8 or 9 blocks. This result is consistent with our previous observation that procedures based on DP block-$g$ adaptively ``interpolate'' between those produced by standard $g$ priors and those generate by GL-$g$. 
\begin{figure}[!h]
    \centering
    \begin{subfigure}[b]{0.49\textwidth}
        \centering
        \includegraphics[width=\textwidth]{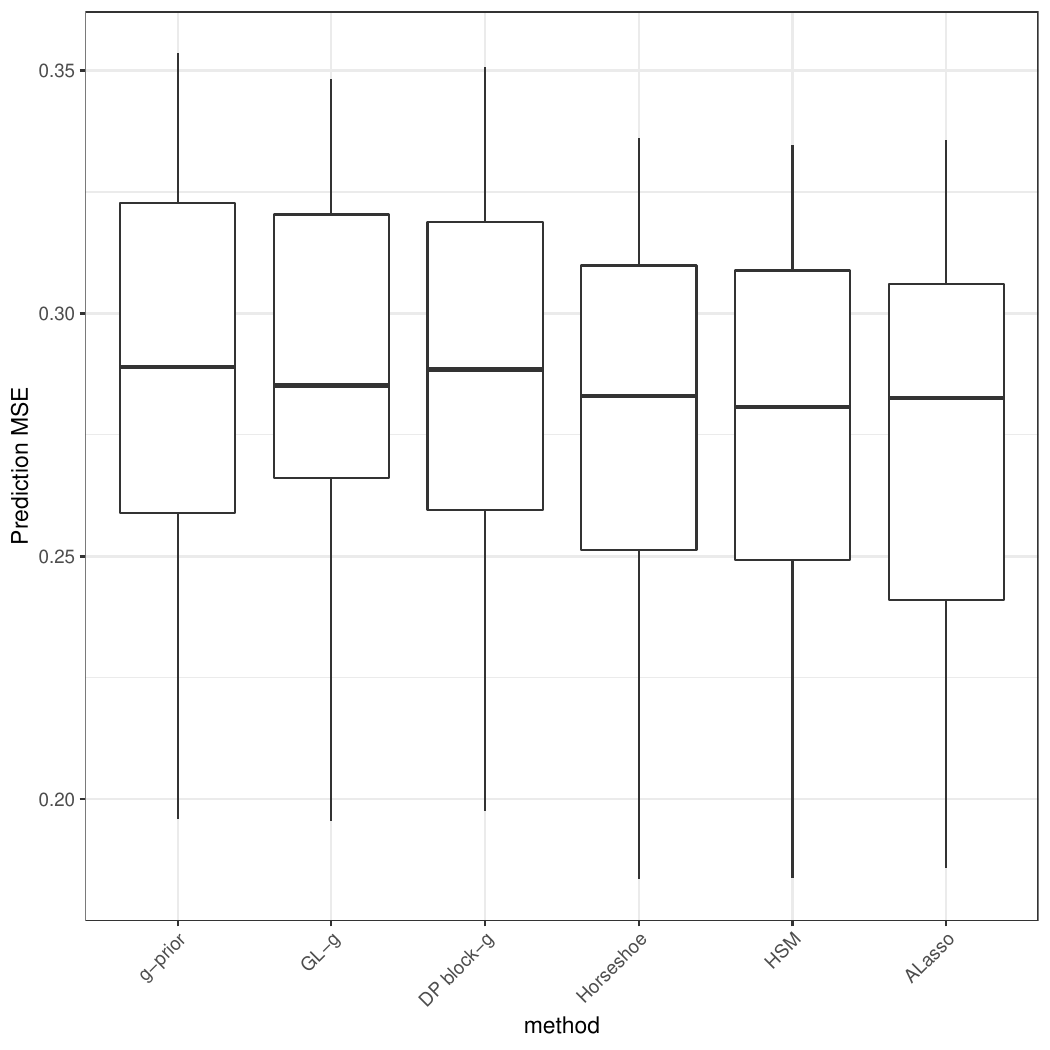}
        \caption{Predictive mean squared error}
        \label{fig:ozone_mse}
    \end{subfigure}
    \hfill
    \begin{subfigure}[b]{0.49\textwidth}
        \centering
        \includegraphics[width=\textwidth]{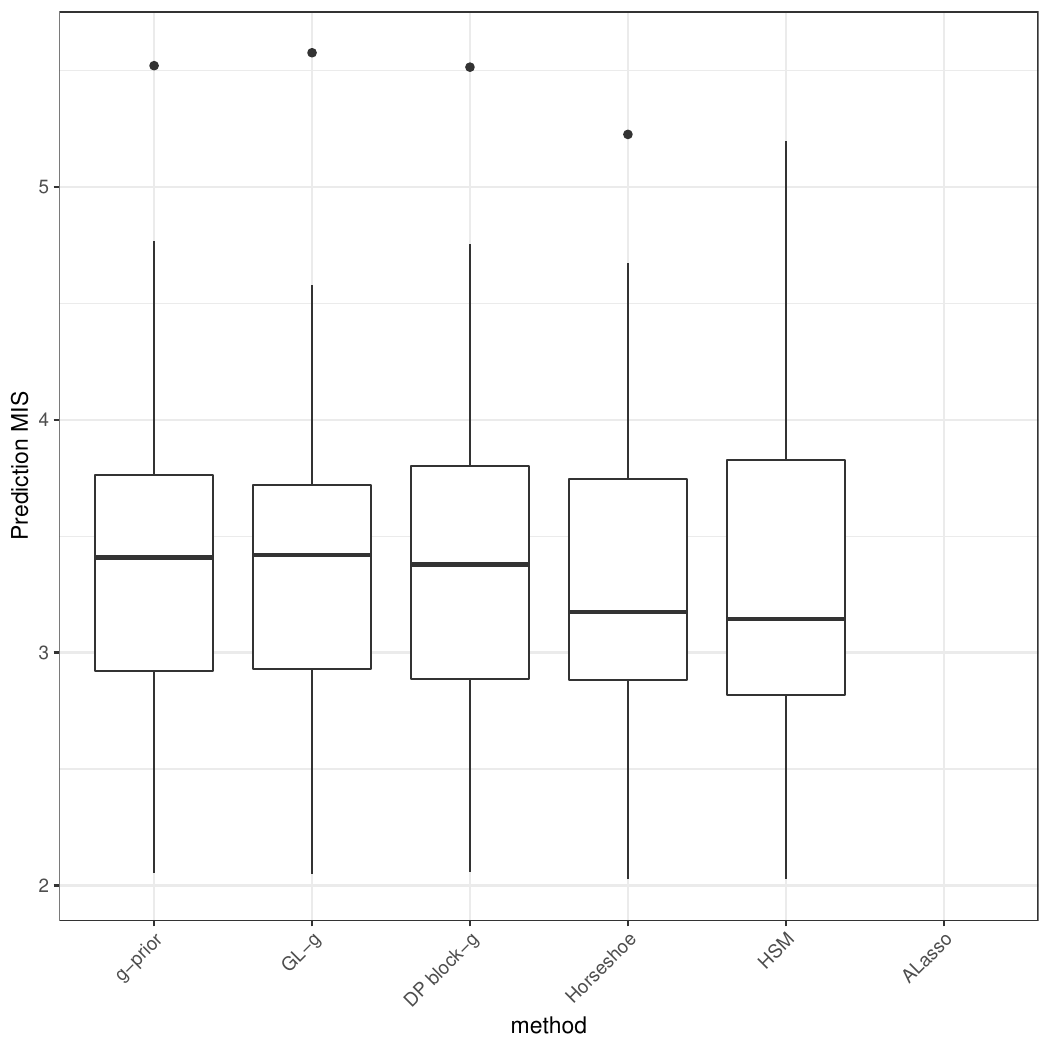}
        \caption{Median Interval Scores}
        \label{fig:ozone_mis}
    \end{subfigure}
    \caption{Predictive mean squared error (MSE) and and median interval scores (MIS) for our crossvalidation exercise for the \texttt{ozone} dataset.  Note that MIS is not readily available for ALasso or other penalized likelihood methods, so it is not included.}
    \label{fig:pred_ozone}
\end{figure}

Finally, Figure \ref{fig:pred_ozone} presents boxplots of the predictive mean squared error (MSE) and 95\% median intervals scores (MIS, \citealp{gneiting2007strictly}) for a crossvalidation exercise in which 20 random 80-20 splits of the data were used to train and then test prediction accuracy. For a variable $z$, the $\alpha \times 100\%$ IS is given by $IS_\alpha(l,u,z)= (u-l) + \frac{2}{1-\alpha}(l-z)\mathbbm{1}\{z<l\}+\frac{2}{1-\alpha}(z -u) \mathbbm{1}\{u<z\}$, where $l$ and $u$ denote the upper and lower bounds of the $\alpha \times 100\%$ posterior intervals of $z$. The first term in this expression rewards narrow predictive intervals, while the second rewards accurate coverage. We do not report the MIS for ALasso because the implementation in the \texttt{glmnet} package does not provide a measure of predictive uncertainty off the shelf. Generally speaking, Horseshoe, HSM and ALasso seem to have a slightly better predictive performance than {procedures based on standard $g$-priors, GL-$g$ priors and DP block $g$ priors}, particularly 
for point prediction. However, the differences are small.

\section{Discussion}\label{sec:discussion}

We introduced novel class of priors that that enable a parsimonious, data driven approach to model selection and prediction in linear models that is free from the so-called conditional Lindley ``paradox'' and that provides a bridge between two strands of the literature (model selection priors and continuous shrinkage priors) that have often been treated as distinct. {The use of a Dirichlet process prior for the distribution of the local shrinkage coefficients enable us to interpret our model in terms of blocks of covariates that receive the same level of shrinkage.  However, the method is not specially reliant to the clustering properties of Dirichlet process. In particular, it does not require that we are able to consistently estimate the ``true'' partition of the coefficients, only that the partitions that receive high probability a posteriori be those that do not mix both ``small'' and ``large'' coefficients.}
%

While this paper focuses on linear regression models, DP mixtures of block $g$ priors can be extended to generalized linear models, and perhaps even non-linear regression, by setting $\bfSigma_{\bfgamma}$ to be an appropriate information matrix, e.g., see \cite{bove2011hyper}, \cite{li2018mixtures} and \cite{porwal2023laplace}. The approach introduced here can also be used to generalize the class of priors introduced in \cite{carvalho2009objective}, leading to a new class of (mixtures of) hyper-inverse Wishart block $g$ priors for model selection in Gaussian graphical models.

{Recently, \cite{lee2020tail} demonstrated that, in the context of continuous shrinkage priors, the behavior of the tail of the distribution of the local shrinkage parameters plays an important role in the performance of the prior, and that different types of tail behavior might be necessary in sparse vs.\ ultra-sparse settings. The use of a non-parametric specification for the distribution of $g_j \mid \bfgamma$ through a Dirichlet process prior enables such tail adaptability in the context of model selection priors.  
}

From a theoretical perspective, there are two additional aspects of our work that are open for extension. First, our results around the conditional Lindley paradox assume that the design matrix is orthogonal. The evidence from our simulations suggested that the results, and in particular the ability of the model to separate ``large'' and ``small'' coefficients into separate clusters, extends to the non-orthogonal case. We believe that theoretical progress in this area can be achieved by developing an asymptotic expansion for the multivariate integral defining the marginal likelihood. Similarly, our model consistency results assume that $p$ is fixed. We believe that it possible to extend the result to the cases where $p$ grows 
with $n$. Both of these directions 
will be explored elsewhere.





\section{Supplementary Materials}

Supplementary materials contain the derivation of marginal likelihood in Equation \eqref{eq:marlikcond}, visualization of the tail behavior of the prior for the bivariate case, proof of theorems in Section \ref{sec:properties}, details of the MCMC algorithm discussed in Section \ref{sec:computation} and additional simulation results for Section \ref{sec:simulations}. \if1\blind
{An implementation of the MCMC algorithm is available at \url{https://github.com/Anupreet-Porwal/DP-mix-block-g-prior}.
} \fi 

\section{Acknowledgment}

We are grateful to the editor, the associate editor, and the referees for their valuable and constructive comments on an earlier version of this article. 
\if1\blind
{We also thank Prof. Merlise Clyde for her insights during the early stages of the project. 

} \fi

\section{Disclosure Statement}

The authors report there are no competing interests to declare. 

\if1\blind
{
\section{Funding}
This work was partially supported by grants NSF-2023495, NSF-2114727 and NSF-2523615.
} \fi 

\appendix

\section{Derivation of Equation \eqref{eq:marlikcond}}\label{ap:eq:marlikcond}

By definition,
\begin{align*}
    f(\bfy \mid \bfgamma,  g_1, \ldots, g_{p_{\bfgamma}}) = \int f(\bfy \mid \bfbeta_{\bfgamma}, \sigma^2)  f( \bfbeta_{\bfgamma} \mid \sigma^2, \bfG_{\bfgamma}, \bfgamma) f(\beta_0, \sigma^2) \dd \beta_0 \dd \bfbeta_{\bfgamma} \dd \sigma^2
\end{align*}

The integral with respect to $\bfbeta_{\bfgamma}$ is trivial to compute using the properties of the multivariate normal distribution, resulting in $\bfy \mid \beta_0, \sigma^2, \bfgamma,  g_1, \ldots, g_{p_{\bfgamma}} \sim \normal \left( \bfones \beta_0, \sigma^2 \bfOmega_{\bfgamma} \right)$, where $\bfOmega_{\bfgamma} = \bfI_n + \bfX_{\bfgamma}\bfG_{\bfgamma}^{1/2}  \bfSigma_{\bfgamma} \bfG_{\bfgamma}^{1/2} \bfX_{\bfgamma}^T$.  Integrating now with respect to $\beta_0$
\begin{align*}
    f(\bfy \mid \bfgamma,  \sigma^2, g_1, \ldots, g_{p_{\bfgamma}}) &= \int (2 \pi \sigma^2)^{-n/2} \left| \bfOmega_{\bfgamma} \right|^{-1/2} \\
    & \;\;\;\;\;\;\;\;\;\exp \left\{ -\frac{1}{2\sigma^2} \left( \bfy - \bfones \beta_0\right)^T \bfOmega_{\bfgamma}^{-1}\left( \bfy - \bfones \beta_0\right)\right\}\dd \beta_0 \\
    & = (2 \pi \sigma^2)^{-n/2} \left| \bfOmega_{\bfgamma} \right|^{-1/2} \exp\left\{ -\frac{1}{2\sigma^2} \left(  \bfy^T \bfOmega_{\bfgamma}^{-1} \bfy - \frac{\left(\bfones_n^{T} \bfOmega_{\bfgamma}^{-1} \bfy \right)^2}{\bfones_n^{T} \bfOmega_{\bfgamma}^{-1} \bfones_n} \right) \right\} \\
    & \;\;\;\;\;\;\;\;\; \int
    \exp\left\{ -\frac{\bfones^T\bfOmega_{\bfgamma}^{-1}\bfones}{2\sigma^2} \left(  \beta_0 - \frac{\bfones_n^{T} \bfOmega_{\bfgamma}^{-1} \bfy }{\bfones_n^{T} \bfOmega_{\bfgamma}^{-1} \bfones_n}\right)^2\right\}
    \dd \beta_0   \\
    & = \left(\frac{1}{2 \pi \sigma^2}\right)^{\frac{n-1}{2}} \frac{\left| \bfOmega_{\bfgamma} \right|^{-1/2}}{\left(\bfones_n^T \bfOmega_{\gamma}^{-1} \bfones_n \right)^{1/2}} \\
    & \;\;\;\;\;\;\;\; \exp\left\{ -\frac{1}{2\sigma^2} \left(  \bfy^T \bfOmega_{\bfgamma}^{-1} \bfy - \frac{\left(\bfones_n^{T} \bfOmega_{\bfgamma}^{-1} \bfy \right)^2}{\bfones_n^{T} \bfOmega_{\bfgamma}^{-1} \bfones_n} \right) \right\} .
\end{align*}
Finally, 
\begin{align*}
    f(\bfy \mid \bfgamma,  \sigma^2, g_1, \ldots, g_{p_{\bfgamma}}) &=  
    \left(\frac{1}{2 \pi}\right)^{\frac{n-1}{2}} \frac{\left| \bfOmega_{\bfgamma} \right|^{-1/2}}{\left(\bfones_n^T \bfOmega_{\gamma}^{-1} \bfones_n \right)^{1/2}} \\
    & \;\;\;\;\;\;\;\; \int
    \left( \frac{1}{\sigma^2} \right)^{\frac{n-1}{2}} 
    \exp\left\{ -\frac{1}{2\sigma^2} \left(  \bfy^T \bfOmega_{\bfgamma}^{-1} \bfy - \frac{\left(\bfones_n^{T} \bfOmega_{\bfgamma}^{-1} \bfy \right)^2}{\bfones_n^{T} \bfOmega_{\bfgamma}^{-1} \bfones_n} \right) \right\} \frac{1}{\sigma^2} \dd \sigma^2 \\
    & = 
    \frac{\Gamma\left(\frac{n-1}{2} \right)}{ \pi^{\frac{n-1}{2}}}
    \frac{\left|  \bfOmega_{\bfgamma} \right|^{-1/2}}
    { \left(\bfones_n^T \bfOmega_{\gamma}^{-1} \bfones_n \right)^{1/2}} \left[ \bfy^T \bfOmega_{\bfgamma}^{-1} \bfy - \frac{\left(\bfones_n^{T} \bfOmega_{\bfgamma}^{-1} \bfy \right)^2}{\bfones_n^{T} \bfOmega_{\bfgamma}^{-1} \bfones_n} \right]^{-\frac{n-1}{2}}.
\end{align*}

Note, however, that since the design matrix has been centered, $\bfones_n^T \bfX = \bfzero$.  This implies $\bfones_n^T \bfOmega_{\gamma}^{-1} \bfones_n = n$ and $\bfones_n^{T} \bfOmega_{\bfgamma}^{-1} \bfy  =  \sum_{i=1}^{n} y_i $, and yields the simplified form in \eqref{eq:marlikcond}.

\section{Scatterplots of realizations of various prior distributions in the bivariate case}\label{ap:scatterplotsbivariagte}

\begin{figure} 
    \centering
    \begin{subfigure}[b]{0.49\textwidth}
        \centering
        \includegraphics[width=\textwidth]{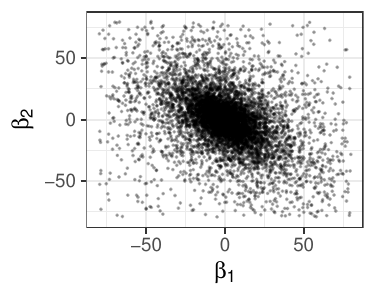}
        \caption{Standard $g$ prior}
        \label{fig:contours_gprior}
    \end{subfigure}
    \hfill
    \begin{subfigure}[b]{0.49\textwidth}
        \centering
        \includegraphics[width=\textwidth]{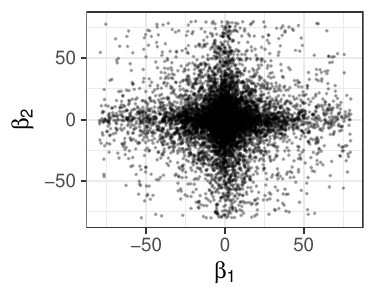}
        \caption{Orthogonal block $g$ prior}
        \label{fig:contours_hs}
    \end{subfigure}
    \centering
    \begin{subfigure}[b]{0.49\textwidth}
        \centering
        \includegraphics[width=\textwidth]{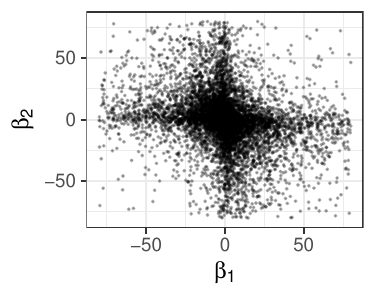}
        \caption{Global-local $g$ prior}
        \label{fig:contours_glp}
    \end{subfigure}
    \hfill
    \begin{subfigure}[b]{0.49\textwidth}
        \centering
        \includegraphics[width=\textwidth]{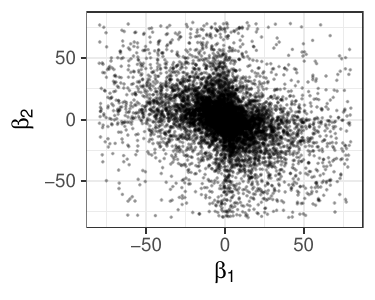}
        \caption{DP mixture of block $g$ priors}
        \label{fig:contours_dpbg}
    \end{subfigure}
\caption[]{Scatterplots of random samples from the Dirchlet mixture of block $g$ priors and some related distributions in the bivariate case under a hyper-$g/n$ distributon for the shrinkage parameter(s).  Panel (a) corresponds to the (elliptical) contours of the standard $g$ prior of \cite{liang2008mixtures}.  Panel (b) shows the density of the prior proposed by \cite{som2016conditional}, which assumes that blocks are orthogonal a priori.  Panel (c) corresponds to a global-local $g$ prior in which each covariate is assigned is own independent shrinkage factor and the prior covariance matrix is proportional to $\left( \bfX^T\bfX\right)^{-1}$.  Panel (d) is our DP mixture of block $g$ priors, which in this case corresponds to a mixture of the distributions in panels (a) and (c).}
\label{fig:contours}
\end{figure}

Figure \ref{fig:contours} shows scatterplots of random samples from the Dirchlet mixture of block $g$ priors and some related distributions in the bivariate case under a hyper-$g/n$ distribution.  In this special case, the DP block $g$ prior in panel (d) is a mixture of the standard $g$ prior in panel (a) and the block $g$ prior in panel (c).  { Note that the contour plots for (\ref{fig:contours}b), (\ref{fig:contours}c) and (\ref{fig:contours}d) are non-elliptical, unlike those of (\ref{fig:contours}a).  These non-elliptical contours indicate that these priors allow for direction-dependent shrinkage, which is what allows them to avoid the conditional Lindley paradox.  Furthermore, note that the axes of symmetry of the contour plots of the global-local $g$ prior and the DP block $g$ prior are not parallel to any of the main axes, a consequence of the fact that both priors account for prior correlation among the explanatory variables (unlike the orthogonal block $g$ prior of \citealp{som2016conditional}).}

\section{Proof of Theorems \ref{th:infoconsistency1} and \ref{th:infoconsistency2}}\label{ap:informationconsistency}

Two results will be useful in what follows.  First, the Woodbury matrix identity implies
\begin{multline*}
   \bfOmega_{\bfgamma}^{-1} = \left[ \mathbf{I}_n + \bfX_{\bfgamma} \bfG_{\bfgamma}^{1/2} \left( \bfX_{\bfgamma}^T \bfX_{\bfgamma} \right)^{-1}\bfG_{\bfgamma}^{1/2} \bfX_{\bfgamma}^T \right]^{-1} = \\
   \mathbf{I}_n - \bfX_{\bfgamma} \bfG_{\bfgamma}^{1/2}
    \left( \bfX_{\bfgamma}^T\bfX_{\bfgamma}  + \bfG_{\bfgamma}^{1/2}\bfX_{\bfgamma}^T\bfX_{\bfgamma}\bfG_{\bfgamma}^{1/2} \right)^{-1}
    \bfG_{\bfgamma}^{1/2} \bfX_{\bfgamma}  .    
\end{multline*}
Secondly, using the matrix determinant lemma
\begin{align*}
    \left|  \bfOmega_{\bfgamma} \right| = \left| \mathbf{I}_n + \bfX_{\bfgamma} \bfG_{\bfgamma}^{1/2} \left( \bfX_{\bfgamma}^T \bfX_{\bfgamma} \right)^{-1}\bfG_{\bfgamma}^{1/2} \bfX_{\bfgamma}^T \right| = \frac{\left| \bfX_{\bfgamma}^T \bfX_{\bfgamma} + \bfG_{\bfgamma}^{1/2} \bfX_{\bfgamma}^T \bfX_{\bfgamma} \bfG_{\bfgamma}^{1/2} \right|}{\left| \bfX_{\bfgamma}^T\bfX_{\bfgamma} \right|} .
\end{align*}

Now, on to the proof of Theorem  \ref{th:infoconsistency1}. 
 From equation \eqref{eq:marlikcond}, we have 
\begin{align*}
    \frac{f(\bfy \mid \bfgamma, \bfG)}{f(\bfy \mid \bfgamma = \bfzero)} &= 
    \left|  \bfOmega_{\bfgamma} \right|^{-1/2}  
    \left[ \frac{\bfy^T\bfy -  n\bar{y}^2}
    {\bfy^T \bfOmega_{\bfgamma}^{-1} \bfy - n\bar{y}^2} \right]^{\frac{n-1}{2}} \\
    & = \frac{\left| \bfX_{\bfgamma}^T\bfX_{\bfgamma} \right|^{1/2}}{\left| \bfX_{\bfgamma}^T \bfX_{\bfgamma} + \bfG_{\bfgamma}^{1/2} \bfX_{\bfgamma}^T \bfX_{\bfgamma} \bfG_{\bfgamma}^{1/2} \right|^{1/2}} \\
    & \;\;\;\; \left[ \frac{\bfy^T\bfy -  n\bar{y}^2}
    {\bfy^T\bfy -  n\bar{y}^2 - \bfy^T  \bfX_{\bfgamma} \bfG_{\bfgamma}^{1/2}
    \left( \bfX_{\bfgamma}^T\bfX_{\bfgamma}  + \bfG_{\bfgamma}^{1/2}\bfX_{\bfgamma}^T\bfX_{\bfgamma}\bfG_{\bfgamma}^{1/2} \right)^{-1}
    \bfG_{\bfgamma}^{1/2} \bfX_{\bfgamma}^T \bfy} \right]^{\frac{n-1}{2}} \\
    & = \frac{\left| \bfX_{\bfgamma}^T\bfX_{\bfgamma} \right|^{1/2}}{\left| \bfX_{\bfgamma}^T \bfX_{\bfgamma} + \bfG_{\bfgamma}^{1/2} \bfX_{\bfgamma}^T \bfX_{\bfgamma} \bfG_{\bfgamma}^{1/2} \right|^{1/2}} \\
    & \;\;\;\; \left[ 1 - R^2_{\bfgamma} \frac{\bfy^T  \bfX_{\bfgamma} \bfG_{\bfgamma}^{1/2}
    \left( \bfX_{\bfgamma}^T\bfX_{\bfgamma}  + \bfG_{\bfgamma}^{1/2}\bfX_{\bfgamma}^T\bfX_{\bfgamma}\bfG_{\bfgamma}^{1/2} \right)^{-1}
    \bfG_{\bfgamma}^{1/2} \bfX_{\bfgamma} ^T\bfy}{\bfy^T \bfX_{\bfgamma} \left( \bfX_{\bfgamma}^T \bfX_{\bfgamma}\right)^{-1} \bfX_{\bfgamma}^T\bfy} \right]^{-\frac{n-1}{2}}   \\
    & = \frac{\left| \bfX_{\bfgamma}^T\bfX_{\bfgamma} \right|^{1/2}}{\left| \bfX_{\bfgamma}^T \bfX_{\bfgamma} + \bfG_{\bfgamma}^{1/2} \bfX_{\bfgamma}^T \bfX_{\bfgamma} \bfG_{\bfgamma}^{1/2} \right|^{1/2}} \\
    & \;\;\;\; \left[ 1 - R^2_{\bfgamma} \frac{
    \hat{\bfbeta}_{\bfgamma}^T
    \left( \bfX_{\bfgamma}^T \bfX_{\bfgamma}  - \left[  \{ \bfX_{\bfgamma}^T \bfX_{\bfgamma} \}^{-1} + \bfG_{\bfgamma}^{1/2}\{ \bfX_{\bfgamma}^T \bfX_{\bfgamma} \}^{-1}\bfG_{\bfgamma}^{1/2}\right]^{-1}  \right)
    \hat{\bfbeta}_{\bfgamma}
    }{\hat{\bfbeta}_{\bfgamma}^T \bfX_{\bfgamma}^T \bfX_{\bfgamma}\hat{\bfbeta}_{\bfgamma}} \right]^{-\frac{n-1}{2}}  
    %
    %
\end{align*}
where $\hat{\bfbeta}_{\bfgamma}$ is the maximum likelihood estimator under model $\bfgamma$ and $R^2_{\bfgamma}$ is the coefficient of determination.  Now, let
$$
\Upsilon (\bfG_{\bfgamma}, \bfy) =  \frac{
    \hat{\bfbeta}_{\bfgamma}^T
    \left( \bfX_{\bfgamma}^T \bfX_{\bfgamma}  - \left[  \{ \bfX_{\bfgamma}^T \bfX_{\bfgamma} \}^{-1} + \bfG_{\bfgamma}^{1/2}\{ \bfX_{\bfgamma}^T \bfX_{\bfgamma} \}^{-1}\bfG_{\bfgamma}^{1/2}\right]^{-1}  \right)
    \hat{\bfbeta}_{\bfgamma}
    }{\hat{\bfbeta}_{\bfgamma}^T \bfX_{\bfgamma}^T \bfX_{\bfgamma}\hat{\bfbeta}_{\bfgamma}} .
$$
Then, the condition required for the Bayes factor to be information consistent in this setting can be written as
\begin{multline*}
   \lim_{\|\hat{\bfbeta}_{\bfgamma}\| \to \infty} \int \left| \bfX_{\bfgamma}^T \bfX_{\bfgamma} + \bfG_{\bfgamma}^{1/2} \bfX_{\bfgamma}^T \bfX_{\bfgamma} \bfG_{\bfgamma}^{1/2} \right|^{-1/2}  \\ 
   \left[ 1 - R^2_{\bfgamma} \Upsilon (\bfG_{\bfgamma}, \bfy) \right]^{-\frac{n-1}{2}} f(g_1, \ldots , g_{p_{\bfgamma}}) \dd g_1 \cdots \dd g_{p_{\bfgamma}} = \infty
\end{multline*}
which, because of dominated convergence, can be written as
\begin{multline}\label{eq:infconcondition}
   \int \left| \bfX_{\bfgamma}^T \bfX_{\bfgamma} + \bfG_{\bfgamma}^{1/2} \bfX_{\bfgamma}^T \bfX_{\bfgamma} \bfG_{\bfgamma}^{1/2} \right|^{-1/2}  \\ 
   \left\{\lim_{\|\hat{\bfbeta}_{\bfgamma}\| \to \infty} 
   \left[ 1 - R^2_{\bfgamma} \Upsilon (\bfG_{\bfgamma}, \bfy) \right]^{-\frac{n-1}{2}} \right\}f(g_1, \ldots , g_{p_{\bfgamma}}) \dd g_1 \cdots \dd g_{p_{\bfgamma}} = \infty
\end{multline}
Recalling that $\lim_{\| \hat{\bfbeta}_{\bfgamma} \| \to \infty} R^2_{\bfgamma} = 1$, \eqref{eq:infconcondition} reduces to 
\begin{multline*}
   \int \left| \bfX_{\bfgamma}^T \bfX_{\bfgamma} + \bfG_{\bfgamma}^{1/2} \bfX_{\bfgamma}^T \bfX_{\bfgamma} \bfG_{\bfgamma}^{1/2} \right|^{-1/2}  \\
   \left[ 1 - \Upsilon^{*} (\bfG_{\bfgamma}) \right]^{-\frac{n-1}{2}} f(g_1 , \ldots , g_{p_{\bfgamma}}) \dd g_1 \ldots \dd g_{p_{\bfgamma}} = \infty .
\end{multline*}
where $\Upsilon^{*} (\bfG_{\bfgamma}) = \lim_{\| \hat{\bfbeta}_{\bfgamma} \| \to \infty} \Upsilon (\bfG_{\bfgamma}, \bfy)$. 

Now, since $\bfX$ is full rank and all $g_j$s are strictly positive, both $\bfX_{\bfgamma}^T\bfX_{\bfgamma}$ and $
\bfX_{\bfgamma}^T \bfX_{\bfgamma}  - \left[  \{ \bfX_{\bfgamma}^T \bfX_{\bfgamma} \}^{-1} + \bfG_{\bfgamma}^{1/2}\{ \bfX_{\bfgamma}^T \bfX_{\bfgamma} \}^{-1}\bfG_{\bfgamma}^{1/2}\right]^{-1}
$ are strictly positive definite matrices.  It follows then that
\begin{multline*}
    0 < \lambda_{-} (\bfG) \left\| \hat{\bfbeta}_{\bfgamma} \right\|^2 \le \\
    \hat{\bfbeta}_{\bfgamma}^T
    \left( \bfX_{\bfgamma}^T \bfX_{\bfgamma}  - \left[  \{ \bfX_{\bfgamma}^T \bfX_{\bfgamma} \}^{-1} + \bfG_{\bfgamma}^{1/2}\{ \bfX_{\bfgamma}^T \bfX_{\bfgamma} \}^{-1}\bfG_{\bfgamma}^{1/2}\right]^{-1}  \right)
    \hat{\bfbeta}_{\bfgamma} \le  \\
    \lambda_{+} (\bfG) \left\| \hat{\bfbeta}_{\bfgamma} \right\|^2 < \infty,
\end{multline*}
and
\begin{align*}
    0 < \nu_{-} \left\| \hat{\bfbeta}_{\bfgamma} \right\|^2 \le \hat{\bfbeta}_{\bfgamma}^T \bfX^T\bfX \hat{\bfbeta}_{\bfgamma} \le \nu_{+} \left\| \hat{\bfbeta}_{\bfgamma} \right\|^2 <\infty ,
\end{align*}
where 
$\lambda_{-} (\bfG)$ and $\lambda_{+} (\bfG)$ are the largest and the smallest eigenvalues of the matrix
$
\bfX_{\bfgamma}^T \bfX_{\bfgamma}  - \left[  \{ \bfX_{\bfgamma}^T \bfX_{\bfgamma} \}^{-1} + \bfG_{\bfgamma}^{1/2}\{ \bfX_{\bfgamma}^T \bfX_{\bfgamma} \}^{-1}\bfG_{\bfgamma}^{1/2}\right]^{-1}
$ and $\nu_{+}$ and $\nu_{-}$ are the largest and smallest eigenvalues of $\bfX_{\bfgamma}^T\bfX_{\bfgamma}$.  Therefore 
\begin{align*}
    0 < \frac{\lambda_{-} (\bfG)}{\nu_{+}} \le \Upsilon (\bfG_{\bfgamma}, \bfy) \le \frac{\lambda_{+}}{\nu_{-}} (\bfG) < \infty .
\end{align*}
These bounds are independent of $\hat{\bfbeta}_{\bfgamma}$ and therefore apply to $\Upsilon^{*} (\bfG_{\bfgamma})$ as well. Hence:
\begin{multline*}
   \int \left| \bfX_{\bfgamma}^T \bfX_{\bfgamma} + \bfG_{\bfgamma}^{1/2} \bfX_{\bfgamma}^T \bfX_{\bfgamma} \bfG_{\bfgamma}^{1/2} \right|^{-1/2}  
   \left[ 1 - \Upsilon^{*} (\bfG_{\bfgamma}) \right]^{-\frac{n-1}{2}} f(g_1 , \ldots , g_{p_{\bfgamma}}) \dd g_1 \ldots \dd g_{p_{\bfgamma}} \ge \\
      \int \left| \bfX_{\bfgamma}^T \bfX_{\bfgamma} + \bfG_{\bfgamma}^{1/2} \bfX_{\bfgamma}^T \bfX_{\bfgamma} \bfG_{\bfgamma}^{1/2} \right|^{-1/2}  
   \left[ 1 - \frac{\lambda_{-} (\bfG)}{\nu_{+}} \right]^{-\frac{n-1}{2}} f(g_1 , \ldots , g_{p_{\bfgamma}}) \dd g_1 \ldots \dd g_{p_{\bfgamma}}.
\end{multline*}
This completes the proof of Theorem \ref{th:infoconsistency1}.  

The proof of Theorem \ref{th:infoconsistency2} relies on the fact that $BF_{\bfgamma, \bfzero} (\bfy)$ under the Dirichlet mixtures of block $g$ prior can be written as a weighted average of Bayes factors conditional on each of the possible partitions of the $p_{\bfgamma}$ coefficients,
\begin{align}\label{eq:BFsum}
    BF_{\bfgamma, \bfzero} (\bfy) = \sum_{\rho} p(\rho) BF_{\bfgamma, \bfzero} (\bfy \mid \rho)
\end{align}
where $p(\rho) = \int p(\rho \mid \alpha) p(\alpha) \dd \alpha$.  Note that one of the terms in the sum corresponds to $\rho = \rho_0 = \left\{ 
 \{ 1,2,3, \ldots, p\} \right\}$, i.e., the Bayes factor under the standard $g$ prior. 
Hence, if the density of the base measure satisfies 
$$
  \int (1 + g)^{(n-1-p_{\bfgamma})/2} f(g \mid \tau^2, a, b) \dd g = \infty
$$
then we know that
$$
\lim_{\| \hat{\bfbeta}_{\bfgamma} \| \to \infty} BF_{\bfgamma, \bfzero} (\bfy \mid \rho_0) = \infty.
$$
But all the other conditional Bayes factors $BF_{\bfgamma, \bfzero} (\bfy \mid \rho)$ in \eqref{eq:BFsum} are non-negative, so we must have $\lim_{\| \hat{\bfbeta}_{\bfgamma} \| \to \infty} BF_{\bfgamma, \bfzero} (\bfy) = \infty$.

\section{Proof of Theorem \ref{th:partition}}\label{ap:cluseringprop}

We start by introducing some notation.  Let $\rho = \{ S_1, \ldots, S_K \}$ be a partition of $\mathcal{I}$, $m_k = |S_k|$ be the number of elements in $S_k$ and, similarly, $m_{1,k} = |S_k \cap \mathcal{I}_1|$ and $m_{2,k} = |S_k \cap \mathcal{I}_2|$.  Clearly, $m_{1,k},m_{2,k} \ge 0$ and $m_{1,k} + m_{2,k} = m_k$.

Consider first the case where $\sigma^2$ is known.  Because $\bfX$ is orthogonal, it is easy to verify that, under \eqref{eq:betaprime}, 
\begin{align*}
    f\left(\bfy(N) \mid \rho, \sigma^2 \right) & \propto \left(\frac{1}{\sigma^2} \right)^{n/2}\prod_{k=1}^{K} \int_{0}^{1} u_k^{b} (1-u_k)^{a + \frac{m_k}{2}} \exp \left\{ \frac{u_k\| \hat{\bfbeta}_{S_k}(N) \|}{2\sigma^2} \right\} \dd u_k \\
    & \propto\left(\frac{1}{\sigma^2} \right)^{n/2} \prod_{k=1}^{K} M\left( b+1, a+ b + \frac{m_k}{2} + 2, \frac{\| \hat{\bfbeta}_{S_k}(N) \|}{2\sigma^2}  \right) 
    %
\end{align*}
where $M$ is Kummer's function \citep{abramowitz1988handbook}, $\| \hat{\bfbeta}_{S_k}(N) \| = \sum_{j \in S_k} \hat{\beta}_j^2(N)$, and $\hat{\beta}_j(N)$ is the maximum likelihood estimator of $\beta_j$ based on $\bfy(N)$.  A well known asymptotic expansion of Kummer's function is $M(a,b,z) \approx \frac{\Gamma(b_0)}{\Gamma(a_0)} z^{a_0 - b_0} \exp\{ z \}$ for large $z$ (see Equation 13.5.1 in page 508 of \citealp{abramowitz1988handbook}). Hence, under the assumptions of the theorem,
\begin{align*}
    f\left(\bfy(N) \mid \rho, \sigma^2 \right) &\approx  \exp \left\{ \mathcal{O}(N) \right\} \mathcal{O} \left( N^{1-a-\sum_{\{k: m_{1,k} > 0 \}} m_k/2}  \right)
\end{align*}
for large $N$.  Now, if $\rho \preceq \rho_0$, then $\sum_{\{k: m_{1,k} > 0 \}} |S_{k}| = p_1$ and therefore
\begin{align*}
    \lim_{N \to \infty} \frac{ f\left(\bfy(N) \mid \rho, \sigma^2 \right)}{ f\left(\bfy(N) \mid \rho_0, \sigma^2 \right)} =  \lim_{N \to \infty} \frac{ \exp \left\{ \mathcal{O}(N) \right\} \mathcal{O} \left( N^{1-a-p_1/2}  \right)}{ \exp \left\{ \mathcal{O}(N) \right\} \mathcal{O} \left( N^{1-a-p_1/2}  \right)} = c_{\rho}
\end{align*}
for some $0 < c_{\rho} < \infty$.

On the other hand, if  $\rho \npreceq \rho_0$ then there exists at least one $S_k$ such that both $m_{1,k} > 0$ and $m_{2,k} > 0$.  Recall that $m_{1,k} + m_{2,k} = m_k$.  Therefore, for a partition $\rho \npreceq \rho_0$, we have $p_1 = \sum_k m_{1,k} = \sum_{\{ k : m_{1,k} > 0 \}} m_{1,k} < \sum_{\{ k : m_{1,k} > 0 \}} m_k$.  Hence, in this case,
\begin{align*}
    \lim_{N \to \infty} \frac{ f\left(\bfy(N) \mid \rho, \sigma^2 \right)}{ f\left(\bfy(N) \mid \rho_0, \sigma^2 \right)} = 0.
\end{align*}

When $\sigma^2$ is unknown, note that \cite{som2014paradoxes} shows that the limit as $N \to \infty$ of the posterior distribution for $\sigma^2$ conditional on the partition is a proper, non-degenerate distribution.  Hence, we have the asymptotic expansion in this case is instead   
\begin{align*}
    f\left(\bfy(N) \mid \rho \right) &\approx   \mathcal{O} \left( N^{1-a-\sum_{\{k: m_{1,k} > 0 \}} m_k/2}  \right).
\end{align*}
for large $N$.  Hence, we again have
$$
\lim_{N \to \infty} \frac{f\left(\bfy(N) \mid \rho\right)}{f\left(\bfy(N) \mid \rho_{0}\right)} = 
\begin{cases}
    0 & \rho \nprec \rho_{0}, \\
    c_{\rho} & \rho \prec \rho_{0}
\end{cases}
$$
for come $0 < c_{\rho} < \infty$.

\section{Proof of Theorem \ref{th:conLindleyparadox}}\label{ap:condLindleyparadox}

As in Theorem \ref{th:partition}, let $\mathcal{I}_1 = \{ j^{(1)}_1, \ldots, j^{(1)}_{p_1} \}$ denote the set of indexes associated with the covariates included in the design matrix $\bfX_1$, $\mathcal{I}_2 = \{ j^{(2)}_1, \ldots, j^{(2)}_{p_2} \}$ be the set associated with $\bfX_2$, and $\mathcal{I} = \mathcal{I}_1 \cup \mathcal{I}_2$.  Also, recall that $\mathcal{I}_1 \cap \mathcal{I}_2 = \emptyset$. 
%
Now, note that
\begin{align*}
BF_{\bfgamma_a, \bfgamma_0}(\bfy(N)) &= \frac{\sum_{\rho } f(\rho \mid \bfgamma_a) f\left(\bfy(N) \mid \bfgamma_a, \rho\right)}
{\sum_{\rho } f(\rho \mid \bfgamma_{0}) f\left( \bfy(N) \mid \bfgamma_0, \rho \right)} \\  
& = \underbrace{ \frac{ f\left(\bfy(N) \mid \bfgamma_a, \rho_a \right)}
{f\left(\bfy(N) \mid \bfgamma_0, \rho_0\right)} }_{A}
\underbrace{ \frac{f(\rho_a \mid \bfgamma_a) + \sum_{\rho \ne \rho_a} f(\rho \mid \bfgamma_a) \frac{f(\bfy(N) \mid \bfgamma_a, \rho)}{f(\bfy(N) \mid \bfgamma_a, \rho_a)} }
{f(\rho_0 \mid \bfgamma_0) + \sum_{\rho \ne \rho_0} f(\rho \mid \bfgamma_0) \frac{f(\bfy(N) \mid \bfgamma_0, \rho)}{f(\bfy(N) \mid \bfgamma_0, \rho_0)}} }_{B}
\end{align*}
where $\rho_0 = \{ \mathcal{I}_1\}$ and $\rho_a = \{ \mathcal{I}_1, \mathcal{I}_2 \}$.  

First focus on the $A$ term.  Because of the orthogonality of the design matrix, from \cite{som2014paradoxes} we know that 
$$
\lim_{N \to \infty} \frac{ f\left(\bfy(N) \mid \bfgamma_a, \rho_0\right)}
{f\left(\bfy(N) \mid \bfgamma_0, \rho_a\right)} > 0 .
$$

Focus now on the $B$ term. From Theorem \ref{th:partition}, the numerator is clearly strictly positive { as long as the prior distribution $f(\rho \mid \bfgamma_a)$ puts non-zero probability on at least one refinement of of $\rho_a$.  The prior distributions on partitions induced by the Dirichlet process obviously satisfies this requirement, as it places positive probability on every possible partition of $\mathcal{I}$}. Similarly, for the denominator, note that the partitions over which we are summing are, by definition, all refinements of $\rho_0$.  Hence, again from Theorem \ref{th:partition}, we know that the limit of each of the terms in the sum is finite and strictly positive.  Hence, $\lim_{N \to \infty} \sum_{\rho \ne \rho_0} f(\rho \mid \bfgamma_0) \frac{f(\bfy(N) \mid \bfgamma_0, \rho)}{f(\bfy(N) \mid \bfgamma_0, \rho_0)}$ is also finite and therefore
$$
\lim_{N \to \infty} \frac{f(\rho_a \mid \bfgamma_a) + \sum_{\rho \ne \rho_a} f(\rho \mid \bfgamma_a) \frac{f(\bfy(N) \mid \bfgamma_a, \rho)}{f(\bfy(N) \mid \bfgamma_a, \rho_a)} }
{f(\rho_0 \mid \bfgamma_0) + \sum_{\rho \ne \rho_0} f(\rho \mid \bfgamma_0) \frac{f(\bfy(N) \mid \bfgamma_0, \rho)}{f(\bfy(N) \mid \bfgamma_0, \rho_0)}} > 0 .
$$
This completes the proof.

\section{Proof of Theorem \ref{th:asympconsistency}}\label{ap:modelselconsistency}

Since 
$$
Pr(\bfgamma = \bfgamma_{T} \mid \bfy)=\frac{1}{1 + \sum_{\bfgamma \ne \bfgamma_T}\frac{f(\bfgamma)}{f(\bfgamma_T)} \frac{f(\bfy \mid \bfgamma)}{f(\bfy \mid \bfgamma_T)}}
$$
and $f(\bfgamma_T) > 0$, it is enough to show that 
$$
\frac{f(\bfy \mid \bfgamma)}{f(\bfy \mid \bfgamma_T)}  \xrightarrow[n \to \infty]{P} 
  0 .
$$ 
for all $\bfgamma \ne \bfgamma_T$.  Now
\begin{align*}
    \frac{f(\bfy \mid \bfgamma)}{f(\bfy \mid \bfgamma_T)} &= \frac{\sum_{\rho}f(\bfy \mid \bfgamma, \rho) f(\rho)}{\sum_{\rho} f(\bfy \mid \bfgamma_T, \rho)f(\rho)} \nonumber \\
    & = \underbrace{\frac{f(\bfy \mid \bfgamma, \rho_0)}{f(\bfy \mid \bfgamma_T, \rho_0)} }_{A} \underbrace{\left[\frac{f(\rho_0) + \sum_{\rho \ne \rho_0} \frac{f(\bfy \mid \bfgamma, \rho)}{f(\bfy \mid \bfgamma, \rho_0)}f(\rho)}{f(\rho_0) + \sum_{\rho \ne \rho_0} \frac{f(\bfy \mid \bfgamma_T, \rho)}{f(\bfy \mid \bfgamma_T, \rho_0)}f(\rho)} \right]}_{B}
\end{align*}
where $\rho_0 = \{ \{ 1, 2, 3, \ldots, p_{\bfgamma}\}\}$, i.e., the partition that assigns all covariates to a single block.

Note that $A$ is the Bayes factor based on the standard $g$ prior.  Since our hyperprior $p(g \mid a,b,\tau^2)$ is a member of the Confluent Hypergeometric (CH) family of distributions and $\tau^2 \sim \mathcal{O}(n)$, this Bayes factor is known to be consistent (e.g., see \citealp{li2018mixtures}).  Hence, 
\begin{align*}
\frac{f(\bfy \mid \bfgamma, \rho_0)}{f(\bfy \mid \bfgamma_T, \rho_0)}  &\xrightarrow[n \to \infty]{P} 
  0     
\end{align*}
for all $\bfgamma \ne \bfgamma_T$.  On the other hand, $f(\bfy \mid \bfgamma, \rho)$ and $f(\bfy \mid \bfgamma, \rho_0 )$ share the same likelihood and differ only on their priors, which are both (approximately) unit information.  Hence,
$$
\frac{f(\bfy \mid \bfgamma, \rho)}{f(\bfy \mid \bfgamma, \rho_0)}  \xrightarrow[n \to \infty]{P} 
  c_{\bfgamma,\rho}, 
$$ 
for all $\bfgamma$, where $0 < c_{\bfgamma,\rho} < \infty$.  Hence, $B$ converges to a finite constant, and the product of A and B converges to zero as desired.


\section{Details of the MCMC algorithm}\label{ap:MCMCM}

As mentioned in Section \ref{sec:computation} of the main manuscript, to construct the MCMC algorithm for our model we take advantage of the conditional conjugacy of the priors and, when possible, we integrate out the intercept $\beta_0$, the vector of regression coefficients $\bfbeta_{\bfgamma}$ and/or the variance $\sigma^2$ when deriving conditional posteriors.  Additionally, we represent the shrinkage coefficients $g_1, \ldots, g_{p_{\bfgamma}}$ in terms of their unique values $\tilde{\bfg}_{\bfgamma} = (\tilde{g}_1, \ldots, \tilde{g}_{K_{\bfgamma}})$ and the group indicators $\bfxi_{\bfgamma} = (\xi_1, \ldots, \xi_{p_{\bfgamma}})$.  The resulting algorithm alternates sampling from the full conditionals $f(\bfgamma, \tilde{\bfg} , \bfxi  \mid \cdots)$, $f(\bfxi \mid \cdots)$, $f(\alpha \mid \cdots)$, $f(\tilde{\bfg} \mid \cdots)$, $f(\beta_0, \bfbeta_{\bfgamma} \mid \cdots)$ and $f(\sigma^2 \mid \cdots)$.  Special cases of our model where either the partition defined by $\bfxi$ and/or the model $\bfgamma$ have been fixed in advance can be handled through slight modifications of the algorithm.  The steps that we use are as follows:

\begin{enumerate}
    \item We sample from the conditional posterior $f(\bfgamma, \tilde{\bfg} , \bfxi  \mid \cdots)$ given by 
\begin{align*}
        f(\bfgamma, \tilde{\bfg}, \bfxi \mid\cdots) \propto f(\bfy \mid \bfgamma,  \tilde{\bfg}, \bfxi)
     f \left(\tilde{\bfg} \mid \bfgamma \right) f(\bfxi \mid \bfgamma, \alpha)  f(\bfgamma) ,
\end{align*}

where $ f(\bfy \mid \bfgamma, \tilde{g}_1, \ldots, \tilde{g}_{K_{\bfgamma}}, \xi_1, \ldots, \xi_{p_{\bfgamma}})$ corresponds to \eqref{eq:marlikcond} with $\bfSigma_{\bfgamma} = \left\{ \bfX_{\bfgamma}^T\bfX_{\bfgamma} \right\}^{-1}$, and $f(\bfgamma)$ is an appropriate prior on the space of models, e.g., a Beta-Binomial prior. We generate samples from the above distribution using a random walk Metropolis Hastings algorithm using a symmetric random walk proposal for $\bfgamma$ similar to equation (46) of \cite{george1997approaches} as follows:
 
 \begin{itemize}
     \item We define a probability vector $p_1=(0.7,0.3)$.
     \item Each time, we decide on one of two types of moves according to the probability vector $p_1$.
     \begin{itemize}
         \item If a move type 1 is selected, then the proposed new model $\bfgamma^{(prop)}$ is generated by randomly flipping one component of $\bgam$.
         \item If a move type 2 is selected, the proposed model $\bfgamma^{(prop)}$ is generated by removing one variable currently included in the model and replacing it with a variable currently excluded, leaving the dimensionality of the model unchanged.  The variables to be added and removed are chosen uniformly at random within each set.
     \end{itemize}

 If a new variable is included in the model $\{i: \bfgamma=0, \bfgamma^{(prop)}=1\}$, draw $\xi_i$ from the following distribution
\begin{align*}
    \Pr(\xi_i =k) \propto \begin{cases} 
m_{\bfgamma,k} & \text{for } k = 1, 2, \dots, K_\gamma, \\
\alpha & \text{for } k = K_\gamma + 1.
\end{cases}
\end{align*}
If necessary, draw $\tilde{g}_{K_{\bfgamma}+1}$ from the centering measure $f(\tilde{g}_j \mid \tau^2, a, b)$.

Similarly, if a variable is removed from the model, remove the corresponding $\xi_i$ and update the number of clusters and partition if a variable that was in a singleton cluster was removed. Update to get $\bfxi^{(prop)},K_{\bfgamma^{(prop)}},\rho_{\bfgamma^{(prop)}}$ and $\tilde{\bfg}^{(prop)}$ accordingly. Then the proposed model is accepted with probability. 
$$
 \min\left\{\frac{f(\bfy \mid \bfgamma^{(prop)},  \tilde{\bfg}^{(prop)}, \bfxi^{(prop)})
     p \left(\tilde{\bfg}^{(prop)} \mid \bfgamma^{(prop)} \right) p(\bfxi^{(prop)} \mid \bfgamma^{(prop)}, \alpha)  p(\bfgamma^{(prop)})}{ f(\bfy \mid \bfgamma,  \tilde{\bfg}, \bfxi)
     p \left(\tilde{\bfg} \mid \bfgamma \right) p(\bfxi \mid \bfgamma, \alpha)  p(\bfgamma)},1\right\}.
$$ 
\end{itemize}

\item Once the model is sampled, we can update $\beta_0$ and $\betag$ by exploiting normal-normal conjugacy as follows: 
     \begin{align*}
         \beta_0 \mid \cdots 
         &\sim \mathcal{N}(\Bar{\bfy},\frac{\sigma^2}{n}),\\
         \betag \mid  \cdots
         &\sim \mathcal{N}(\bm_{\bgam,\bfxi},\bV_{\bgam,\bfxi})
     \end{align*}
     where 
     \begin{align*}
    \bV_{\bgam,\bfxi}= \sigma^2\left\{\frac{{\bfG_{\bgam}}^{-1/2}\bfSigma_{\bgam}^{-1}{\bfG_{\bgam}}^{-1/2}}{\tau^2} + \bfX_{\bgam}^{T}\bfX_{\bgam}\right\}^{-1} ,\quad\quad 
\bm_{\bgam,\bfxi}=\frac{\bV_{\bgam,\bfxi}\bfX_{\bgam}^T\bfy}{\sigma^2}.
    \end{align*}
\item  We can sample sample variance as 
\begin{align*}
    \sigma^2\mid  \cdots\sim \text{Inverse-Gamma}\left(\frac{n-1}{2}, \frac{\bfy^T(\bfI+\tau^2 \bfX_{\bgam}{\bfG_{\bgam}}^{1/2}\bfSigma_{\bgam}{\bfG_{\bgam}}^{1/2}\bfX_{\bgam}^T)^{-1}\bfy -n\Bar{\bfy}^2}{2}\right)
\end{align*}

\item Conditional on the current model and the observational variance $\sigma^2$, sequentially sample $\xi_i$ for variables included in the model i.e. $\mathcal{I} = \{i: \gamma_i=1\}$ similar to Algorithm 8 of \cite{neal2000markov}:   Let $K_{\bfgamma}^{-}$ be the number of distinct $\xi_j$ for $j\neq i$ and let $h=K_{\bfgamma}^{-}+d$. We choose $d$ to be 20, by default. Label these $\xi_j$ with values $\{1,\dots, K_{\bfgamma}^{-}\}$. If $\xi_i=\xi_j$, for some $j\neq i$, draw values independently from the base measure given by \eqref{eq:betaprime} for those $\tilde{g}_k$ for which $1\leq k\leq K_{\bfgamma}^{-}$. If $\xi\neq\xi_j$ for all $j\neq i$, let $\xi_i$ have the label $K_{\bfgamma}^{-}+1$ and draw independently from the base measure for those $\tilde{g}_k$ for which $K_{\bfgamma}^{-}+1 < k \leq h$. The, draw a new value of $\xi_i$ from $\{1,\dots,h\}$ with probabilities
\begin{align*}
    \Pr(\xi_i=k\mid \cdot) \propto \begin{cases}
        m_{\bfgamma,k}^{-i} \phi\left(\bfbeta_{\bfgamma} \mid \bfzero, \sigma^2 \bfG_{\bfgamma,k}^{\star 1/2} 
    \left\{ \bfX_{\bfgamma}^T \bfX_{\bfgamma}\right\}^{-1} \bfG_{\bfgamma,k}^{\star 1/2} \right) & \text{ for } 1\leq k\leq K_{\bfgamma}^{-} \\
    \frac{\alpha}{d} \phi\left(\bfbeta_{\bfgamma} \mid \bfzero, \sigma^2 \bfG_{\bfgamma,k}^{\star 1/2} 
    \left\{ \bfX_{\bfgamma}^T \bfX_{\bfgamma}\right\}^{-1} \bfG_{\bfgamma,k}^{\star 1/2} \right) & \text{ for } K_{\bfgamma}^{-}+1 < k \leq h
    \end{cases} 
\end{align*}
where $m_{\bfgamma,k}^{-i} $ is the number of $\xi_j$ for $j\neq i$ that are equal to k  and $\bfG_{\bfgamma,k}^{\star}$ is same as $\bfG_{\bfgamma}$ except for the fact that $g_{\xi_i}$ replaced by $g_{k}$.  

\item Using equation (2) and (3) of \cite{rodriguez2013jeffreys}, the posterior distribution of the concentration parameter $\alpha$ can be written as
\begin{align*}
    f(\alpha \mid  \cdots) &\propto f(\bfxi \mid \bfgamma, \alpha) f(\alpha \mid \bfgamma)\\
    &\propto \frac{\Gamma(\alpha)}{\Gamma(\alpha+p_{\bfgamma})}\alpha^{K_{\bfgamma}}\prod_{k=1}^{K_{\bfgamma}}\Gamma(m_{\bfgamma,k}) \sqrt{\frac{1}{\alpha}\sum_{j=1}^{p_{\bfgamma}-1} \frac{j}{(\alpha + j)^2} }
\end{align*}

To sample from the above density, we employ a random walk Metropolis-Hasting algorithm with Gaussian proposals for $\log \alpha$; the default variance was the proposal was 0.05 but this needs to be tuned, depending on the dataset to achieve an average acceptance rate of 40-50\%. 

\item The conditional posterior distribution of $\tilde{g}_k$ for $k=1,\dots,K_{\bfgamma}$ is given by 
\begin{align*}
    f(\tilde{g}_k\mid \cdot)&\propto 
     \phi\left(\bfbeta_{\bfgamma} \mid \bfzero, \sigma^2 \bfG_{\bfgamma}^{1/2} 
    \left\{ \bfX_{\bfgamma}^T \bfX_{\bfgamma}\right\}^{-1} \bfG_{\bfgamma}^{1/2} \right) f(\tilde{g}_k \mid \tau, a, b) \\
    &= \phi\left(\bfbeta_{\bfgamma} \mid \bfzero, \tau^2\sigma^2 \bfG_{\bfgamma}^{1/2} 
    \left\{ \bfX_{\bfgamma}^T \bfX_{\bfgamma}\right\}^{-1} \bfG_{\bfgamma}^{1/2} \right) f(\tilde{g}_k \mid \tau = 1, a, b) 
\end{align*}

We can simplify the above conditional posterior as
\begin{align*}
    f(\tilde{g}_k\mid .)\propto {(\tilde{g}_k)}^{b-\frac{m_{\bgam,k}}{2}}(1+\tilde{g}_k)^{-a-b-2}\exp\left(-\frac{v_k}{\tilde{g}_k}-\frac{w_k}{\sqrt{\tilde{g}_k}}\right), 
\end{align*}
where 
\begin{align*}
  v_k=\frac{1}{2\sigma^2\tau^2}\sum_{\substack{j\in S_{\bgam,k}\\ i\in S_{\bgam,k} }} \bfSigma_{\bgam,jj}^{-1}\beta_{\bgam,j}\beta_{\bgam,i}\quad \quad 
  w_k=\frac{1}{\sigma^2\tau^2}\sum_{\substack{j\in S_{\bgam,k}\\ i\notin S_{\bgam,k} } }\frac{\bfSigma_{\bgam,ji}^{-1}\beta_{\bgam,j}\beta_{\bgam,i}}{\sqrt{g_{\xi_i}}}.  
\end{align*}

Using the transformation $t_k=\frac{v_k}{\tilde{g}_k}$, we can re-parametrize this density as
\begin{align*}
    f(t_k\mid .)\sim t_k^{a+\frac{m_{\bgam,k}}{2}} \left(1+\frac{t_k}{v_k}\right)^{-a-b-2}\exp\left(-t_k-\frac{w_k}{\sqrt{v_k}}\sqrt{t_k}\right).
\end{align*}

Introduce the auxiliary variable $u_k$. Then, we can use slice sampling in conjunction with a modification of rejection sampler developed by \cite{liu2012rejection} to sample $t_k$ from a truncated extended gamma distribution  to sample $t_k$ as follows
\begin{align*}
    u_k\mid t_k&\sim \mathcal{U}\left(0, \left(\frac{v_k}{v_k+t_k}\right)^{a+b+2}\right),\\
    t_k \mid  u_k,. &\sim \text{Truncated-Extended-Gamma}\left(a+\frac{m_{\bgam,k}}{2}+1, \frac{w_k}{2\sqrt{v_k}}, v_k(u_k^{\frac{-1}{a+b+2}}-1)\right),
\end{align*}
where Truncated-Extended-Gamma distribution is given by 
\begin{align*}
    f(t\mid a,b,c)\propto t^{a-1}\exp(-t-2\sqrt{t}b)\mathbbm{1}_{\{0<t<c\}}, \quad t>0,
\end{align*}
for $a>0$ and $b\in\reals$.  Note that rejection sampler for untruncated extended gamma distributions developed by \cite{liu2012rejection} can be modified in a straightforward manner by using truncated proposals at the truncation level $c$. This can then be used to efficiently draw from truncated versions of extended Gamma distribution.

\end{enumerate}

\section{Additional results for our second simulation study}\label{ap:MSEest}

In this section we supplement the results shown in Section 6.2 of the main paper by exploring the ability of the DP block $g$-prior to identify clusters of coefficients with similar shrinkage coefficients, as well as the estimation accuracy of the various procedures discussed in the paper.

\begin{figure} 
    \centering
    \begin{subfigure}[b]{0.49\textwidth}
        \centering
        \includegraphics[width=\textwidth]{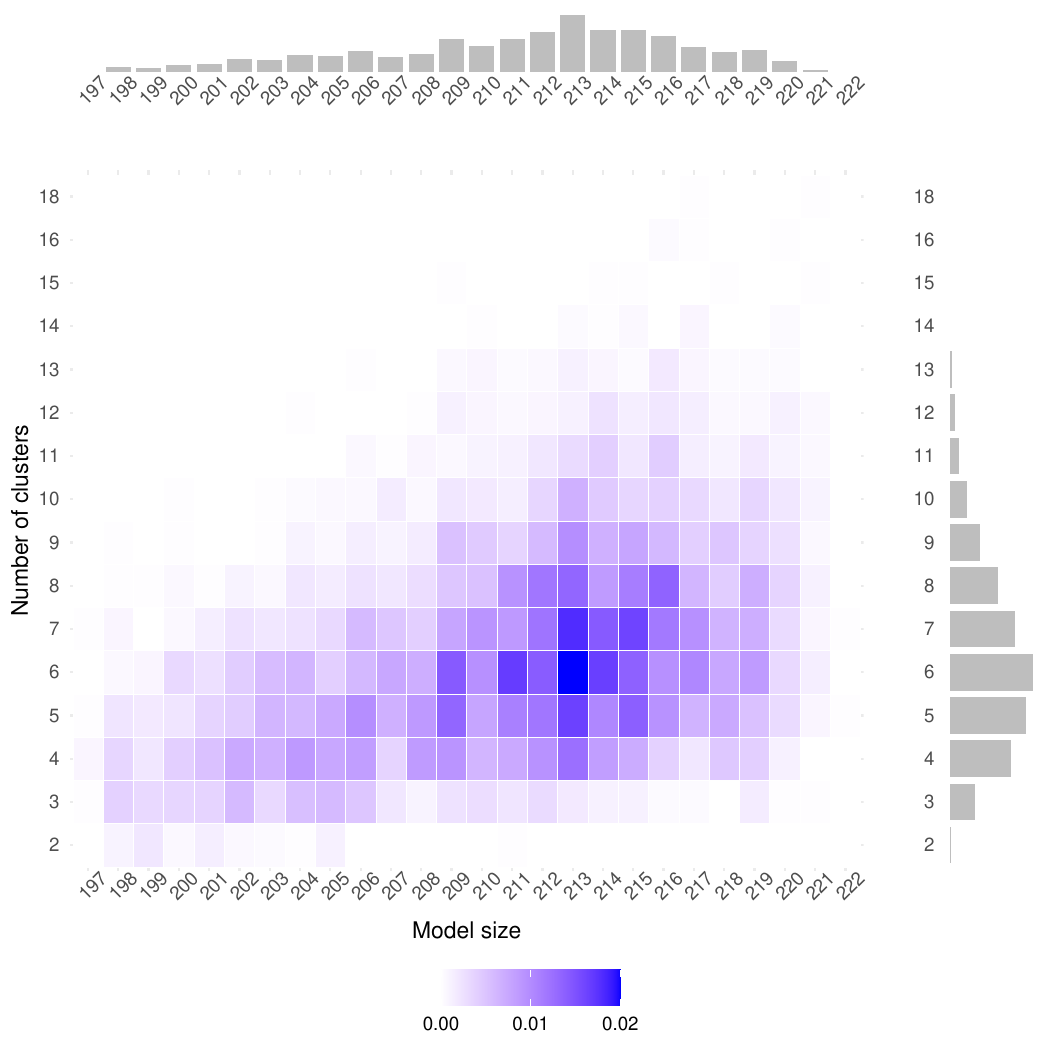}
        \caption{Dataset 31}
        \label{fig:joint_p_K_sim_31}
    \end{subfigure}
    \hfill
    \begin{subfigure}[b]{0.49\textwidth}
        \centering
        \includegraphics[width=\textwidth]{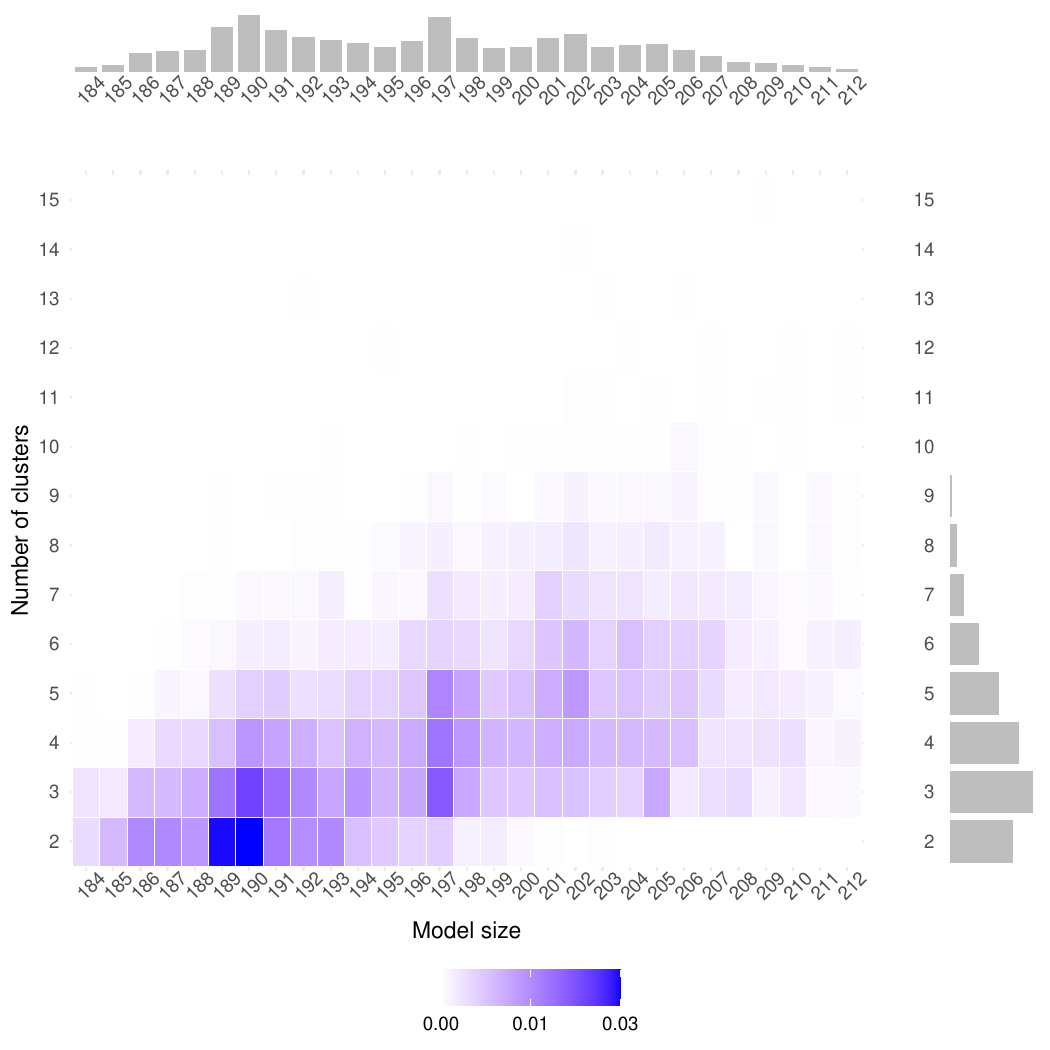}
        \caption{Dataset 63}
        \label{fig:joint_p_K_sim_63}
    \end{subfigure}
\caption[]{Joint and marginal posterior distributions for $p_{\gamma}$ and $K_{\bfgamma}$ in two representative examples of our simulated datasets in the $p=250$ and $\eta=0$ scenario.}
\label{fig:joint_p_K_sim}
\end{figure}

{ First, to provide some intuition on the role of clustering in the performance of Dirichlet process mixtures of block-$g$ priors, we present in Figure \ref{fig:joint_p_K_sim}, the joint and marginal posterior distributions for $p_{\gamma}$ and $K_{\bfgamma}$ for the DP block $g$ ($\tau^2 = n$) in two representative datasets.  Note that, in both cases, the posterior distribution puts all of its mass in at least two clusters of coefficients, with a (marginal) mode of 3 and maximum of 9 in the Dataset 63, and mode of 6 and a maximum of 13 in Dataset 31.  These results are in agreement with the theory developed in Theorem \ref{th:partition}.}

\begin{figure}[H]
    \centering
    \includegraphics[width = 0.95\linewidth]{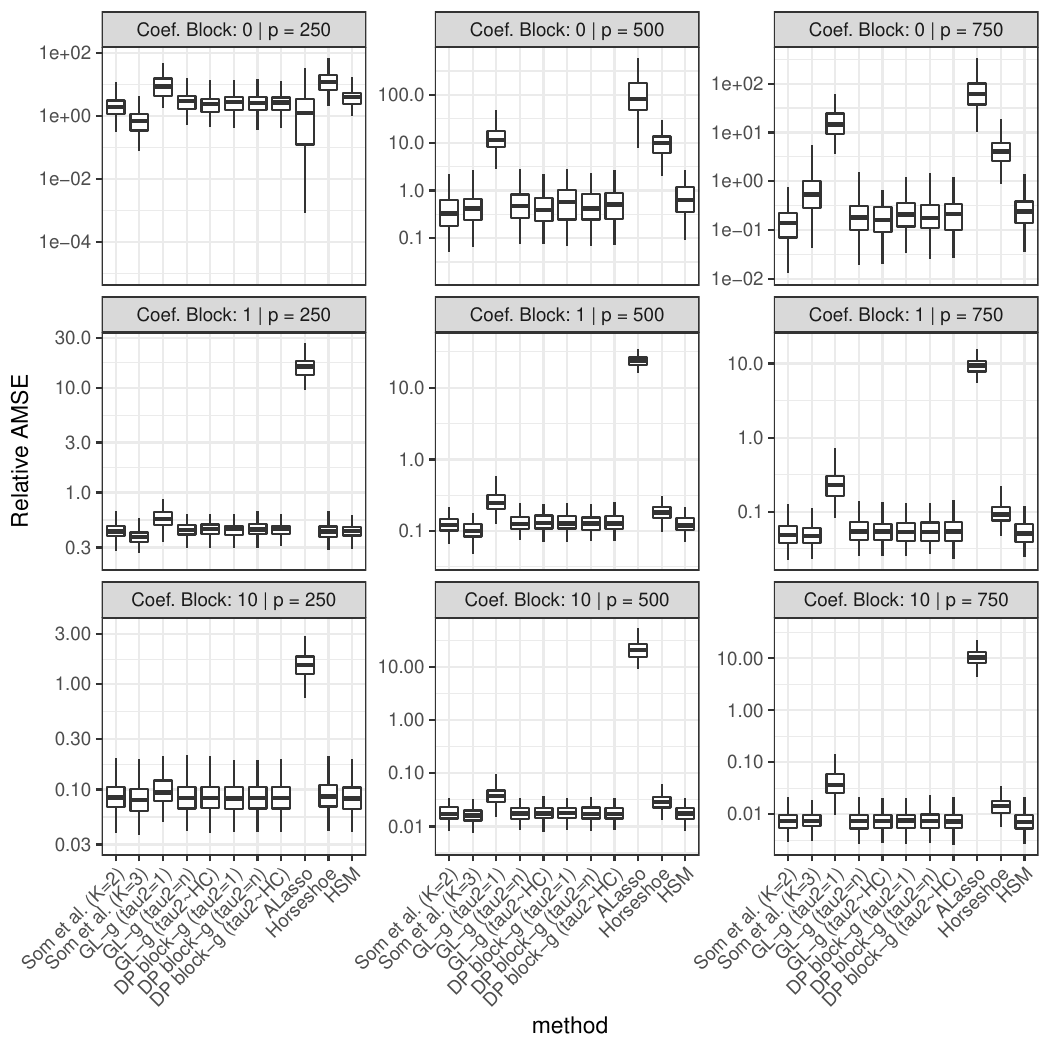}
    \caption{Relative mean squared error of the coefficients for $\eta = 0$, broken down by coefficient block.  Results are shown in the log scale because of the poor performance of ALasso.}
\label{fig:normMSE_beta_rho0}
\end{figure}

\begin{figure}[H]
    \centering
    \includegraphics[width = 0.95\linewidth]{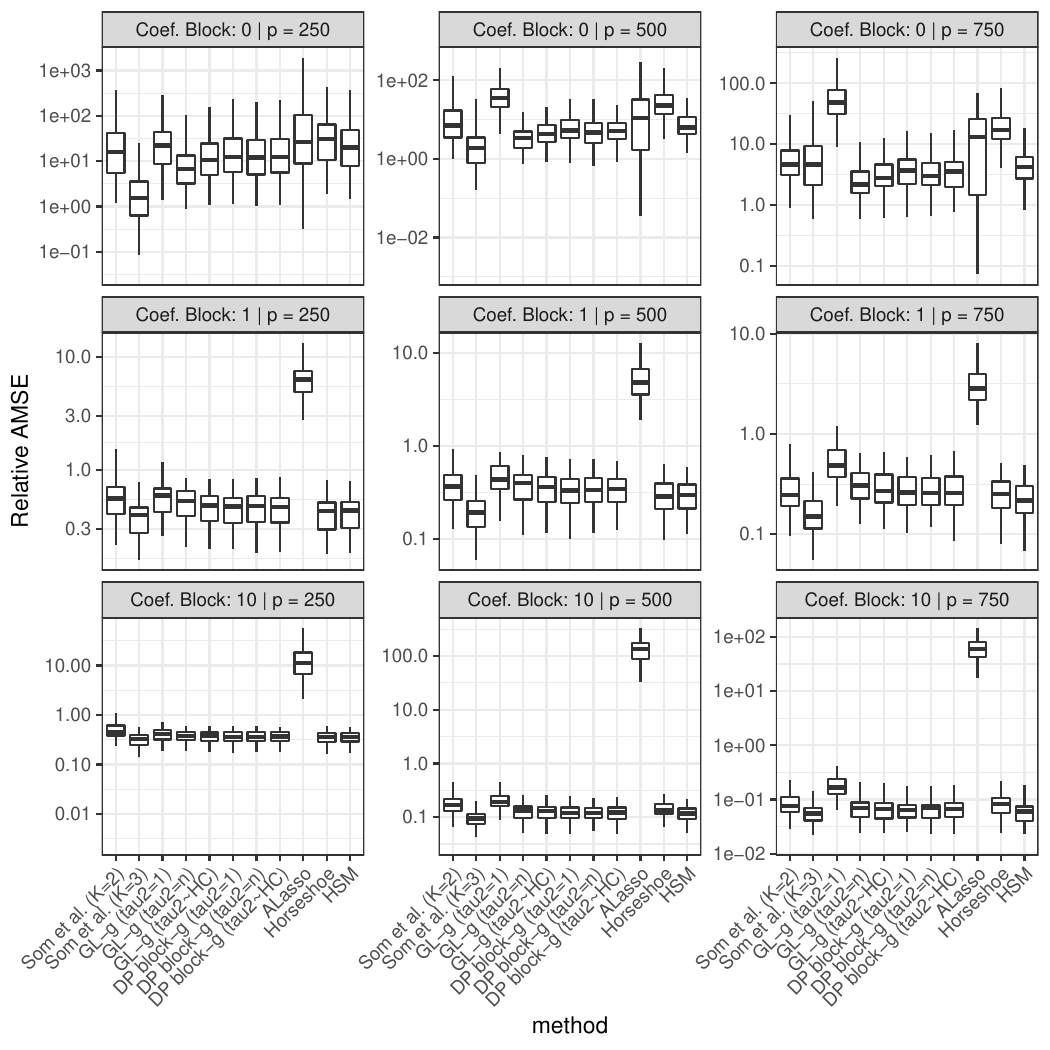}
    \caption{Relative mean squared error of the coefficients for $\eta = 0.9$, broken down by coefficient block.  Results are shown in the log scale because of the poor performance of ALasso.}
\label{fig:normMSE_beta_rho5}
\end{figure}

{ Next, Table \ref{tab:errortypes05} below is analogous to Table 1 of the main manuscript, but contains results for $\eta = 0.5$ (a middle ground between the cases $\eta=0$ and $\eta=0.9$).  These results are aligned with those presented in the main manuscript.  Note that the difference between the GL $g$ and DP block prior when $\eta=0.5$ is very small, which is similar to what we obtained for $\eta=0$. This suggests that the differences between these two methods arise for relatively high correlations among the covariates.

\begin{table}[]
    \centering
    \begin{subfloat}
    \centering 
    \tiny
    \scalebox{1}{
    \begin{tabular}{lccc} 
       &  \makecell{Power \\ (``large'' coeffs)}  &
        \makecell{Power \\ (``small'' coeffs)}  &
       \makecell{ Type I error \\(null coeffs)} \\ \cline{1-4}
       & \multicolumn{3}{c}{$p = 250$} \\ \cline{1-4}
       & \multicolumn{3}{c}{$\eta = 0.5$}\\ \cline{1-4}
g-prior ($\tau^2=$n) & 0.979 & 0.793 & 0.005 \\ 
  Som et al. (K=2) & 0.980 & 0.876 & 0.046 \\ 
  Som et al. (K=3) & 0.990 & 1.000 & 1.000 \\ 
  \midrule
  GL-$g$ ($\tau^2=$1) & 1.000 & 1.000 & 1.000 \\ 
  GL-$g$ ($\tau^2=$n) & 0.987 & 0.869 & 0.060 \\ 
  GL-$g$ ($\tau^2\sim$HC) & 0.987 & 0.865 & 0.056 \\ 
  DP block-g ($\tau^2=$1) & 0.990 & 0.903 & 0.246 \\ 
  DP block-g ($\tau^2=$n) & 0.988 & 0.867 & 0.056 \\ 
  DP block-g ($\tau^2\sim$HC) & 0.988 & 0.876 & 0.102 \\ 
  \midrule
    ALasso & 0.942 & 0.534 & 0.119 \\ 
  Horseshoe & 0.985 & 0.844 & 0.026 \\ 
  HSM & 0.982 & 0.813 & 0.006 \\ 

 \cline{1-4}
       & \multicolumn{3}{c}{$p = 500$} \\ \cline{1-4}
       & \multicolumn{3}{c}{$\eta = 0.5$}\\ \cline{1-4}
  g-prior ($\tau^2=$n) & 0.968 & 0.721 & 0.000 \\ 
  Som et al. (K=2) & 0.974 & 0.819 & 0.007 \\ 
  Som et al. (K=3) & 0.981 & 0.901 & 1.000 \\ 
  \midrule
  GL-$g$ ($\tau^2=$1) & 1.000 & 0.999 & 0.990 \\ 
  GL-$g$ ($\tau^2=$n) & 0.979 & 0.812 & 0.008 \\ 
  GL-$g$ ($\tau^2\sim$HC) & 0.979 & 0.811 & 0.008 \\ 
  DP block-g ($\tau^2=$1) & 0.980 & 0.826 & 0.025 \\ 
  DP block-g ($\tau^2=$n) & 0.979 & 0.814 & 0.009 \\ 
  DP block-g ($\tau^2\sim$HC) & 0.980 & 0.817 & 0.011 \\ 
\midrule
  ALasso & 0.581 & 0.050 & 0.028 \\ 
  Horseshoe & 0.974 & 0.781 & 0.011 \\ 
  HSM & 0.974 & 0.765 & 0.001 \\

 \cline{1-4}
       & \multicolumn{3}{c}{$p = 750$} \\ \cline{1-4}
       & \multicolumn{3}{c}{$\eta = 0.5$}\\ \cline{1-4}
  g-prior ($\tau^2=$n) & 0.942 & 0.479 & 0.000 \\ 
  Som et al. (K=2) & 0.975 & 0.793 & 0.006 \\ 
  Som et al. (K=3) & 0.980 & 0.844 & 0.502 \\ 
  \midrule
  GL-$g$ ($\tau^2=$1) & 0.990 & 0.896 & 0.474 \\ 
  GL-$g$ ($\tau^2=$n) & 0.980 & 0.783 & 0.004 \\ 
  GL-$g$ ($\tau^2\sim$HC) & 0.979 & 0.781 & 0.004 \\ 
  DP block-g ($\tau^2=$1) & 0.980 & 0.796 & 0.008 \\ 
  DP block-g ($\tau^2=$n) & 0.980 & 0.789 & 0.006 \\ 
  DP block-g ($\tau^2\sim$HC) & 0.980 & 0.788 & 0.007 \\ 
  \midrule
    ALasso & 0.553 & 0.041 & 0.023 \\ 
  Horseshoe & 0.973 & 0.725 & 0.005 \\ 
  HSM & 0.976 & 0.739 & 0.000 \\ 

 \cline{1-4}
    \end{tabular}}
    \end{subfloat}
    \vspace{2mm}
    \caption{Estimates of power for ``small'' (generated from a $\mathcal{N}(0,1)$ distribution) and ``large'' (generated from a $\mathcal{N}(0,10)$ distribution) coefficients, 
    and of type I error for null coefficients ($\beta=0$) in our second simulation study, for $\eta=0.5$.  For the purpose of this table, coefficients are considered ``significant'' is their posterior inclusion probability is greater than 0.5.}\label{tab:errortypes05}
\end{table}

Next, Figures \ref{fig:normMSE_beta_rho0} and \ref{fig:normMSE_beta_rho5} present the average mean squared error (AMSE) for the point estimators of the regression coefficients under various approaches (the posterior mean in the case of Bayesian procedures, and the argument of the penalized likelihood procedure for ALasso).  As was the case for prediction MSEs in the main document, all results are shown relative to the AMSE associated with the standard $g$ prior.  As before, ALasso is the worst performer, Horseshoe tends to perform poorly in ``large $p$'' scenarios, and the remaining Bayesian procedures consistently outperform the $g$ prior (except, perhaps, for the null coefficients on some of the datasets).  Perhaps surprisingly at first sight, Som et al.\ ($K=3$) outperforms the other procedures. While this result contrasts with what happened when carrying out variable selection (where Som et al.\ ($K=3$) was one of the worst performers), it is not surprising.  The overfitting of the shrinkage factors for null coefficients that led to issues in model selection becomes an advantage when it comes to the accuracy of point estimators.}

{ Finally, we study the sensitivity of procedures based on global-local $g$-priors and Dirichlet process mixtures of block $g$-priors to the hyperpriors on the concentration parameter $\alpha$ and on the vector of inclusion indicators $\bfgamma$.  To this effect, Table \ref{tab:errortypessensitivity} shows the power and type I error associated with DP block-$g$ ($\tau^2 = n$) and DP block-$g$ ($\tau^2 \sim \mathsf{HC}$) where $\alpha$ is given an Exponential distribution with mean 1, as well as results for GL-$g$ ($\tau^2=n$) and DP block-$g$ ($\tau^2 = n$) under a uniform prior on model space.  Note that the prior on $\alpha$ seems to have no effect on the power or type I errors.  On the other hand, the use of a uniform prior on model space seems to lead to slightly higher power for detecting smaller coefficients when $p \ll n$ and slightly lower power when $p \ge n$.  On average, this results in lower $F_1$ scores under a uniform prior on model space.
}

\begin{table}[hbt!]
    \centering
    \begin{subfloat}
    \centering 
    \scriptsize
    \scalebox{0.8}{
    \begin{tabular}{lccc|ccc} 
       &  \makecell{Power \\ (large coeffs)}  &
        \makecell{Power \\ (small coeffs)}  &
       \makecell{ Type I error \\(null coeffs)} &
              \makecell{Power \\ (large coeffs)} &
       \makecell{Power \\ (small coeffs)} &
       \makecell {Type I error \\ (null coeffs)}\\ \cline{1-7}
       & \multicolumn{6}{c}{$p = 250$} \\ \cline{1-7}
       & \multicolumn{3}{c|}{$\eta = 0$} & \multicolumn{3}{c}{$\eta = 0.9$} \\ \cline{1-7}
  DP block-g ($\tau^2=$n) Gamma & 0.990 & 0.900 & 0.043 & 0.974 & 0.755 & 0.111 \\ 
  DP block-g ($\tau^2/n\sim$HC) Gamma & 0.990 & 0.902 & 0.045 & 0.976 & 0.772 & 0.159 \\ 
  \midrule
  GL-$g$ ($\tau^2=$n) Uniform & 0.987 & 0.867 & 0.008 & 0.962 & 0.626 & 0.018 \\ 
  DP block-g ($\tau^2=$n) Uniform & 0.987 & 0.867 & 0.007 & 0.964 & 0.638 & 0.022 \\ 
 \cline{1-7}
       & \multicolumn{6}{c}{$p = 500$} \\ \cline{1-7}
       & \multicolumn{3}{c|}{$\eta = 0$} &  \multicolumn{3}{c}{$\eta = 0.9$}  \\ \cline{1-7}
  DP block-g ($\tau^2=$n) Gamma & 0.984 & 0.867 & 0.006 & 0.958 & 0.611 & 0.018 \\ 
  DP block-g ($\tau^2/n\sim$HC) Gamma & 0.984 & 0.868 & 0.005 & 0.958 & 0.614 & 0.019 \\ 
  \midrule
  GL-$g$ ($\tau^2=$n) Uniform & 0.985 & 0.874 & 0.010 & 0.959 & 0.635 & 0.024 \\ 
  DP block-g ($\tau^2=$n) Uniform & 0.986 & 0.879 & 0.018 & 0.963 & 0.661 & 0.047 \\ 
 \cline{1-7}
       & \multicolumn{6}{c}{$p = 750$} \\ \cline{1-7}
       & \multicolumn{3}{c|}{$\eta = 0$}  & \multicolumn{3}{c}{$\eta = 0.9$}  \\ \cline{1-7}
  DP block-g ($\tau^2=$n) Gamma & 0.985 & 0.854 & 0.003 & 0.950 & 0.534 & 0.009 \\ 
  DP block-g ($\tau^2/n\sim$HC) Gamma & 0.985 & 0.853 & 0.003 & 0.949 & 0.537 & 0.010 \\ 
  \midrule
  GL-$g$ ($\tau^2=$n) Uniform & 0.987 & 0.875 & 0.017 & 0.962 & 0.631 & 0.058 \\ 
  DP block-g ($\tau^2=$n) Uniform & 0.988 & 0.882 & 0.085 & 0.965 & 0.654 & 0.132 \\ 

 \cline{1-7}
    \end{tabular}}
    \end{subfloat}
    \vspace{2mm}
    \caption{Sensitivity analyses.  Estimates of power for small (generated from a $\mathcal{N}(0,1)$ distribution) and large (generated from a $\mathcal{N}(0,10)$ distribution) coefficients, 
    and of type I error for null coefficients ($\beta=0$) in our second simulation study.  For the purpose of this table, coefficients are considered ``significant'' is their posterior inclusion probability is greater than 0.5.}\label{tab:errortypessensitivity}
\end{table}

\newpage

\begingroup{
\singlespacing
\bibliographystyle{bka}
\bibliography{uwthesis}

@article{maclehose2010bayesian,
  title={Bayesian semiparametric multiple shrinkage},
  author={MacLehose, Richard F and Dunson, David B},
  journal={Biometrics},
  volume={66},
  number={2},
  pages={455--462},
  year={2010},
  publisher={Oxford University Press}
}

@article{sethuraman1994constructive,
  title={A constructive definition of Dirichlet priors},
  author={Sethuraman, Jayaram},
  journal={Statistica sinica},
  pages={639--650},
  year={1994},
  publisher={JSTOR}
}

@article{tipping2001sparse,
  title={Sparse {B}ayesian learning and the relevance vector machine},
  author={Tipping, Michael E},
  journal={Journal of Machine Learning Research},
  volume={1},
  number={Jun},
  pages={211--244},
  year={2001}
}

@article{boss2023group,
  title={Group Inverse-Gamma Gamma Shrinkage for Sparse Linear Models with Block-Correlated Regressors},
  author={Boss, Jonathan and Datta, Jyotishka and Wang, Xin and Park, Sung Kyun and Kang, Jian and Mukherjee, Bhramar},
  journal={Bayesian Analysis},
  volume={1},
  number={1},
  pages={1--30},
  year={2023},
  publisher={International Society for Bayesian Analysis}
}

@article{griffin2005alternative,
  title={Alternative prior distributions for variable selection with very many more variables than observations},
  author={Griffin, JE and Brown, PJ},
  journal={University of Kent Technical Report},
  year={2005}
}

@article{leng2014bayesian,
  title={Bayesian adaptive lasso},
  author={Leng, Chenlei and Tran, Minh-Ngoc and Nott, David},
  journal={Annals of the Institute of Statistical Mathematics},
  volume={66},
  number={2},
  pages={221--244},
  year={2014},
  publisher={Springer}
}

@article{bhattacharya2015dirichlet,
  title={Dirichlet--{L}aplace priors for optimal shrinkage},
  author={Bhattacharya, Anirban and Pati, Debdeep and Pillai, Natesh S and Dunson, David B},
  journal={Journal of the American Statistical Association},
  volume={110},
  number={512},
  pages={1479--1490},
  year={2015},
  publisher={Taylor \& Francis}
}

@article{finegold2014robust,
  title={Robust {B}ayesian graphical modeling using {D}irichlet t-distributions},
  author={Finegold, Michael and Drton, Mathias},
  journal={Bayesian Analysis},
  volume={9},
  number={3},
  pages={521--550},
  year={2014},
  publisher={International Society for Bayesian Analysis}
}

@article{piironen2017sparsity,
  title={Sparsity information and regularization in the horseshoe and other shrinkage priors},
  author={Piironen, Juho and Vehtari, Aki and others},
  journal={Electronic Journal of Statistics},
  volume={11},
  number={2},
  pages={5018--5051},
  year={2017},
  publisher={The Institute of Mathematical Statistics and the Bernoulli Society}
}

@article{breiman1985estimating,
  title={Estimating optimal transformations for multiple regression and correlation},
  author={Breiman, Leo and Friedman, Jerome H},
  journal={Journal of the American statistical Association},
  volume={80},
  number={391},
  pages={580--598},
  year={1985},
  publisher={Taylor \& Francis}
}

@article{zellner1980posterior,
  title={Posterior odds ratios for selected regression hypotheses},
  author={Zellner, Arnold and Siow, Aloysius},
  journal={Trabajos de Estad{\'i}stica y de Investigaci{\'o}w Operativa},
  volume={31},
  number={1},
  pages={585--603},
  year={1980},
  publisher={Springer}
}

@article{hans2009bayesian,
  title={Bayesian lasso regression},
  author={Hans, Chris},
  journal={Biometrika},
  volume={96},
  number={4},
  pages={835--845},
  year={2009},
  publisher={Oxford University Press}
}

@article{carvalho2010horseshoe,
  title={The horseshoe estimator for sparse signals},
  author={Carvalho, Carlos M and Polson, Nicholas G and Scott, James G},
  journal={Biometrika},
  volume={97},
  number={2},
  pages={465--480},
  year={2010},
  publisher={Oxford University Press}
}

@article{liu2012rejection,
  title={Rejection sampling for an extended Gamma distribution},
  author={Liu, Ying and Wichura, Michael J and Drton, Mathias},
  journal={Unpublished manuscript},
  year={2012}
}

@misc{abramowitz1988handbook,
  title={Handbook of mathematical functions with formulas, graphs, and mathematical tables},
  author={Abramowitz, Milton and Stegun, Irene A and Romer, Robert H},
  year={1988},
  publisher={American Association of Physics Teachers}
}

@article{porwal2022effect,
  title={Effect of Model Space Priors on Statistical Inference with Model Uncertainty},
  author={Porwal, Anupreet and Raftery, Adrian E},
  journal={The New England Journal of Statistics in Data Science},
  pages={1--10},
  year={2022},
  publisher={New England Statistical Society}
}

@article{porwal2023laplace,
  title={Laplace Power-Expected-Posterior Priors for Logistic Regression},
  author={Porwal, Anupreet and Rodr{\'\i}guez, Abel},
  journal={Bayesian Analysis},
  volume={1},
  number={1},
  pages={1--24},
  year={2023},
  publisher={International Society for Bayesian Analysis}
}

@article{som2016conditional,
  title={A conditional {L}indley paradox in {B}ayesian linear models},
  author={Som, Agniva and Hans, Christopher M and MacEachern, Steven N},
  journal={Biometrika},
  volume={103},
  number={4},
  pages={993--999},
  year={2016},
  publisher={Oxford University Press}
}

@phdthesis{som2014paradoxes,
  title={Paradoxes and Priors in Bayesian Regression},
  author={Som, Agniva},
  year={2014},
  school={The Ohio State University}
}

@article{gneiting2007strictly,
  title={Strictly proper scoring rules, prediction, and estimation},
  author={Gneiting, Tilmann and Raftery, Adrian E},
  journal={Journal of the American Statistical Association},
  volume={102},
  number={477},
  pages={359--378},
  year={2007},
  publisher={Taylor \& Francis}
}

@Manual{Horseshoe,
    title = {horseshoe: Implementation of the Horseshoe Prior},
    author = {Stephanie {van der Pas} and James Scott and Antik Chakraborty and Anirban Bhattacharya},
    year = {2019},
    note = {R package version 0.2.0},
    url = {https://CRAN.R-project.org/package=horseshoe},
  }

@article{george1997approaches,
  title={Approaches for {B}ayesian variable selection},
  author={George, Edward I and McCulloch, Robert E},
  journal={Statistica Sinica},
  volume={7},
  number={2},
  pages={339--373},
  year={1997},
}

@article{green1995reversible,
  title={Reversible jump {M}arkov chain {M}onte {C}arlo computation and {B}ayesian model determination},
  author={Green, Peter J},
  journal={Biometrika},
  volume={82},
  number={4},
  pages={711--732},
  year={1995},
  publisher={Oxford University Press}
}

@article{kass1995reference,
  title={A reference {B}ayesian test for nested hypotheses and its relationship to the {S}chwarz criterion},
  author={Kass, Robert E and Wasserman, Larry},
  journal={Journal of the American Statistical Association},
  volume={90},
  number={431},
  pages={928--934},
  year={1995},
  publisher={Taylor \& Francis}
}

@article{Forte2018,
  title={Methods and tools for {B}ayesian variable selection and model averaging in normal linear regression},
  author={Forte, Anabel and Garcia-Donato, Gonzalo and Steel, Mark F. J.},
  journal={International Statistical Review},
  volume={86},
  number={2},
  pages={237--258},
  year={2018},
  publisher={Wiley Online Library}
}

@article{park2008bayesian,
  title={The {B}ayesian lasso},
  author={Park, Trevor and Casella, George},
  journal={Journal of the American Statistical Association},
  volume={103},
  number={482},
  pages={681--686},
  year={2008},
  publisher={Taylor \& Francis}
}

@article{bayarri2012criteria,
  title={Criteria for {B}ayesian model choice with application to variable selection},
  author={Bayarri, Maria J and Berger, James O and Forte, Anabel and Garc{\'\i}a-Donato, G and others},
  journal={The Annals of Statistics},
  volume={40},
  number={3},
  pages={1550--1577},
  year={2012},
  publisher={Institute of Mathematical Statistics}
}

@article{polson2012local,
  title={Local shrinkage rules, L{\'e}vy processes and regularized regression},
  author={Polson, Nicholas G and Scott, James G},
  journal={Journal of the Royal Statistical Society: Series B (Statistical Methodology)},
  volume={74},
  number={2},
  pages={287--311},
  year={2012},
  publisher={Wiley Online Library}
}

@article{fouskakis2015power,
  title={Power-expected-posterior priors for variable selection in {G}aussian linear models},
  author={Fouskakis, Dimitris and Ntzoufras, Ioannis and Draper, David},
  journal={Bayesian Analysis},
  volume={10},
  number={1},
  pages={75--107},
  year={2015},
  publisher={International Society for Bayesian Analysis}
}

@article{li2017variable,
  title={Variable selection using shrinkage priors},
  author={Li, Hanning and Pati, Debdeep},
  journal={Computational Statistics \& Data Analysis},
  volume={107},
  pages={107--119},
  year={2017},
  publisher={Elsevier}
}

@article{carvalho2009objective,
  title={Objective {B}ayesian model selection in {G}aussian graphical models},
  author={Carvalho, Carlos M and Scott, James G},
  journal={Biometrika},
  volume={96},
  number={3},
  pages={497--512},
  year={2009},
  publisher={Oxford University Press}
}

@article{neal2000markov,
  title={Markov chain sampling methods for Dirichlet process mixture models},
  author={Neal, Radford M},
  journal={Journal of computational and graphical statistics},
  volume={9},
  number={2},
  pages={249--265},
  year={2000},
  publisher={Taylor \& Francis}
}

@article{li2018mixtures,
  title={Mixtures of g-priors in generalized linear models},
  author={Li, Yingbo and Clyde, Merlise A},
  journal={Journal of the American Statistical Association},
  volume={113},
  number={524},
  pages={1828--1845},
  year={2018},
  publisher={Taylor \& Francis}
}

@article{liang2008mixtures,
  title={Mixtures of g-priors for {B}ayesian variable selection},
  author={Liang, Feng and Paulo, Rui and Molina, German and Clyde, Merlise A and Berger, Jim O},
  journal={Journal of the American Statistical Association},
  volume={103},
  number={481},
  pages={410--423},
  year={2008},
  publisher={Taylor \& Francis}
}

@article{bove2011hyper,
  title={Hyper-$g$ priors for generalized linear models},
  author={Bov{\'e}, Daniel Saban{\'e}s and Held, Leonhard},
  journal={Bayesian Analysis},
  volume={6},
  number={3},
  pages={387--410},
  year={2011},
  publisher={International Society for Bayesian Analysis}
}

@article{lee2020continuous,
  title={Continuous shrinkage prior revisited: a collapsing behavior and remedy},
  author={Lee, Se Yoon and Pati, Debdeep and Mallick, Bani K},
  journal={arXiv preprint arXiv:2007.02192},
  year={2020}
}

@article{bai2018beta,
  title={On the Beta Prime Prior for Scale Parameters in High-Dimensional {B}ayesian Regression Models},
  author={Bai, Ray and Ghosh, Malay},
  journal={arXiv preprint arXiv:1807.06539},
  year={2018}
}

@article{brown2010inference,
  title={Inference with normal-gamma prior distributions in regression problems},
  author={Brown, Philip J and Griffin, Jim E},
  journal={Bayesian Analysis},
  volume={5},
  number={1},
  pages={171--188},
  year={2010},
  publisher={International Society for Bayesian Analysis}
}

@article{lee2020tail,
  title={Tail-adaptive {B}ayesian shrinkage},
  author={Lee, Se Yoon and Zhao, Peng and Pati, Debdeep and Mallick, Bani K},
  journal={Electronic Journal of Statistics},
  volume={18},
  number={2},
  pages={4667--4723},
  year={2024},
  publisher={The Institute of Mathematical Statistics and the Bernoulli Society}
}

@Manual{BAS,
    title = {{BAS}: Bayesian Variable Selection and Model Averaging using Bayesian
Adaptive Sampling},
    author = {Merlise Clyde},
    year = {2020},
    note = {R package version 1.5.5},
  }

@article{denti2021horseshoe,
  title={A horseshoe mixture model for Bayesian screening with an application to light sheet fluorescence microscopy in brain imaging},
  author={Denti, Francesco and Azevedo, Ricardo and Lo, Chelsie and Wheeler, Damian G and Gandhi, Sunil P and Guindani, Michele and Shahbaba, Babak},
  journal={The Annals of Applied Statistics},
  volume={17},
  number={3},
  pages={2639--2658},
  year={2023},
  publisher={Institute of Mathematical Statistics}
}

@article{consonni2018prior,
  title={Prior distributions for objective {B}ayesian analysis},
  author={Consonni, Guido and Fouskakis, Dimitris and Liseo, Brunero and Ntzoufras, Ioannis and others},
  journal={Bayesian Analysis},
  volume={13},
  number={2},
  pages={627--679},
  year={2018},
  publisher={International Society for Bayesian Analysis}
}

@article{scott2010bayes,
  title={Bayes and empirical-{B}ayes multiplicity adjustment in the variable-selection problem},
  author={Scott, James G and Berger, James O},
  journal={The Annals of Statistics},
  pages={2587--2619},
  year={2010},
  publisher={JSTOR}
}

@article{johnson2012bayesian,
  title={Bayesian model selection in high-dimensional settings},
  author={Johnson, Valen E and Rossell, David},
  journal={Journal of the American Statistical Association},
  volume={107},
  number={498},
  pages={649--660},
  year={2012},
  publisher={Taylor \& Francis Group}
}

@article{johnson2010use,
  title={On the use of non-local prior densities in {B}ayesian hypothesis tests},
  author={Johnson, Valen E and Rossell, David},
  journal={Journal of the Royal Statistical Society: Series B (Statistical Methodology)},
  volume={72},
  number={2},
  pages={143--170},
  year={2010},
  publisher={Wiley Online Library}
}

@incollection{Zellner1986,
  author  = "Zellner, Arnold",
  title   = "On Assessing Prior Distributions and {B}ayesian Regression Analysis With g-Prior Distributions",
  year    = 1986,
  booktitle = "Bayesian Inference and Decision Techniques: Essays in Honor of Bruno de Finetti",
  editor  = "P. K. Goel and A. Zellner",
  publisher = "North-Holland/Elsevier",
  address   = "Amsterdam",
  pages   = "233--243"
}

@incollection{berger1996linear,
  author  = "Berger, James O. and Pericchi, Luis R.",
  title   = "The intrinsic {B}ayes factor for linear models",
  year    = 1996,
  booktitle = "Bayesian Statistics 5",
  editor  = "J. M. Bernardo, J. O. Berger, A. P. Dawid and A. F. M. Smith",
  publisher = "Oxford Univ. Press",
  pages   = "25--44"
}

@article{berger1998bayes,
  title={Bayes factors and marginal distributions in invariant situations},
  author={Berger, James O and Pericchi, Luis R and Varshavsky, Julia A},
  journal={Sankhy{\=a}: The Indian Journal of Statistics, Series A},
  pages={307--321},
  year={1998},
  publisher={JSTOR}
}

@article{andrade2011bayesian,
  title={Bayesian robustness modelling of location and scale parameters},
  author={Andrade, Jose Ailton Alencar and O'Hagan, Anthony},
  journal={Scandinavian Journal of Statistics},
  volume={38},
  number={4},
  pages={691--711},
  year={2011},
  publisher={Wiley Online Library}
}

@techreport{gordy1998generalization,
  author={Gordy, Michael B},
  title={A generalization of generalized {B}eta distributions},
  year={1998},
  institution = {Division of Research and Statistics, Division of Monetary Affairs, Federal Reserve}
}

@book{berger1985statistical,
  title={Statistical decision theory and Bayesian analysis},
  author={Berger, James O},
  year={1985},
  publisher={Springer Science \& Business Media},
  address={New York}
}

@article{ferguson1973bayesian,
  title={A Bayesian analysis of some nonparametric problems},
  author={Ferguson, Thomas S},
  journal={The annals of statistics},
  pages={209--230},
  year={1973},
  publisher={JSTOR}
}

@article{rodriguez2013jeffreys,
  title={On the Jeffreys prior for the multivariate Ewens distribution},
  author={Rodr{\'\i}guez, Abel},
  journal={Statistics \& Probability Letters},
  volume={83},
  number={6},
  pages={1539--1546},
  year={2013},
  publisher={Elsevier}
}

@article{blackwell1973ferguson,
  title={Ferguson distributions via P{\'o}lya urn schemes},
  author={Blackwell, David and MacQueen, James B},
  journal={The annals of statistics},
  volume={1},
  number={2},
  pages={353--355},
  year={1973},
  publisher={Institute of Mathematical Statistics}
}

@article{antoniak1974mixtures,
  title={Mixtures of Dirichlet processes with applications to Bayesian nonparametric problems},
  author={Antoniak, Charles E},
  journal={The annals of statistics},
  pages={1152--1174},
  year={1974},
  publisher={JSTOR}
}

@book{bertoin2006random,
  title={Random fragmentation and coagulation processes},
  author={Bertoin, Jean},
  volume={102},
  year={2006},
  publisher={Cambridge University Press}
}

@article{berger2009formal,
  title={The formal definition of reference priors},
  author={Berger, James O and Bernardo, Jos{\'e} M and Sun, Dongchu},
  year={2009},
  journal={Annals of Statistics},
  volume={37},
  pages={905-938}
}

@article{huang2008adaptive,
  title={Adaptive Lasso for sparse high-dimensional regression models},
  author={Huang, Jian and Ma, Shuangge and Zhang, Cun-Hui},
  journal={Statistica Sinica},
  pages={1603--1618},
  year={2008},
  publisher={JSTOR}
}

@article{casella2006objective,
  title={Objective Bayesian variable selection},
  author={Casella, George and Moreno, Elias},
  journal={Journal of the American Statistical Association},
  volume={101},
  number={473},
  pages={157--167},
  year={2006},
  publisher={Taylor \& Francis}
}
}
\endgroup

\end{document}